\documentclass[twoside,12pt]{article}
\usepackage{epsfig}
\usepackage{graphics}
\usepackage{latexsym}
\usepackage{amsmath}
\usepackage{units}

\newif\iflanl\lanltrue

\def\Journal#1#2#3#4{{#1} {\bf #2} (#4) #3 }

\def\PLA{{\em Phys. Lett.} A}
\def\PLB{{\em Phys. Lett.} B}

\def\PRL{\em Phys. Rev. Lett.}
\def\PRL{\em Phys. Rev. Lett.}
\def\PREV{\em Phys. Rev.}

\def\PRA{{\em Phys. Rev.} A}
\def\PRD{{\em Phys. Rev.} D}

\def\RMP{{\em Rev. Mod. Phys.}}

\def\IJMPD{{\em Int. J. Mod. Phys.} D}
\def\APJ{\em Astroph. J.}
\def\ASJ{\em Astron. J.}
\def\APJL{\em Astroph. J. Lett.}
\def\ASTAST{\em Astron. \&\ Astroph.}
\def\CQG{\em Class. Quant. Grav.}
\def\NAR{\em New Astron. Rev.}
\def\NAT{\em Nature}
\def\NATP{\em Nature Phys.}
\def\NIMA{{\em Nuc. Inst. Meth.} A}
\def\SOVA{\em Sov. Astron.}
\def\MNRAS{\em Mon. Not. Roy. Astron. Soc}
\def\SCI{\em Science}
\def\PTRSLA{{\em Phil. Trans. R. Soc. Lon.} A}
\def\RPP{\em Rep. Prog. Phys.}
\def\PNAS{\em Proc. Nat. Acad. Sci.}
\def\PCPS{\em  Proc. Cam. Phil. Soc.}
\def\ARAA{\em  Ann. Rev. Astron. Astroph.}
\def\NJP{\em New J. Phys.}
\def\PT{\em Phys. Today}
\def\OL{\em Opt. Lett.}
\def\PASP{\em Pubs. Astron. Soc. Pacif.}
\def\PASA{\em Pubs. Astron. Soc. Australia}
\def\PSPIE{\em Proc. SPIE}
\def\IEEE{\em IEEE Proc.}

\def\JPCS{\em J. Phys. Conf. Ser.}
\def\AJP{\em Amer. J. Phys.}
\def\RSI{\em Rev. Sci. Inst.}
\def\APB{{\em Appl. Phys.} B}
\def\APP{\em Astropart. Phys.}
\def\JMO{\em J. Mod. Opt.}
\def\AO{\em Appl. Opt.}
\def\LP{\em Las. Phys.}
\def\JAP{\em J. Appl. Phys.}
\def\SJETP{\em Sov. J. Exp. \&\ Theor. Phys.}
\def\JETPL{\em J. Exp. \&\ Theor. Phys. Lett.}
\def\JGR{\em J. Geophys. Res.}

\def\kb{k_{\rm B}}
\newcommand{\be}{\begin{equation}}
\newcommand{\ee}{\end{equation}}
\newcommand{\bea}{\begin{eqnarray}}
\newcommand{\eea}{\end{eqnarray}}

\def\etal{{\it et al.}}
\def\ie{{\it i.e.}}
\def\eg{{\it e.g.}}
\def\vs{{\it vs.}}
\def\etc{{\it etc.}}

\def\apriori{{\it a priori}}
\def\vv{{\it vice-versa}}

\def\aba{J.~Abadie \etal}
\def\bab{B.~Abbott \etal}

\def\msolar{M_\odot}
\def\hrss{h_{\textrm{rss}}}

\def\galfbet{g_{\alpha\beta}}
\def\aalfbet{a_{\alpha\beta}}
\def\galfbetcart{g_{\alpha\beta}^{\rm Cart.}}
\def\galfbetschwarz{g_{\alpha\beta}^{\rm Schwarz.}}
\def\chrisabg{\Gamma^\alpha_{\beta\gamma}}
\def\xbold{{\bf x}}
\def\Talfbet{T_{\alpha\beta}}
\def\halfbet{h_{\alpha\beta}}
\def\etaalfbet{\eta_{\alpha\beta}}
\def\hplus{h_+}
\def\tcoal{t_{\rm coal}}
\def\tdelay{t_{\rm delay}}
\def\fgw{f_{\rm GW}}
\def\hcross{h_\times}
\def\mchirp{M_{\rm chirp}}
\def\taubar{\bar\tau}
\def\Islash{{\lower.5ex\hbox{$\mathchar'26$}\mkern-9muI}}
\def\fstatistic{$F$-statistic}

\begin{document}

\title{ \vspace{1cm} Gravitational Waves: Sources, Detectors and Searches}
\author{K.\ Riles,$^1$
\\
$^1$Physics Department, University of Michigan}
\iflanl
\date{Published in {\it Progress in Particle \&\ Nuclear Physics 68 (2013) 1}\\
\strut Originally submitted July 1, 2012\\
This revision: January 31, 2013}
\fi
\maketitle
\begin{abstract} 

Gravitational wave science should transform in this decade from
a study of what has {\it not} been seen to
a full-fledged field of astronomy in which detected signals reveal the
nature of cataclysmic events and exotic objects.
The LIGO Scientific Collaboration and Virgo Collaboration
have recently completed joint data runs of unprecedented sensitivities
to gravitational waves. So far, no gravitational radiation has
been seen (although data mining continues). It seems likely that the
first detection will come from 2nd-generation LIGO and Virgo interferometers now
being installed. These new detectors are expected to detect $\sim$40 coalescences 
of neutron star binary systems per year at full sensitivity.
At the same time, research and development continues on 3rd-generation
underground interferometers and on space-based interferometers.
In parallel there is a vigorous effort in
the radio pulsar community to detect $\sim$several-nHz gravitational waves via
the timing residuals from an array of pulsars at different
locations in the sky. As the dawn of gravitational wave astronomy
nears, this review, intended primarily for interested particle and nuclear
physicists, describes what we have learned to date and the prospects
for direct discovery of gravitational waves.
\end{abstract}
\section{Introduction}
\label{sec:intro}

Einstein's General Theory of Relativity (hereafter: general relativity) predicts the existence of
gravitational waves, disturbances of space-time itself that propagate at the
speed of light and have two transverse quadrupolar 
polarizations~\cite{bib:thorne300}. Scientists have searched for these
waves for several decades without success, but with the ongoing 
installation and commissioning of Advanced LIGO and Advanced Virgo detectors,
direct discovery of gravitational waves appears to be only a few
years away (or perhaps sooner if deep mining of initial LIGO and Virgo
data succeeds). There is also a possibility that radio astronomers will
succeed first in direct detection of gravitational waves -- at extremely
low frequencies ($\sim$ several nHz) via their influence on apparent pulsar timing.
This review article is intended for an audience of nuclear and particle
physicists already conversant in special relativity and electrodynamics 
who want to understand why gravitational waves are interesting,
what technical and analytic methods are used in direct searches for this
predicted radiation, and what the prospects are for discovery in this
decade. Particular attention is given to the evolving data analysis techniques
in this rapidly developing field.

The search for gravitational waves has many motivations. First is simple, fundamental scientific 
curiosity about new phenomena. More prosaically, one can use gravitational radiation to
test general relativity. For example, one can test the predicted transverse and quadrupolar nature of
the radiation, and one can test whether or
not the radiation travels at the speed of light, as one would expect for a massless graviton.
One can also directly probe highly relativistic phenomena, such as
black-hole formation. Perhaps more intriguing, though, is the entirely new view one gains of the universe. 
Gravitational waves cannot be appreciably absorbed by dust or stellar envelopes, 
and most detectable sources are some of the most interesting and least understood
objects in the universe. More generally,
gravitational wave astronomy opens up an entirely new non-electromagnetic spectrum.
Astronomy has found many surprises since the mid 20th-century, as non-optical light bands have
been explored, from the radio to gamma rays. New surprises likely await
in the exploration of the gravitational spectrum.
We do not yet know the sky distribution of detectable sources, but it is likely to
include both isotropic components from sources at cosmological distances and local components
dominated by our own galactic plane.

Strong indirect evidence already exists for gravitational wave emission.
The famous Hulse-Taylor binary system, consisting of an observed pulsar
with 17-Hz radio emission in an 8-hour orbit with an unseen neutron star
companion, has shown a small but unmistakable quadratic decrease in orbital period
($\sim$40 seconds over 30 years), in remarkably good agreement with
expectation from gravitational wave energy loss~\cite{bib:weisbergtaylor}. 
The 1993 Nobel Prize in Physics was awarded to Taylor
and Hulse for the discovery and use of the PSR B1913+16 
system to test general relativity,
in particular, for the verification of an orbital decay rate consistent with
that expected from gravitational wave energy emission.
Its presumed gravitational wave emission frequency ($\sim$70 $\mu$Hz) is far too 
low to be observed directly by present gravitational wave detectors, but
if we were to wait about 300 million years, the system would
eventually spiral into a spectacular coalescence easily ``heard'' with
present gravitational wave detectors.

Perhaps our best hope for gravitational wave discovery lies with corresponding
binary systems in the numerous galaxies far away from us, but there are large 
uncertainties in estimated coalescence rates for compact binary 
systems containing neutron stars (NS) and/or black holes (BH). For example, a
recent compilation of rates~\cite{bib:cbcratespaper} estimates a ``plausible''
range from $2\times10^{-4}$ to 0.2 per year for initial LIGO detection of 
a NS-NS coalescence, $7\times10^{-5}$ to 0.1 per year for a NS-BH coalescence, 
and $2\times10^{-4}$ to 0.5 per year for BH-BH coalescence,
assuming 1.4-$\msolar$ NS and $\sim$10-$\msolar$ BH ($\msolar$ $\equiv$ solar mass = $2.0\times10^{30}$ kg).
As discussed below, these predicted rates increase dramatically
for 2nd-generation detectors. For example, a realistic estimate
for Advanced LIGO at full sensitivity is 40 detected NS-NS coalescences per year.
Estimates for NS-BH and BH-BH inspiral rates have especially
large uncertainties~\cite{bib:cbcratespaper}, but such systems could be observed
at farther distances because their larger masses give rise to large 
gravitational wave amplitudes in the final stages of the inspiral.
Similarly, the waveform shapes for inspiraling binary systems are thought to be well understood
for systems with two neutron star systems, while larger uncertainties apply to
systems with one or two black holes. Nonetheless, there has been recent dramatic progress
in numerical-relativity calculations of expected waveforms in these complicated
systems.

Other candidate transient sources of gravitational waves include supernovae
and gamma ray bursts (some of which may well be coalescing binary systems).
If we are fortunate, electromagnetic transients will be seen simultaneously
by other astronomers, allowing more confident gravitational wave detection
with lower signal-to-noise ratio (SNR) and yielding greater understanding
of the sources. Potential non-transient gravitational wave 
sources include rapidly spinning neutron stars in our own galaxy, emitting 
long-lived continuous waves, or a cosmological background of stochastic
gravitational waves, analogous to the cosmic microwave background radiation.
Results from searches for both transient and long-lived
gravitational-wave sources will be discussed below.

In the following review, the primary focus will be upon current
ground-based gravitational wave interferometers and on 2nd-generation interferometers now
under construction, with brief mention of future, 3rd-generation underground interferometers. 
Attention will also be given to searches for extremely
low-frequency waves via pulsar timing. 
Longer-term searches using space-based
interferometers will be discussed only briefly.
There will be no discussion of attempts to detect gravitational
waves indirectly via their primordial imprint upon the cosmic microwave background~\cite{bib:cmbrgw}
or of nascent ideas for direct detection based upon matter-wave interferometry~\cite{bib:matterwaveifo}.

In addition to the references cited below for specific topics, there
exist many informative books and review articles. Comprehensive texts specific
to gravitational waves include
Saulson~\cite{bib:saulsontext}, Maggiore~\cite{bib:maggioretext} 
Creighton \&\ Anderson~\cite{bib:jolienwarren}, and Jaranowski and Krol\'ak~\cite{bib:jktext}.
Texts on gravitation and general relativity with treatments of gravitational radiation 
include (among others) Hartle~\cite{bib:hartletext}, 
Misner, Thorne \&\ Wheeler~\cite{bib:mtwtext}, and
Schutz~\cite{bib:schutztext}.
Early influential review
articles concerning gravitational waves include Tyson \&\ Giffard~\cite{bib:tysongiffard},
Thorne~\cite{bib:thorne300} and the collection of articles
in Blair~\cite{bib:blairbook1}.  Reviews within the last few years include
Sathyaprakash \&\ Schutz (2009)~\cite{bib:sathyaschutz},
Pitkin \etal\ (2011)~\cite{bib:pitkin}, and Freise \&\ Strain (2010)~\cite{bib:freisestrain}. 
A very recent volume (2012) contains detailed articles on plans for 2nd-generation
and 3rd-generation detectors~\cite{bib:blairbook2}.

One technical note: to minimize confusion, physical units in the following
will be S.I. primarily, with occasional cgs conversions shown where useful
(apologies in advance to those who prefer 
streamlined equations with G = c = $\hbar$ = $k_B$ $\equiv$ 1).

\section{Gravitational Wave Sources}
\label{sec:gwsources}

\subsection{Gravitational wave generation and properties}
\label{sec:gwgeneration}

Strictly speaking, gravitational waves which describe fluctuations in the curvature
of space cannot be rigorously separated from other curvature caused, for example, by 
a nearby star. Nonetheless, 
one can usefully (and accurately) apply a short-wave 
formalism to separate rapid variations from a slowly varying background~\cite{bib:thorne300,bib:isaacson}
by taking the background as an average over many wavelengths of the wave.

Measuring curvature requires two or more
separated test objects. A classic example is that of an astronaut's observations while orbiting the
Earth in a windowless spacecraft. Observing the slow relative drift of two test masses
placed initially at rest w.r.t. each other at a nominal initial separation allows
the astronaut to detect the tidal influence of the Earth upon local space-time.
The same principle applies to detecting a gravitational wave; in the case of
the LIGO interferometers, for example, 
one measures via light propagation time the influence of gravitational
waves on pairs of test masses (mirrors) separated by 4 km.

The following brief and simplified summary of the generation and propagation
of gravitational waves borrows heavily from the treatment of the Hartle text~\cite{bib:hartletext}
in which analogs with electromagnetic radiation are made manifest.

First, let's review some fundamental concepts from general relativity.
The differential line element $ds$ at space-time point $\xbold$ has the form:
\begin{equation}
ds^2\quad = \quad \galfbet(\xbold)\,dx^\alpha\, dx^\beta
\end{equation}
where $\galfbet$ is the symmetric metric tensor, and repeated indices imply summation. Two examples are a
flat Cartesian-coordinate metric ($\alpha = (t,x,y,z)$):
\begin{equation}
\galfbetcart(\xbold) = 
\left(\begin{array}{cccc} -1 & 0 & 0 & 0 \\ 0 & 1 & 0 & 0 \\ 0 & 0 & 1 & 0 \\ 0 & 0 & 0 & 1\end{array}\right)
\label{eqn:minkowskimetric}
\end{equation}
and a curved, spherical-coordinate Schwarzschild 
metric ($\alpha=(t,r,\theta,\phi)$) 
exterior to a spherically symmetric mass distribution of total mass $M$:
\begin{equation}
\galfbetschwarz(\xbold) = 
\left(\begin{array}{cccc} -(1-2\,GM/ c^2r) & 0 & 0 & 0 \\ 0 & (1+2GM/ c^2r)^{-1} & 0 & 0\\
0 & 0 & r^2 & 0 \\ 0 & 0 & 0 & r^2\sin^2(\theta) \end{array}\right).
\end{equation}
One way to quantify the curvature of a metric is via the covariant equation of
motion for a test particle:
\begin{equation}
\frac{d^2x^\alpha}{d\tau^2} \quad = \quad -\chrisabg \, \frac{dx^\beta}{d\tau}\,\frac{dx^\gamma}{d\tau}
\end{equation}
where $\tau$ is proper time and $\chrisabg$ is the Christoffel symbol defined by
\begin{equation}
g_{\alpha\delta}\,\Gamma^\delta_{\beta\gamma}\quad = \quad {1\over2}\left(
{\partial g_{\alpha\beta}\over\partial x^\gamma}+
{\partial g_{\alpha\gamma}\over\partial x^\beta}+
{\partial g_{\beta\gamma}\over\partial x^\alpha}\right)
\end{equation}
and from which one can define the Riemann curvature tensor:
\begin{equation}
R^\alpha_{\beta\gamma\delta} \quad = \quad {\partial\Gamma^\alpha_{\beta\delta}\over\partial x^\gamma}
-{\partial\Gamma^\alpha_{\beta\gamma}\over\partial x^\delta}
+\Gamma^\alpha_{\gamma\epsilon}\Gamma^\epsilon_{\beta\delta}
-\Gamma^\alpha_{\delta\epsilon}\Gamma^\epsilon_{\beta\gamma}\,.
\end{equation}
\noindent Contracting two of the indices of the Riemann tensor leads to
the Ricci tensor:
\begin{equation}
R_{\beta\delta} \quad = \quad R^\alpha_{\beta\alpha\delta}\>,
\end{equation}
which appears in the famous Einstein Equation:
\begin{equation}
\label{eqn:einsteinequation}
R_{\alpha\beta} - {1\over2}\galfbet R \quad = \quad {8\pi G\over {\rm c}^4}\, \Talfbet,
\end{equation}
where $\Talfbet$ is the stress-energy tensor which can
be regarded as having the following qualitative form:
\begin{equation}
\Talfbet = \left(\begin{array}{c|c} {\rm Energy\> Density} & {1\over{\rm c}}\strut({\rm Energy\> Flux}) \\\noalign{\vskip2pt} \cline{1-2}\\ {1\over{\rm c}}({\rm Momentum} & {\rm Stress} \\ {\rm Density}) & {\rm Tensor}\\ \\ \end{array}\right).
\end{equation}
Specifically, $T^{tt}(\xbold)$ is the local energy density,
$T^{ti}(\xbold)$ is the flux of energy in the $x^i$ direction,
$T^{it}$ is the density of momentum in the $x^i$ direction
(note: $T^{ti}=T^{it}$), and $T^{ij}$ is the $i$th component
of the force per unit area exerted across a surface
with normal in direction $x^j$. Diagonal elements
$T^{ii}$ represent pressure components, and off-diagonal
elements represent shear stresses. Local energy and momentum
conservation (in flat space-time) can be represented by the equation:
\begin{equation}
\label{eqn:epconservation}
{\partial\Talfbet\over\partial x^\beta}\quad =\quad 0\,.
\end{equation}

\begin{figure}[tb]
\begin{center}
\epsfig{file=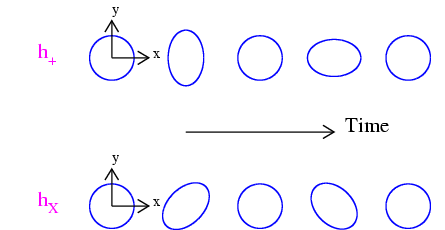,scale=0.5}
\caption{Illustration of the effects of a (strong!) gravitational wave
passage upon a ring of free test masses.
\label{fig:ringofmasses}}
\end{center}
\end{figure}

The Einstein Equation quantifies how energy density leads to 
curvature and, in turn, how curvature influences energy density.
Generation of gravitational waves is implicit in these equations.
To see why, consider a region far from a source, a 
nearly flat region where the gravitational wave perturbs
a flat Cartesian metric by only a small amount $\halfbet$:
\begin{equation}
\galfbet(\xbold) \quad = \quad \etaalfbet + \halfbet(\xbold),
\end{equation}
where $\etaalfbet$ is the Minkowsi metric given in equation~\ref{eqn:minkowskimetric},
and $|\halfbet|<<1$. 
In this {\it linearized gravity}, 
the left side of the Einstein Equation can
be greatly simplified by keeping only
first order terms in $\halfbet$ and applying
the Lorenz gauge condition (analogous to that of electrodynamics):
\begin{equation}
\partial_\beta h^\beta_\alpha(\xbold) - {1\over2}\partial_\alpha h^\beta_\beta(\xbold) \quad = \quad 0.
\end{equation}
In vacuum ($\Talfbet=0$), one obtains the
homogeneous wave equation:
\begin{equation}
\Box \halfbet(\xbold) \quad = \quad 0\,,
\end{equation}
where $\Box\equiv -{1\over c^2}{\partial^2\over\partial t^2}+\nabla^2$. This equation
has solutions with familiar space and time dependence, but describes
a tensor perturbation. For example, a solution with fixed wave vector
$\vec k$ can be written as
\begin{equation}
\halfbet(\xbold) \quad = \quad \aalfbet \,e^{i[\vec{k}\cdot\vec{x}-\omega t]}\,
\end{equation}
where $\aalfbet$ is a symmetric  4$\times$4 matrix of
constants and where $\omega = kc$. Imposing the gauge condition above and additional
gauge freedom~\cite{bib:hartletext} ({\it transverse-traceless gauge}) and choosing the $z$ axis to lie
along $\vec k$ leads to the relatively simple form:
\begin{equation}
\halfbet(\xbold) \quad = \quad \left(
\begin{array}[c]{cccc} 0 & 0 & 0 & 0 \\ 0 & \hplus & \hcross & 0 \\ 0 & \hcross & -\hplus & 0 \\ 0 & 0 & 0 & 0 \end{array}\right) e^{i[kz-\omega t]}\,,
\end{equation}
where $\hplus$ and $\hcross$ are constant amplitudes.
For illustration, figure~\ref{fig:ringofmasses}
depicts the quadrupolar nature of these two polarizations (``$+$'', ``$\times$'') 
as gravitational waves propagating along the $z$-axis impinge upon a ring of test masses in free-fall in the $x$-$y$ plane.

The relation of gravitational waves to their {\it source} 
is found from the inhomogeneous Einstein Equation [equation (\ref{eqn:einsteinequation})]
where, again, we assume weak amplitudes in a nearly flat space-time.
Choosing Cartesian spatial coordinates and the transverse-traceless gauge,
one has an inhomogeneous wave equation:
\begin{equation}
\label{eqn:hwaveeqn}
\Box \halfbet(\xbold) \quad = \quad -{16\,\pi\,G\over{\rm c}^4}\,\Talfbet,
\end{equation}
which is analogous to the wave equation for relativistic electrodynamic
fields:
\begin{equation}
\Box A^\alpha \quad = \quad -\mu_0 \,J^\alpha\,,
\end{equation}
where $A^\alpha = (\Phi/c,\vec A)$ contains the scalar and vector potential
functions and where $J^\alpha = (c\rho, \vec J)$ contains the
electric scalar charge and current density. As for electrodynamics,
Green function formalism can be fruitfully applied to derive
solutions. As a reminder, for example, the electrodynamic vector potential solution can be
written as an integral over a source volume:~\cite{bib:jacksontext}
\begin{equation}
\vec A(t,\vec x)\quad  = \quad {\mu_0\over4\,\pi}\int d^3x'\,{[\vec J(\vec x\,',t')]_{\rm ret}\over|x-x'|},
\end{equation}
where $[...]_{\rm ret}$ indicates evaluation 
at the {\it retarded time} defined by $t'\equiv t-|\vec x-\vec x'|/c$.
Similarly, the solution to equation~(\ref{eqn:hwaveeqn}) can be
written as
\begin{equation}
\halfbet(t,\vec x) \quad = \quad {4G\over c^4}\int d^3x'\,\,{[\Talfbet(t',\vec x')]_{\rm ret}\over|x-x'|}.
\end{equation}

To gain an intuitive understanding of this solution,
consider a source that varies harmonically with time
with characteristic angular frequency $\omega$ and
wavelength $\lambda$ and make two approximations:
1) the long-wavelength approximation such that 
$\lambda\gg R_{\rm source}$ and 2) the distant-source
approximation $r\gg R_{\rm source}$. Here $R_{\rm source}$
is the outermost radius of the source, and $r$ is the
distance from the observer to the source.
In this limit (weak gravitational waves), the above
solution for $\halfbet$ reduces to 
\begin{equation}
\halfbet \quad \approx \quad {4G\over rc^4}\int  d^3x'\,\Talfbet(t-r/c,\vec x').
\end{equation}
Applying local energy/momentum conservation [see equation (\ref{eqn:epconservation})]
and integrating by parts (see Hartle~\cite{bib:hartletext} for details)
leads to 
\begin{equation}
\int d^3x\,T^{ij}(x) \quad = \quad {1\over2}\,{1\over{\rm c}^2}\,{d^2\over dt^2}\left[\int d^3x\> x^ix^j\,T^{tt}(x)\right].
\end{equation}
If one further restricts the source to one dominated by
its rest-mass density $\mu$ (non-relativistic internal velocities),
then 
\begin{equation}
\label{eqn:hquadformula}
h^{ij}(t,\vec x)\quad\approx\quad {2\,G\over rc^4}\,{d^2\over dt^2}\left[I^{ij}(t-r/c)\right],
\end{equation}
where $I^{ij}$ is the 2nd mass moment:
\begin{equation}
\label{eqn:idef}
I^{ij} \quad \equiv \quad \int d^3x\,\mu(t,\vec x)x^ix^j.
\end{equation}
Hence, to lowest order, gravitational radiation is a quadrupolar
phenomenon, in contrast to electrodynamics, for which electric
dipole and magnetic dipole radiation are supported.
As monopole electromagnetic radiation is prohibited by electric charge
conservation and monopole gravitational radiation is prohibited by
energy conservation, electric and magnetic dipole gravitational radiation are prohibited by
translational momentum and angular momentum conservation, respectively. Note that,
as for electrodynamics, gravitational radiation intensity is not
spherically symmetric (isotropic) about the source. 

The fact that the constant $2G/ c^4$  in front of equation~(\ref{eqn:hquadformula}) is 
so small in SI units ($1.7\times10^{-44}$ s$^2\cdot$kg$^{-1}\cdot$m$^{-1}$) is sobering when contemplating the
detection of gravitational radiation. 
The source quadrupole's 2nd time derivative must be
enormous to give detectable effects far from the source, 
implying large masses ($\sim\msolar$) 
with high characteristic velocities.

As a classic illustration from Saulson~\cite{bib:saulsontext}, 
consider a pair of 1.4-$\msolar$ neutron stars 15 Mpc away (\eg, near the center of the
Virgo galactic cluster) in a circular orbit of 20-km radius
(with coalescence imminent!) which have an orbital frequency of 400 Hz and emit gravitational waves
at 800 Hz with an amplitude (Newtonian, point mass approximation):
\begin{equation}
h \quad \approx \quad {10^{-21}\over (r/15\textrm{\ Mpc})},
\end{equation}
where $h\sim10^{-21}$ is a characteristic amplitude for transient sources
detectable by the LIGO and Virgo detectors described below.

Qualitatively, sensitive gravitational wave detectors have large characteristic
length scales $L$ (4 km for LIGO, 3 km for Virgo) in order to gain precision 
on the dimensionless strain $\Delta L/L$  induced by a gravitational
wave, for a given precision on $\Delta L$ determined by instrumental
and environmental noise.

It is useful to consider the energy flux implicit in gravitational waves.
The energy required to distort space is analogous to that required
to induce an elastic deformation of steel, but to a much greater degree,
which is to say, space is extremely stiff, as quantified below. 
Defining gravitational wave energy
flux is most straightforward in a spatial volume encompassing many wavelengths,
but small in dimension compared to the characteristic
radius of curvature of the background space.
In that regime, for example, the energy flux of
a sinusoidal, linearly polarized wave of amplitude $\hplus$ 
and angular frequency $\omega$ is~\cite{bib:hartletext}
\begin{equation}
\label{eqn:energyflux}
\mathcal{F} \quad = \quad {1\over32\,\pi}{c^3\over G} \hplus^2\omega^2
\end{equation}
For a 100-Hz sinusoidal wave of amplitude $\hplus=10^{-21}$, one obtains
a flux of 1.6 mW$\cdot$m$^{-2}$ (1.6 erg$\cdot$s$^{-1}\cdot$cm$^{-2}$). As one comparison, the total radiated
energy flux in the 2-10 keV X-ray band from the Crab nebula is $2.4\times10^{-11}$ W$\cdot$m$^{-2}$ 
($2.4\times10^{-8}$ erg$\cdot$s$^{-1}\cdot$cm$^{-2}$).
As another comparison, the radiation energy flux
bathing the earth from the Sun is about 1400 W/m$^2$. Hence during the brief
moment when the waves of a coalescing binary neutron star system
in the Virgo cluster pass the Earth, the implicit energy flux 
is more than a millionth that from the Sun! As we shall see below, however, 
detecting the passage of this energy flux is a formidable experimental challenge.

A general result~\cite{bib:hartletext} for the total energy luminosity for 
waves in the radiation zone depends on the
third time derivative of a modified inertia tensor $I^{ij}$:
\begin{equation}
\label{eqn:luminosity}
\mathcal{L} \quad = \quad {G\over5\,c^5}\,\left<\dddot\Islash_{ij}\dddot\Islash^{ij}\right>,
\end{equation}
where $<>$ represents an average over several cycles, and $\Islash$ is the traceless {\it quadrupole tensor}:
\begin{equation}
\label{eqn:islash}
\Islash^{ij} \quad \equiv \quad I^{ij}-{1\over3}\,\delta^{ij}I_k^k.
\end{equation}

\begin{figure}[tb]
\begin{center}
\epsfig{file=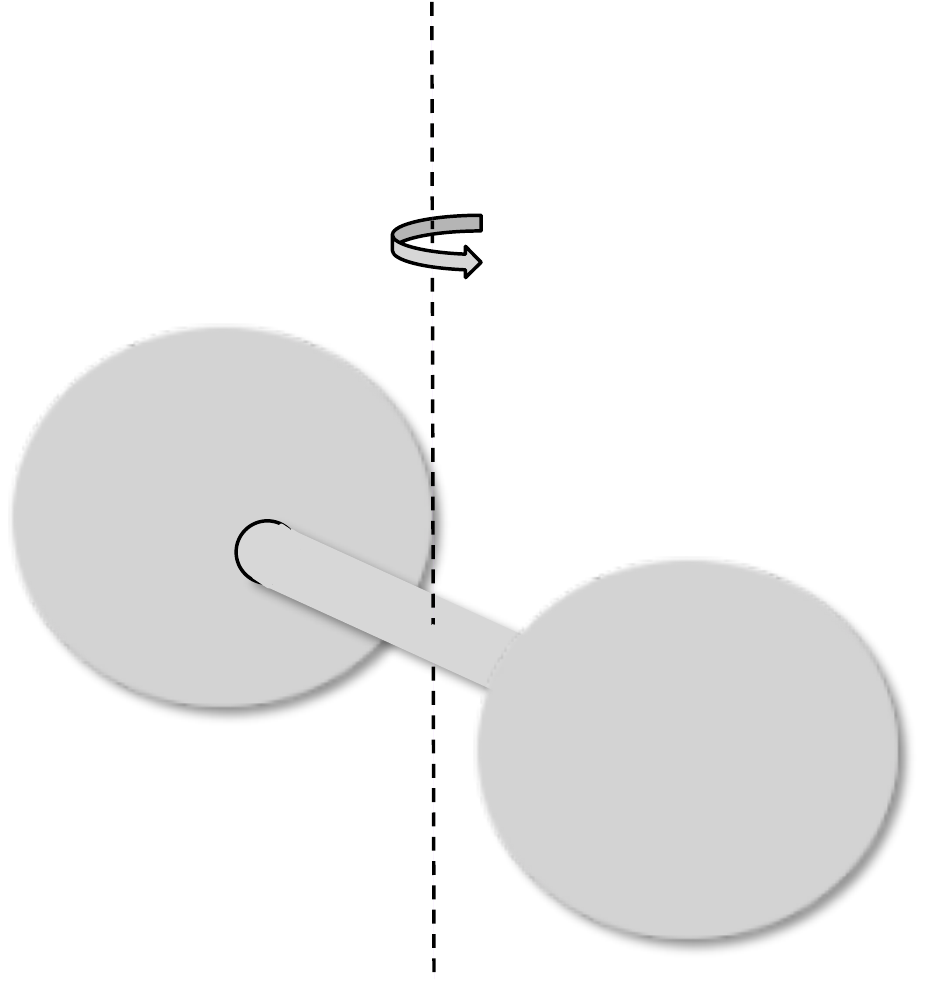,scale=0.5}
\caption{Rotating dumbbell model for a system with a changing quadrupole moment
\label{fig:dumbbell}}
\end{center}
\end{figure}

Before turning to likely sources of detectable gravitational radiation,
it is useful to consider additional comparisons with electromagnetic 
radiation:
\begin{itemize}
\item Most naturally emitted electromagnetic radiation is an incoherent
      superposition of light from sources much larger than the
      light's wavelength, while in contrast,
      gravitational radiation likely to be detectable ($<$ few kHz) comes from sources
      with sizes comparable to the wavelength. Hence the signal reflects
      coherent motion of extremely massive objects.
\item Because the detectable gravitational wave frequencies are so low, graviton energies ($\hbar\omega$)
      are (presumably) tiny, making detection of individual quanta even more 
      difficult than the already daunting challenge of detecting classical radiation.
\item In passing through ordinary matter, gravitational radiation suffers no more
      than a tiny absorption or scattering (although, like light, it is subject to gravitational lensing
      by large masses). As a result, gravitational waves can carry to us
      information about violent processes, for example, deep within stars or behind dust 
      clouds. As discussed below in the context of detection, even neutrinos have large 
      scattering cross sections, in comparison. 
\item It appears to be utterly impractical with current technology to
      detect manmade gravitational waves. To borrow another classic example from Saulson~\cite{bib:saulsontext},
      imagine a dumbbell consisting of two 1-ton compact masses with their centers separated by 2 meters 
      and spinning at 1 kHz about a line bisecting and orthogonal to their symmetry axis, as shown in 
      figure~\ref{fig:dumbbell}. For an observer 300-km away (in the radiation zone), one obtains
      an amplitude of $h\sim10^{-38}$ (setting aside the impracticality of such fast dumbbell rotation).
\end{itemize}

Finally, it bears emphasizing that the above linearized-gravity
approximations are useful for qualitative understanding, and
in many cases of interest, should be accurate. But for highly relativistic
sources, such as the merger of two rapidly spinning black holes,
detailed numerical calculation is necessary. Even for 
mildly relativistic systems, post-Newtonian perturbative approaches
(discussed in section~\ref{sec:cbcwaveforms}) are necessary.

\subsection{Overview of likely sources}
\label{sec:overviewsources}

As will be discussed in more detail in section~\ref{sec:searchoverview}, 
for purposes of detection, one can usefully classify sources
in four broad categories~\cite{bib:dawhitepaper2011}: 1) short-lived and well defined, for which coalescence
of a compact binary system is the canonical example; 2) short-lived and
{\it a priori} poorly known, for which a supernova explosion is the
canonical example; 3) long-lived and well defined, \eg, continuous
waves from spinning neutron stars; and 4) long-lived and stochastic,
\eg, primordial gravitational waves from the Big Bang.
For existing and upcoming terrestrial detectors, the most promising
category is the first. Detectable event rates for compact binary coalescence (CBC)
can be estimated with the greatest confidence and imply
highly likely discovery by Advanced LIGO and Virgo detectors.

For future spaced-based detectors,
which can probe to lower frequencies, the pre-coalescence phase of 
galactic binaries NS-NS is accessible, at the same time that coalescence
of binary super-massive black holes (SMBHs), \eg, from galaxy mergers should
be detectable~\cite{bib:sathyaschutz}. Similarly, pulsar timing arrays
can potentially detect a stochastic astrophysical background from
the superposition of cosmologically distant SMBH binary 
systems~\cite{bib:ptareview} at still lower frequencies ($\sim$nHz).

\subsection{Compact binary coalescence}
\label{sec:cbcsources}

Although binary star systems are common in our galaxy, only a tiny fraction
experience an evolution that arrives at two compact objects in an orbit
tight enough to lead to compact binary coalescence (CBC) in a Hubble time. That end results
requires both stars to be massive enough to undergo collapse to a
compact object without destroying its companion, without shedding so 
much mass that the orbit is no longer bound, and without undergoing
a ``birth kick'' that disrupts the bound orbit. 

Two distinct approaches (but with some common observational constraints)
have been used to estimate the average rates at which NS-NS coalescences occur
in the local region of the Universe. 
The first method~\cite{bib:clarketal,bib:brown} is based on {\it a priori}
calculations of binary star evolution, including the evolution of each
star in the presence of the other, where a common envelope phase
is not unusual. This general approach can be used to estimate rates for
NS-BH and BH-BH coalescence, too. 

The second estimation method~\cite{bib:phinney,bib:narayanetal}
is based largely on extrapolation from
observed double-neutron-star systems in our local galaxy,
albeit systems far from coalescence. Only a handful of binary systems with
two neutron stars are known, including the Hulse-Taylor binary~\cite{bib:weisbergtaylor} 
mentioned above and the double-pulsar system (J0737-3039) discovered
in 2003~\cite{bib:doublepulsar}. 

It is beyond the scope of this article to describe in detail these
calculations or their assumptions. A recent joint publication~\cite{bib:cbcratespaper} by the
LIGO Scientific Collaboration and Virgo collaboration summarizes the
recent literature and provides convenient tables of both estimated coalescence
rates and resulting expected coalescence detection rates for initial and
advanced detectors. 

In summary, these estimates for NS-NS coalescence yield ``realistic'' rates
of once every 10$^4$ years in a galaxy the size of the Milky Way (``Milky Way Equivalent Galaxy'' - MWEG). 
with ``plausible'' rates ranging from once every 10$^6$ years to once every
$10^3$ years. The corresponding rates for a NS-BH system are once per 300,000 years (``realistic''),
with a plausible range from once per 20 million years to once per 10$^4$ years.
For a BH-BH system (stellar masses), the corresponding rates are once per 
2.5 million years (realistic) with a plausible range from once per 100 million
years to once per 30,000 years.  Table~\ref{tab:cbcrates} summarizes
these estimates more compactly in units of MWEG$^{-1}$ Myr$^{-1}$.
An alternative rate unit is in terms of
coalescences per Mpc$^3$ per Myr. A rough conversion rate (for the local Universe)
is 0.01 MWEG/Mpc$^3$, giving estimated realistic rates of 1, 0.03 and 0.005 coalescences Mpc$^{-3}\cdot$Myr$^{-1}$
for NS-NS, NS-BH and BH-BH, respectively~\cite{bib:cbcratespaper}.

\begin{table}
\begin{center}
\begin{tabular}{l|ccc}
Source & $R_{\rm Low}^{\rm Plausible}$ & $R^{\rm Realistic}$ & $R_{\rm High}^{\rm Plausible}$ \\
\hline\hline
NS-NS & 1 & 100 & 1000 \\
NS-BH & 0.05 & 3 & 100 \\
BH-BH & 0.01 & 0.4 & 30 \\
\hline\hline
\end{tabular}
\caption{Summary of estimated coalescence rates (MWEG$^{-1}$ Myr$^{-1}$) for NS-NS, NS-BH
and BH-BH binary systems from the compilations in ref.~\cite{bib:cbcratespaper}.}
\label{tab:cbcrates}
\end{center}
\end{table}

Corroborating evidence for these estimates comes from the rate of 
observed short hard gamma ray bursts (GRBs). While long soft GRBs ($>$2 s) are thought to
arise primarily from the death of massive stars, 
short hard GRBs are widely thought to arise primarily from coalescence of NS-NS 
(or NS-BH) systems. Although the large correction for average 
beaming effects remains uncertain, the short hard GRB rate per unit
volume appears to be roughly consistent with the above range of 
estimates~\cite{bib:obk} for a variety of galactic evolution 
models. 

Conversion of coalescence rates into {\it detected} coalescence rates
depends, of course, on details of frequency-dependent detector 
sensitivity and on averaging over stellar orientations and sky positions.
Resulting estimated  detection rates for 1st- and 2nd-generation
interferometers will be presented in sections~\ref{sec:cbcsearches} and
\ref{sec:summary}.

Detection of CBC events will provide an unprecedented view of
strong-field gravity and offer demanding tests of general relativity,
especially in the case of detection by multiple detectors, allowing
disentanglement of waveform polarization.
The coalescence of two compact massive objects (neutron stars and
black holes) into a single final black hole 
can be divided into three reasonably distinct stages: inspiral, merger and
ringdown. During the inspiral stage, analytic expressions (perturbative post-Newtonian
approximations) for gravitational waveforms are expected to be accurate.
In the merger stage, strongly relativistic effects require numerical relativity 
calculations. In the ringdown of the final resulting black hole, however, simplicity
is once again expected. 

The inspiral stage lends itself to a natural perturbative approach~\cite{bib:clarkeardley}.
To illustrate with a simple, concrete example, consider two stars of
equal mass $M$ in an circular orbit of instantaneous radius $R(t)$
and angular velocity $\omega(t)$ (assumed slowly changing),
where the stars are treated as point masses far enough apart that
tidal effects can be neglected. From simple Newtonian mechanics, we obtain Kepler's 3rd Law:
\begin{equation}
\label{eqn:forcebalance}
M\omega^2R \quad = \quad {GM^2\over(2R)^2} \qquad \Longrightarrow \qquad \omega^2 \quad = \quad {GM\over4\,R^3}
\end{equation}
The total energy of this system (potental + kinetic) is 
\begin{equation}
E \quad = \quad -{GM^2\over4R},
\end{equation}
and the decrease in $E$ with time is
\begin{equation}
{dE\over dt}\quad = \quad{GM^2\over4R^2} {dR\over dt},
\end{equation}
as the orbit shrinks.

For convenience define the origin at the orbit's center and the $x-y$ plane to coincide with the 
orbital plane, with one star at $x_1=R$ at time $t=0$:
\begin{equation}
x_1(t) = -x_2(t) = R\cos(\omega t); \qquad y_1(t) = -y_2(t) = R\sin(\omega t); \qquad z_1 = z_2 = 0.
\end{equation}
from which one obtains [using equations~(\ref{eqn:idef}) and~(\ref{eqn:islash})]:
\begin{equation}
\dddot\Islash \quad= \quad MR^2\,(2\,\omega)^3 \left(
\begin{array}{ccc}\sin(2\,\omega t) & -\cos(2\,\omega t) & 0 \\
-\cos(2\,\omega t) & -\sin(2\,\omega t) & 0 \\
0 & 0 & 0\end{array}
\right),
\end{equation}
and a total radiated luminosity [using equation~(\ref{eqn:luminosity})]:
\begin{equation}
\mathcal{L} \quad = \quad {128\over5} {GM^2\over c^5} R^4\omega^6.
\end{equation}
Setting $dE/dt = -\mathcal{L}$ and using equation~(\ref{eqn:forcebalance}), one obtains
a differential equation for $R$:
\begin{equation}
R^3\,{dR\over dt}\quad = \quad -{8\over5} {G^3M^3\over c^5}.
\end{equation}
Integrating from a present time $t$ to a future coalescence time $\tcoal$ when $R\rightarrow0$,
one finds the orbital radius 
\begin{equation}
R(t) \quad = \quad \left[{32\over5}{G^3M^3\over c^5}(\tcoal-t)\right]^{1\over4},
\end{equation}
from which the gravitational wave frequency [$\fgw = 2\,\omega/2\pi$] is
derived (via equation~\ref{eqn:forcebalance}):
\begin{equation}
\label{eqn:fvstimeequalmass}
\fgw \quad = \quad{1\over8\,\pi} 
       \left[2\cdot5^3\right]^{1\over8}
       \left[{c^3\over GM}\right]^{5/8}
       {1\over(\tcoal-t)^{3\over8}}.
\end{equation}
As expected, the frequency diverges as $t\rightarrow\tcoal$. Now consider the
amplitude $h_0$ of the circularly polarized wave observed a distance $r$ away along the
orbital axis of rotation. From equation~(\ref{eqn:hquadformula}), one has:
\begin{equation}
\label{eqn:hvstimeequalmass}
h_0(t) \quad = \quad {1\over r}
         \left[{5\,G^5M^5\over2\,c^{11}}\right]^{1\over4}
         {1\over(\tcoal-t)^{1\over4}}.
\end{equation}
Substituting sample $M$ and $r$ values from the binary neutron star 
example in section~\ref{sec:gwgeneration} and defining the time
remaining until coalescence detection as $\tau$, one has for the gravitational
wave frequency and amplitude:
\begin{equation}
\fgw(t)\quad = \quad(1.9\>{\rm Hz}) \left({1.4\>\msolar\over M}\right)^{5\over8}\left({1\>{\rm day}\over\tau}\right)^{3\over8}
\end{equation}
and
\begin{equation}
h_0(t)\quad = \quad(1.7\times10^{-23})\left({15\>{\rm Mpc}\over r}\right)\left({1\>{\rm day}\over\tau}\right)^{1\over4}\left({M\over1.4\>\msolar}\right)^{5/4}.
\end{equation}
The increase in frequency with $\tau^{-{3\over8}}$ and in amplitude with $\tau^{-{1\over4}}$ leads to
a characteristic ``chirp'' in the gravitational waveform. 
Note that if the distance to the source is known, the common stellar mass of this system can be
derived from either the frequency or amplitude evolution.
Expressions for an unequal-mass binary will be presented in section~\ref{sec:cbcsearches}. 

Thus the early phases of the inspiral stage should provide a 
well understood post-Newtonian system, from which stellar masses (and perhaps spins) can be
determined. With these parameters determined (to some precision), one can
then make detailed comparisons of observations in
the merger stage with numerical predictions for those parameters. 
The ringdown mode frequencies and damping times are primarily governed by the total mass and spin of the final
black hole, allowing clean and analytic comparisons to the inspiral stage,
largely independent of the merger uncertainties.

There has been a flurry of work in the last two decades to improve
the numerical relativity calculations, to permit detailed comparisons
between observation and theory during the difficult merger phase.
A number of technical breakthroughs~\cite{bib:pretorius,bib:campanelli,bib:baker} have
led to dramatic progress in this area. In parallel, there is a coordinated
effort (NINJA = Numerical INJection Analysis~\cite{bib:ninja}) to produce families of detailed waveform
templates and evaluate algorithms for detecting them, 
not only for making comparisons after detection, but also
to increase the chances of detection via matched-filter algorithms.

Coalescences involving neutron stars offer the potential
for probing the neutron star equation of state via distortions
of the detected waveform away from that expected for two
point masses, because of tidal disruption of one or both 
stars~\cite{bib:tidaldisruption}.
The effects are expected to be small, however, and their detection
dependent on the detector sensitivity at the highest frequencies
in the detector bands.

Very distant coalescences also offer interesting cosmological
measurements via their role as ``standard candles'', analogous to 
Type 1A supernovae~\cite{bib:cbccandle}. Since the masses of the system can be
determined from the waveform shape (assuming polarization
has been determined via coincidence detection in multiple
detectors), the luminosity distance to the system
can be determined (assuming the correctness of general relativity).
If the redshift of the source's host galaxy can be determined from electromagnetic
measurements, \eg, simultaneous detection of a short GRB or of
an afterglow, then one obtains an independent measure of the
Hubble constant. 

Note, however, that the gravitational waves
are subject to the same redshift as electromagnetic radiation, 
which leads to an ambiguity in determining the redshift directly from the gravitational
waveform. For example, the reduction of the wave amplitude
due to luminosity distance (correlated to redshift) can be compensated 
by changes to the assumed rest-frame
mass of the system. Recently it has been appreciated, however,
that for coalescences involving a neutron star, the tidal disruption
can provide an independent clue to the stellar masses (in their
local frame), allowing the use of the standard candle even in the absence
of a known host galaxy~\cite{bib:readmessenger}. Similarly, the
statistical distribution of neutron star masses provides another
means to calibrate the standard candle~\cite{bib:taylorgair}.
And if the host galaxy 
{\it is} known, then one has a valuable cross check on the relation
between luminosity distance and redshift distance. 

While stellar spin is thought to be unimportant in searches for NS-NS inspirals,
it can be important for coalescences involving one or two black holes,
for which high spin can create significant waveform distortions over
a spinless assumption, where the maximum allowed angular momentum in general
relativity is $GM_{\rm BH}^2/c$ for a black hole of mass $M_{\rm BH}$~\cite{bib:hartletext}. 
Both amplitude and phase can be modulated, making
the parameter space over which one must search much larger than for
the NS-NS case, as discussed below in section~\ref{sec:cbcsearches}.

\subsection{Bursts}
\label{sec:burstsources}

Gravitational wave bursts customarily refer to transients of
poorly known or unknown phase evolution. Although the algorithms
used to search for bursts (described in section~\ref{sec:burstsearches}) are
sensitive to high-SNR, well predicted waveforms such as from
NS-NS coalescence, they are necessarily less sensitive than
matched-filter approaches, where known phase evolution can be exploited.
More generic transient algorithms must be
used for burst sources, such as supernovae, because of uncertain dynamics
in these violent processes and because of uncertain but almost
certainly varying initial stellar conditions. 

As a reminder, a spherically symmetric explosion (or implosion) does
not lead to gravitational waves in general relativity (no monopole term).
To be detected via gravitational waves then, a supernova presumably
needs to exhibit some asymmetry. The fact that many pulsars formed 
in supernovae have large measured speeds relative to their neighbors
(high ``birth kicks'')~\cite{bib:birthkicks} strongly suggests that
some supernovae do exhibit substantial non-spherical motion, perhaps
as a result of dynamical instabilities in rapidly rotating, massive
progenitor stars.
One recently appreciated mechanism for potentially strong 
gravitational wave emission during core-collapse supernovae is
hydrodynamical oscillation of the protoneutron star core~\cite{bib:nsconvection}.

With gravitational wave detection now on the horizon,
much work has gone into detailed simulations of the supernova
process, to predict possible resulting waveforms. As one might
imagine, this violent process, in which strong magneto-hydrodynamics, 
nuclear physics and general relativity are all important,
is a formidable challenge to simulate. Indeed, it has proven
challenging to reproduce in these simulations the spectacular explosions
that we associate with supernovae~\cite{bib:sathyaschutz}. Nonetheless, this recent
work has given new insights into the strength and spectral
content one might expect from supernovae. 
Unfortunately, predictions of strength remain subject
to large uncertainties. 

For scale, consider a supernova a distance $r$ away in our galaxy 
that emits  energy $E$ in gravitational waves, with a
characteristic duration $T$ and characteristic frequency $f$.
One expects~\cite{bib:sathyaschutz} a detectable strain amplitude
at the Earth of about
\begin{equation}
\label{eqn:hvsenergy}
h \quad \sim \quad 6\times10^{-21}\left({E\over10^{-7}\msolar {\rm c}^2}\right)^{1\over2}\left({1\ {\rm ms}\over T}\right)
\left({1\ {\rm kHz}\over f}\right)\left({10\ {\rm kpc}\over r}\right).
\end{equation}
For the nominal (but not necessarily accurate)
values of $r$, $E$, $T$ and $f$ in this expression,
the initial LIGO and Virgo interferometers should have been able to detect a galactic
supernova in gravitational waves. But no supernova was detected electromagnetically
in our galaxy during initial LIGO and Virgo data taking, which is unsurprising,
giving their expected low occurrence rate. With the expected order of
magnitude improvement in sensitivity of the advanced detectors, 
a galactic supernova with 100 times smaller energy or a supernova with
the same energy ten times further away would be accessible. Note, however,
that until one reaches the Andromeda galaxy ($\sim$780 kpc),
there is relatively little additional stellar mass beyond the edge
of the Milky Way, with nearby dwarf galaxies contributing only
a few percent additional mass. (Nonetheless, the most recent known
nearby supernova -- SN1987A -- was in the Large Magellanic Cloud at $\sim$50 kpc.)

One intriguing scenario in which a core collapse supernova could
be seen in gravitational waves to much larger distances is
via a bar mode instability~\cite{bib:barmodeinstability}, in which
differential rotation in a collapsing star leads to a large, rapidly
spinning quadrupole moment, generating waves detectable from well outside
our own galaxy~\cite{bib:thorne300}.
Another type of instability ($r$-mode) may develop in the birth of
a neutron star, but its lifetime is expected to be long enough, that it will
be discussed below in the category of continuous wave sources.

Another potential transient source of poorly known gravitational waveform shape is 
the sudden release of energy from a highly magnetized neutron star
(magnetar). Although ``ordinary'' neutron stars are characterized by 
extremely strong surface magnetic fields ($\sim$10$^{12}$ G), many magnetars appear to have
fields $\sim$100-1000 times still stronger, implying enormous pent-up magnetic energy.
It is thought that soft gamma ray repeaters (SGRs) and anomalous X-ray pulsars (AXPs)
are different observational manifestations of the same underlying system - a
highly magnetized star which sporadically converts magnetic field energy
into radiation~\cite{bib:kaspi}. Whether this process involves rupture of the neutron star
crust, vortex rearrangement in a core superconducting fluid, or some other
process, is not yet well understood. Especially dramatic instances are
superflares, such as the December 2004 flare from SGR 1806-20, in which
$\sim$10$^{39}$ J (10$^{46}$ erg) of electromagnetic energy was released~\cite{bib:sgr1806}. 
This radiation release from $\sim$10 kpc away
disturbed the Earth's ionosphere sufficiently to disrupt some radio communications~\cite{bib:ionospheredisruption}.
How much gravitational wave energy might be released in such events is unclear,
although it has been proposed that the energy released into
neutron star crustal vibrations could be comparable to that
released electromagnetically~\cite{bib:corsiowen}, 
in which case gravitational radiation
due to those vibrations could be substantial.
For scale, the magnetic energy stored in
a neutron star with surface field of $10^{15}$ G is O(10$^{40}$ J = 10$^{47}$ erg), assuming an
internal field no larger than the surface field. If the star had
still stronger internal fields, the energy would be still larger.
Given the uncertainties in the mechanism leading to these enormous radiation
releases, it is hard to be confident of predicted waveforms. Hence generic
GW transient algorithms are appropriate in searching for flares, as discussed 
below in section~\ref{sec:burstsearches}.

Another possible transient source is emission of bursts of gravitational radiation
from ``cosmic string cusps''~\cite{bib:cosmicstring}. 
Cosmic strings might be defects remaining
from the electroweak (or earlier) phase transition or possibly primordial
superstrings redshifted to enormous distances. In either model, kinks 
in these strings would travel at the speed of light with an 
isotropic distribution of directions, generating a model-dependent gravitational
wave spectrum that is collimated along the direction of cusp travel. According
to this idea, one would expect a cosmological background of GW bursts,
that might be detected individually. As discussed below, this same model
could lead to a steady-state, lower-level stochastic background from cusp radiation 
from more distant reaches of the universe.

A general consideration in burst searches is the energy release implicit
for a given source distance and detectable strain amplitude. 
As the distance of the source increases, the energy required
for its waves to be detectable on Earth increases as the 
square of the distance. Specifically, rewriting equation~(\ref{eqn:hvsenergy}),
one obtains the relation:
\begin{equation}
E \quad \sim \quad (3\times10^{-3}\msolar c^2)\left({h\over10^{-21}}\right)^2\left({T\over1\>{\rm ms}}\right)\left({f\over1\>{\rm kHz}}\right)
\left({r\over10\>{\rm Mpc}}\right)^2
\end{equation}
Hence for a source distance much beyond 10 Mpc and for initial
LIGO/Virgo sensitivities to transients, one needs sources emitting
significant fractions of a solar mass in gravitational radiation in
frequency bands accessible to terrestrial detectors, such as
is expected in the case of coalescing binary systems.

\subsection{Continuous waves}
\label{sec:cwsources}

Continuous gravitational waves refer to those that are long-lasting
and nearly monochromatic. In the frequency band of present and
planned terrestrial detectors, the canonical sources are
galactic, non-axisymmetric neutron stars spinning fast enough that
twice their rotation frequency is in the detectable band. (For
future space-based gravitational wave detectors, the early stages
of coalescing binaries provide another continuous-wave source,
where the orbital decay leads to only a small secular departure
from monochromaticity.)

Several different mechanisms have been proposed by which spinning
neutron stars could generate detectable gravitational waves.
Isolated neutron stars may have intrinsic non-axisymmetry from residual crustal
deformation (\eg, from cooling \&\ cracking of the crust)~\cite{bib:crustdeformation}, or from
non-axisymmetric distribution of magnetic field energy trapped
beneath the crust~\cite{bib:buriedbfieldasymmetry}. 

An isolated star may also exhibit normal modes of oscillations, including
$r$-modes in which quadrupole mass currents emit gravitational waves~\cite{bib:rmodes}.
These $r$-modes can be inherently unstable, arising from azimuthal
interior currents that are retrograde in the star's rotating frame, but
are prograde in an external reference frame. As a result, the
quadrupolar gravitational wave emission due to these currents
leads to an {\it increase} in the strength of the current. This
positive-feedback loop leads to an intrinsic instability.
The frequency of such emission is expected to be approximately 4/3 the rotation 
frequency~\cite{bib:rmodes}.  Serious
concerns have been raised~\cite{bib:rmodesdoubts}, however, about the 
detection utility of this effect for young 
isolated neutron stars (other than truly newborn stars in our
galaxy), where mode saturation appears 
to occur at low $r$-mode amplitudes because
of various dissipative effects.
This notion of a runaway rotational instability was
first appreciated for high-frequency $f$-modes~\cite{bib:cfs}, (Chandrasekhar-Friedman-Schutz
instability), but realistic viscosity effects seem likely to 
suppress the effect~\cite{bib:cfskiller}. 

In addition, as discussed below, a binary neutron star may experience non-axisymmetry
from non-isotropic accretion (also possible for an isolated 
young neutron star that has experienced fallback accretion shortly
after birth). 

The detection of continuous gravitational waves
from a spinning neutron star should yield precious information
on neutron star structure and the equation of state of nuclear matter 
at extreme pressures when combined with electromagnetic measurements
of the same star.

In principle, there should be O(10$^{8-9}$) neutron stars in our galaxy~\cite{bib:nspopulation},
out of which only about 2000 have been detected, primarily as radio pulsars.
The small fraction of detections is understandable, given several considerations.
Radio pulsations appear empirically to require the combination of the magnetic field and
rotation frequency to satisfy the approximate 
relation $B\cdot f_{\rm rot}^2>1.7\times10^{11}$ G$\cdot$(Hz)$^2$~\cite{bib:deathline}.
As a result, isolated pulsars seem to have lifetimes of $\sim10^7$ years~\cite{bib:pulsarastronomy},
after which they are effectively radio-invisible.
On this timescale, they also cool to where X-ray emission is difficult
to detect. There remains the possibility of X-ray emission
from steady accretion of interstellar medium (ISM), but it appears that the
kick velocities from birth highly suppress such accretion~\cite{bib:bondi}
which depends on the inverse cube of the star's velocity through
the ISM.

A separate population of pulsars and non-pulsating neutron stars
can be found in binary systems. In these systems accretion from a 
non-compact companion star can lead to ``recycling,'' in which
a spun-down neutron star regains angular momentum from the infalling matter.
The rotation frequencies achievable through this spin-up are
impressive -- the fastest known rotator is J1748-2446ad at 716 Hz~\cite{bib:hessels}.
One class of such systems is the set of low mass X-ray binaries (LMXBs)
in which the neutron star ($\sim$1.4 $\msolar$) has a much
lighter companion ($\sim$0.3 $\msolar$)~\cite{bib:pulsarastronomy} that
overfills its Roche lobe, spilling material onto an accretion
disk surrounding the neutron star or possibly spilling material
directly onto the star, near its magnetic polar caps. 
When the donor companion star eventually shrinks and decouples
from the neutron star, the neutron star can retain a large
fraction of its maximum angular momentum and rotational energy.
Because the neutron star's magnetic field decreases during
accretion (through processes that are not well understood),
the spin-down rate after decoupling can be very small.

Equating rotational energy loss rate to magnetic dipole
radiation losses, leads to the relation~\cite{bib:pacini}:
\begin{equation}
\left({dE\over dt}\right)_{\rm mag} \quad = \quad {\mu_0M_\perp^2\omega^4\over6\pi c^3},
\end{equation}
where $M_\perp$ is the component of the star's magnetic dipole
moment perpendicular to the rotation axis: $M_\perp=M\sin(\alpha)$,
with $\alpha$ the angle between the axis and north magnetic pole.
In a pure dipole moment model, the magnetic pole field strength
at the surface is $B_0 = \mu_0M\,/\,2\pi R^3$.
Equating this energy loss to that of the (Newtonian) rotational
energy ${1\over2}I_{\rm zz}\omega^2$ leads to the prediction:
\begin{equation}
{d\omega\over dt} \quad = \quad {\mu_0R^6\over6\,\pi c^3I_{\rm zz}}B_\perp^2\omega^3.
\end{equation}
Note that the spindown rate is proportional to the square of $B_\perp=B_0\sin(\alpha)$
and to the cube of the rotation frequency. The cubic dependence of
$d\omega/dt$ on $\omega$ leads to a relation between the
present day rotational frequency $f$ and the star's spindown age $\tau$:
\begin{equation}
\label{eqn:spindownagefull3}
\tau \quad = \quad -\left[{f\over 2\,\dot f}\right]\,\left[1-\left({f\over f_0}\right)^2\right],
\end{equation}
where $f_0$ was the frequency a time $\tau$ in the past (assuming magnetic
dipole radiation has dominated rotational energy loss). 
In the limit $f\ll f_0$, this reduces simply to
\begin{equation}
\label{eqn:spindownage3}
\tau\quad =\quad -\left[{f\over 2\,\dot f}\right]\,.
\end{equation}
More generally, for a star spinning down with dependence:
\begin{equation}
\label{eqn:genericbrakingindex}
\left({d\omega\over dt}\right)\quad =\quad K \omega^n,
\end{equation}
for some constant $K$,
equation~\ref{eqn:spindownagefull3} becomes (assuming $n\ne1$):
\begin{equation}
\tau \quad = \quad -\left[{f\over (n-1)\,\dot f}\right]\,\left[1-\left({f\over f_0}\right)^{(n-1)}\right],
\end{equation}
and equation~(\ref{eqn:spindownage3}) becomes:
\begin{equation}
\label{eqn:approxageindex}
\tau \quad = \quad -\left[{f\over (n-1)\,\dot f}\right]\,.
\end{equation}
Assuming $n$ (often called the ``braking index'' and derived
from the ratio $f\ddot f/\dot f^2$ when $\ddot f$ is measurable) is three
(as would be the case for a rotating magnetic dipole), leads
to approximate inferred ages for many binary radio pulsars 
in excess of $10^9$ and even well over $10^{10}$ years~\cite{bib:atnfdb}.
A recent calculation suggests that 
this surprising result can be explained by
reverse-torque spindown during the Roche lobe
decoupling phase~\cite{bib:taurisrldp}. In fact,
measured braking indices for even young pulsars tend to be less than three,
suggesting that the model of a neutron star spinning down with
constant magnetic field is inaccurate~\cite{bib:pulsarastronomy}.
(See refs.~\cite{bib:palomba} for discussions of spindown evolution
in the presence of both gravitational wave and electromagnetic
torques.)

\begin{figure}[tb]
\begin{center}
\epsfig{file=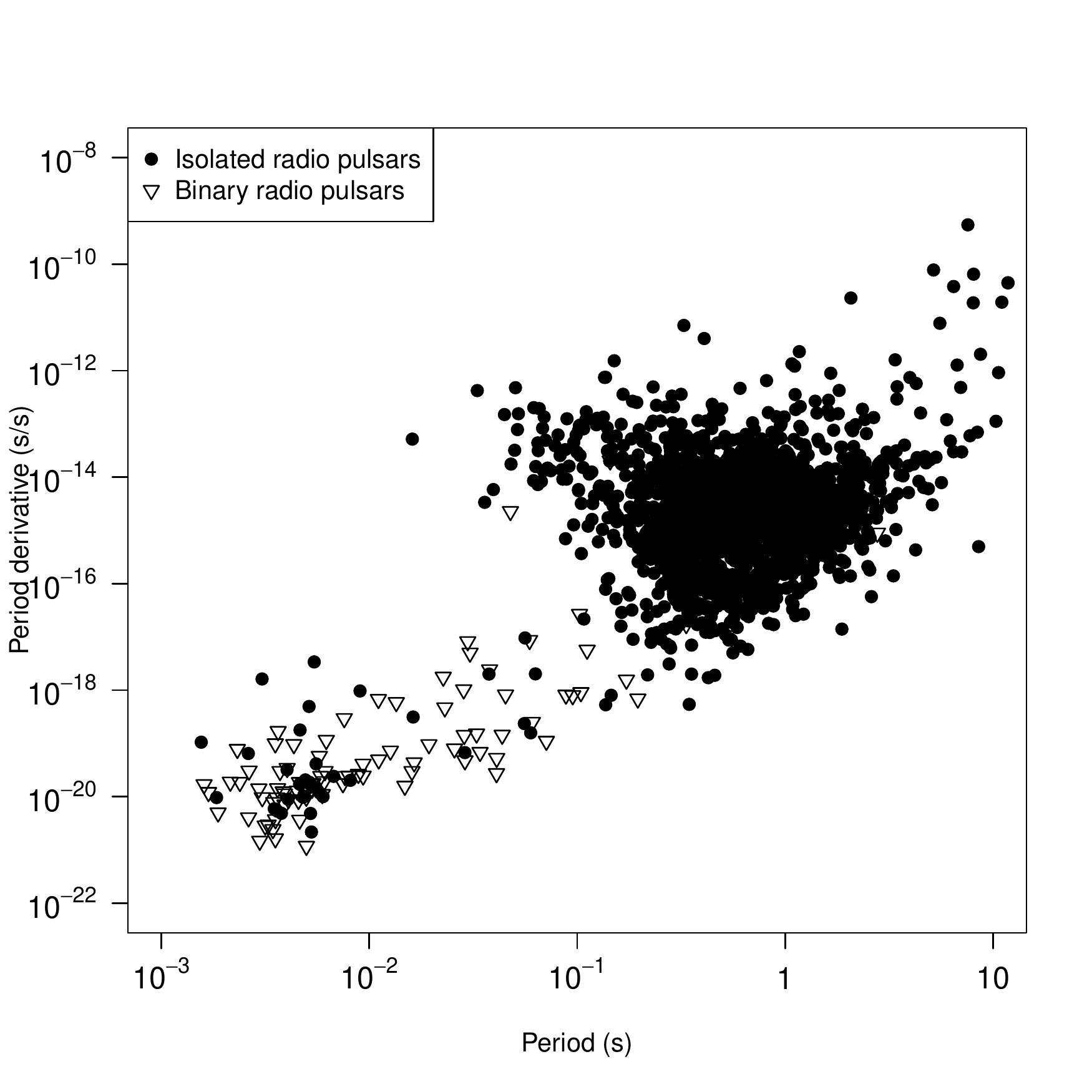,scale=0.7}
\caption{Measured periods and period derivatives for known radio
pulsars. Closed circles indicate isolated stars. Open triangles
indicate binary stars.
\label{fig:ppdotdiagram}}
\end{center}
\end{figure}

In summary, there are at least three distinct populations
of neutron stars potentially detectable via continuous
gravitational waves: relatively young, isolated stars with spin
frequencies below $\sim$50 Hz, such as the Crab pulsar; 
actively accreting stars in binary systems; and recycled
``millisecond'' stars for which accretion has ceased and which generally
have rotation frequencies above 100 Hz. In some cases
the companion donor has disappeared, \eg, via ablation, 
leaving an isolated neutron star, but most known millisecond
pulsars remain in binary systems, as is clear 
from figure~\ref{fig:ppdotdiagram}, based on data
from the Australia National Telescope Facility's 
pulsar database~\cite{bib:atnfdb}

Let's now consider the gravitational radiation one might
expect from these stars. If a star at a distance $r$ away has a quadrupole
asymmetry, parametrized by its ellipticity:
\begin{equation}
\label{eqn:ellipticity}
\epsilon \quad \equiv \quad {I_{xx}-I_{yy}\over I_{zz}}, 
\end{equation}
and if the star is spinning about the approximate symmetry axis of rotation ($z$),
(assumed optimal -- pointing toward the Earth), then the expected intrinsic strain amplitude $h_0$ is
\begin{equation}
\label{eqn:hexpected}
h_0 \quad = \quad {4\,\pi^2GI_{\rm zz}f_{\rm GW}^2\over c^4r}\,\epsilon 
\quad = \quad (1.1\times10^{-24})\left({I_{\rm zz}\over I_0}\right)\left({f_{\rm GW}\over1\>{\rm kHz}}\right)^2
\left({1\>{\rm kpc}\over r}\right)\left({\epsilon\over10^{-6}}\right)\,
\end{equation}
where $I_0=10^{38}$ kg$\cdot$m$^2$ (10$^{45}$ g$\cdot$cm$^2$) is a nominal quadrupole moment of
a neutron star, and the gravitational radiation is emitted at frequency $f_{\rm GW}=2\,f_{\rm rot}$.
(Reduction of detectable strain amplitude by non-optimal star
orientation will be discussed in section~\ref{sec:cwtargeted}.)
The total power emission in gravitational waves from
the star (integrated over all angles) is 
\begin{equation}
\label{eqn:powerloss}
{dE\over dt} \quad = \quad - {32\over5} {G\over C^5}\,I_{\rm zz}^2\, \epsilon^2\, \omega^6 
\quad = \quad- (1.7\times10^{33}\>{\rm J/s})\left({I_{\rm zz}\over I_0}\right)^2
\left({\epsilon\over10^{-6}}\right)^2
\left({f_{\rm GW}\over1\>{\rm kHz}}\right)^6\,.
\end{equation}
For an observed neutron star of measured $f$ and $\dot f$,
one can define the ``spindown limit'' on maximum detectable
strain by equating the power loss in equation~(\ref{eqn:powerloss})
to the time derivative of the (Newtonian) rotational kinetic
energy: ${1\over2}I\omega^2$, as above for magnetic dipole radiation. 
One finds:
\begin{eqnarray}
\label{eqn:spindownlimit}
h_{\rm spindown}\quad & = & \quad {1\over r}\sqrt{-{5\over4}{G\over c^3}I_{\rm zz}{\dot f_{\rm GW}\over f_{\rm GW}}} \nonumber \\
& = & \quad (2.5\times10^{-25})\left({1\>{\rm kpc}\over r}\right)\sqrt{\left({1\>{\rm kHz}\over f_{\rm GW}}\right)
\left({-\dot f_{\rm GW}
\over10^{-10}\>{\rm Hz/s}}\right)\left({I_{\rm zz}\over I_0}\right)}
\end{eqnarray}
Hence for each observed pulsar with a measurable spindown and
well determined distance $r$,
one can determine whether energy conservation even permits detection
of gravitational waves in an optimistic scenario. Unfortunately,
nearly all known pulsars have strain spindown limits below what
could be detected by the initial LIGO and Virgo detectors, as discussed
below. 

A similarly optimistic limit based only on the age of a known neutron 
star of unknown spin frequency can also be derived. If one assumes a star is spinning down entirely
due to gravitational radiation, then the energy loss for this {\it gravitar} satisfies equation~(\ref{eqn:genericbrakingindex})
with a braking index of five.
Assuming a high initial spin frequency,
the star's age then satisfies:
\begin{equation}
\tau_{\rm gravitar}\quad = \quad -{f\over 4\,\dot f}\,.
\end{equation}
If one knows the distance and the age of the star, \eg, from
the expansion rate of its visible nebula, then
under the assumption that the star has been
losing rotational energy since birth primarily
due to gravitational wave emission, then one
can derive the following frequency-independent
age-based limit on strain~\cite{bib:agebasedlimit}:
\begin{equation}
h_{\rm age} \quad = \quad (2.2\times10^{-24})\left({1\>{\rm kpc}\over r}\right)\sqrt{\left({1000\>yr\over\tau}\right)
\left({I_{\rm zz}\over I_0}\right)}
\end{equation}
A notable example is the Compact Central Object (CCO)
in the Cassiopeia A supernova remnant. Its birth
aftermath may have been observed by Flamsteed~\cite{bib:casabirth} in
1680, and the expansion of the visible shell is consistent
with that date. Hence Cas A, which is visible in X-rays 
but shows no pulsations, is almost certainly a very young 
neutron star at a distance of about 3.4 kpc. From the above equation,
one finds an age-based strain limit of $1.2\times10^{-24}$, which is accessible to
initial LIGO and Virgo detectors in their most sensitive band.

A simple steady-state argument by Blandford~\cite{bib:thorne300} led to 
an early estimate of the maximum detectable strain amplitude expected from a population of
isolated gravitars of a few times 10$^{-24}$, independent of typical ellipticity values,
in the optimistic scenario that most neutron stars become gravitars. 
A later detailed numerical simulation~\cite{bib:knispelallen}
revealed, however, that the steady-state assumption does not generally hold, leading
to ellipticity-dependent expected maximum amplitudes that can be 2-3 orders
of magnitude lower in the LIGO band for ellipticities as low as 10$^{-9}$ and a few
times lower for ellipticity of about $10^{-6}$.

Yet another approximate strain limit can be defined
for accreting neutron stars in binary systems,
such as Scorpius X-1. The X-ray luminosity  from
the accretion is a measure of mass accumulation at
the surface. As the mass rains down on the surface
it can add angular momentum to the star, which in 
equilibrium may be radiated away in gravitational waves.
Hence one can derive a torque-balance limit~\cite{bib:wagoner,bib:papapringle,bib:bildsten}:
\begin{equation}
\label{eqn:torquebalance}
h_{\rm torque}\quad = \quad(5\times10^{-27})
\sqrt{\left({600\>{\rm Hz}\over f_{\rm GW}}\right)
\left({\mathcal{F}_{\rm x}\over10^{-8}\>{\rm erg/cm}^2/{\rm s}}\right)}
\end{equation}
where $\mathcal{F}_{\rm x}$ is the observed energy flux at the Earth of
X-rays from accretion. Note that this limit is independent
of the distance to the star.

The notion of gravitational wave torque equilibrium is potentially important,
given that the maximum observed rotation frequency of neutron
stars in LMXBs is substantially lower than one might expect from
calculations of neutron star breakup rotation speeds ($\sim$1400 Hz)~\cite{bib:breakupspeed}.
It has been suggested~\cite{bib:speedlimit} that there is a ``speed limit''
governed by gravitational wave emission that governs the maximum
rotation rate of an accreting star. In principle, the distribution
of frequencies could have a quite sharp upper frequency cutoff,
since the angular momentum emission is proportional to the 
5th power of the frequency. For example, for 
an equilibrium frequency corresponding to a particular accretion rate,
doubling the accretion rate would increase the equilibrium frequency
by only about 15\%. 

A number of mechanisms have been proposed by which the accretion
leads to gravitational wave emission. The simplest is localized accumulation
of matter, \eg, at the magnetic poles (assumed offset from the rotation axis), 
leading to a non-axisymmetry.
One must remember, however, that matter can and will diffuse into
the crust under the star's enormous gravitational field. This diffusion of
charged matter can be slowed by the also-enormous magnetic fields in
the crust, but detailed calculations~\cite{bib:vigeliusmelatos} indicate the
slowing is not dramatic. Another proposed mechanism is excitation of
$r$-modes in the fluid interior of the star~\cite{bib:rmodes},
with both steady-state emission and cyclic spinup-spindown 
possible~\cite{bib:rmodeslmxb,bib:rmodesdoubts}.

\subsection{Stochastic waves}
\label{sec:stochsources}

Stochastic gravitational waves arise from a superposition
of incoherent sources. While a cosmological background from
primordial gravitational waves created in the Big Bang are a
natural possible source~\cite{bib:grishchuk}, other isotropic possibilities are
from cosmic strings and from very distant mergers of neutron stars or of 
supermassive black holes (accessible to space-based detectors).
Non-isotropic sources in the band of terrestrial detectors
could include the superposition of pulsar radiation from,
say, the Virgo Cluster. Over very long time scales, 
gravitational radiation from an accreting neutron star
could also appear stochastic, as the phase of the narrowband
signal wanders.

A primordial isotropic gravitational wave background is predicted
by most cosmological theories, although the predicted strengths of the
background vary enormously. It is customary~\cite{bib:thorne300,bib:allen} to
parametrize the background strength \vs\ frequency $f$ 
by its energy density per unit logarithm normalized to
the present-day critical energy density $\rho_\textrm{crit} = 3H_0^2c^2\,/\,8\pi G$ of the universe,
where $H_0$ is Hubble's constant, taken here to be 70.5 km/s/Mpc~\cite{bib:pdg}:
\begin{equation}
\label{eqn:omegadefinition}
\Omega_\textrm{gw}(f) \quad = \quad {1\over \rho_\textrm{crit}}\, 
{d\rho_\textrm{gw}(f) \over d\>ln(f)}
\end{equation}
The associated power spectral density can be written~\cite{bib:allenromano}:
\begin{equation}
S_{\rm GW} \quad = \quad {3H_0^2\over10\pi^2}\>f^{-3}\>\Omega(f).
\end{equation}
Note that, as for the Cosmic Microwave Background Radiation (CMBR), 
the primordial gravitational waves would be highly
redshifted from the expansion of the universe, but likely to a much greater degree,
since they would have decoupled from matter at vastly earlier times.

A more convenient reformulation in amplitude spectral density can
be written as~\cite{bib:stochsearchs1}
\begin{equation}
h(f) \quad \equiv \quad \left[S_{\rm GW}(f)\right]^{1\over2}\quad =\quad (5.6\times10^{-22})\,h_{100}\left(\Omega(f)\right)^{1\over2}\left({100\>{\rm Hz}\over f}\right)^{3\over2}\>{\rm Hz}^{-{1\over2}},
\end{equation}
where $h_{100}\equiv H_0/({\rm 100 km/s/Mpc}$).
 
\begin{figure}[tb]
\begin{center}
\epsfig{file=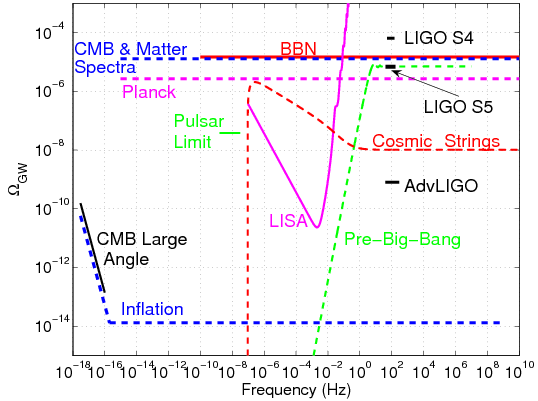,scale=0.75}
\caption{Comparison of different stochastic gravitational wave background measurements 
and models~\cite{bib:stochsearchs5}. Results for LIGO S4 and S5 searches are shown in
the frequency band around 100 Hz, along with projected Advanced LIGO
sensitivity. The indirect bounds due to BBN and CMBR / matter power spectra apply to the integral
of $\Omega_{\rm GW}(f)$ over the frequency bands denoted by the corresponding dashed curves. 
Projected sensitivities of the satellite-based
Planck CMBR experiment and LISA GW detector are also shown. The pulsar bound is based on the fluctuations
in the pulse arrival times of millisecond pulsars and applies at frequencies around 10$^{-8}$ Hz. 
Measurements of the CMBR at
large angular scales constrain the possible redshift of CMBR photons due to a stochastic
gravitational wave background, 
and therefore limit the amplitude of that background at largest wavelengths 
(smallest frequencies). Examples of inflationary, cosmic strings, and
pre-big-bang models are also shown 
(the amplitude and the spectral shape in these models can vary significantly as a
function of model parameters).
\label{fig:stochasticlandscape}}
\end{center}
\end{figure}

A key question is what range of values are expected for $\Omega(f)$?
Figure~\ref{fig:stochasticlandscape} shows a range of
expectations \vs\ frequency (28 orders of magnitude in frequency and 12 in $\Omega$).
The bottom curve is a rough estimate expected from standard
inflationary scenarios~\cite{bib:inflation1,bib:inflation2}. This graph also shows direct limits on
gravitational wave energy density from comparison of observed
abundances of elements with predictions from Big Bang 
nucleosynthesis (BBN)~\cite{bib:bbnlimits}, in addition to
limits derived from measurements of anisotropies in the CMBR~\cite{bib:cmbrlimits}.
For reference, the normalized total energy density of
the CMBR itself is about 
$\Omega_{CMBR} = 5\times10^{-5}$, and the energy density from
primordial neutrinos is estimated to be bounded by $\Omega_{\nu\bar\nu}<0.014$~\cite{bib:pdg}.

As discussed below, the $\Omega(f)$ sensitivity of the initial LIGO and Virgo
detectors to this isotropic background is O(several $\times$ 10$^{-6}$), with an 
expected improvement of more than three orders of magnitude for advanced detectors.
From figure~\ref{fig:stochasticlandscape}, it is clear, though, that
even advanced detectors fall far short of the sensitivity needed
to probe standard inflation.

There are other Big Bang scenarios, however, that permit much higher
primordial gravitational wave energy densities, In particular,
the curve labeled ``Pre-BigBang'' in figure~\ref{fig:stochasticlandscape} shows
an upper range expected in certain pre-Big Bang models~\cite{bib:stochsearchs5}.
The advanced detectors can address the upper range of this region.

A completely different source of cosmological, isotropic stochastic
waves could come from the cosmic strings discussed 
above in section~\ref{sec:burstsources} as a potential
source of gravitational wave burst radiation. Figure~\ref{fig:stochasticlandscape}
shows a range of predictions of stochastic radiation energy density in this model
for a range in assumed string tension $\mu$ and string reconnection probability $p$~\cite{bib:stochsearchs5}.
Part of this region can be addressed by the initial LIGO and Virgo
detectors, with more parameter space accessible to advanced detectors.

A more conventional source of isotropic, stochastic gravitational
waves is the superposition of radiation from many distant events,
such as binary coalescences from compact stars too far away to be seen
individually~\cite{bib:phinneystoch}. 
In the terrestrial band these
coalescences could be from stellar NS-NS, NS-BH and BH-BH systems. 
A recent detailed analysis~\cite{bib:stochbnsade}
suggests that this background could well be detectable by 2nd-generation
detectors and could present a significant background for 3rd-generation searches.
In the space-based
detector band ($\sim10^{-4}$--1 Hz for the original LISA design~\cite{bib:lisa}), the coalescences could be
from supermassive black hole mergers, \eg, from galaxy collisions.
As discussed in section~\ref{sec:pta}, binary super-massive black hole (SMBH) systems are
a serious prospect for pulsar timing arrays in the several-nHz band.

\section{Gravitational Wave Detectors}
\label{sec:gwdetectors}

\subsection{Overview of gravitational wave detection}
\label{sec:detoverview}

Until the mid 20th century there remained some
question as to whether or not gravitational waves
were truly predicted by general relativity~\cite{bib:saulsonhistory}.
It was not obvious that what appeared to be a wave
phenomenon could not be explained away as an
artifact of coordinate/gauge transformations (recall discussion
in section~\ref{sec:gwgeneration}). The reality of gravitational
wave prediction was confirmed, however, by the realization that energy could
be extracted from the waves, \ie, it was possible, in principle,
to build a detector that could register their passage~\cite{bib:saulsonhistory}.

The earliest manmade gravitational wave detectors were based on
a simple gedanken experiment: if two masses on a spring
are momentarily stretched apart and then compressed by a 
gravitational wave, potential energy is imparted to the
spring, independent of how coordinates are defined.
If the characteristic frequency of the wave is near
the resonance frequency of the mechanical system, the
response to the wave is magnified, not unlike an
LRC antenna circuit's response to a passing electromagnetic
wave. One early approach was to search for excitations
of the Earth's crustal vibrational normal modes 
($\sim$sub-mHz and higher harmonics)~\cite{bib:earthmodes},
a technique useful for setting upper limits,
but large earthquakes made it unattractive
for detection.

\begin{figure}[tb]
\begin{center}
\epsfig{file=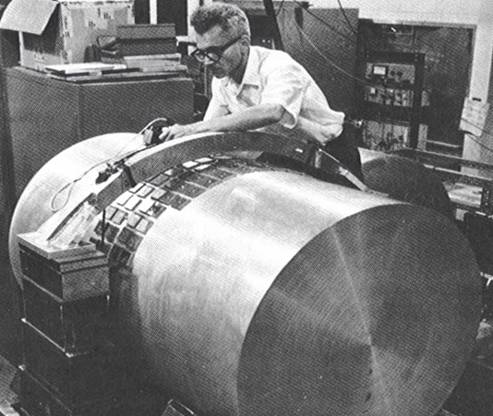,scale=0.5}
\caption{Joe Weber working on an early gravitational wave bar with
piezo-electric transducers as strain sensors (circa 1965).
Credit: University of Maryland.
\label{fig:weberbar}}
\end{center}
\end{figure}

In practice, since it is the elastic energy that matters,
the first gravitational wave detectors were simple metal cylinders,
where the energy converted to longitudinal oscillations of 
the bar was measured via piezoelectric transducers near the
``waist'' of the bar, as shown in figure~\ref{fig:weberbar}.
One looked for a sudden change in the amplitude of nominally
thermal motion of the bar~\cite{bib:weberbar}. Joe Weber of the University of Maryland
pioneered this detector design and implementation; he  also 
reported anomalies attributed to gravitational waves, such
as coincident transients in geographically separated pairs
of bars~\cite{bib:weberdiscovery}, but subsequent 
experiments with comparable or more sensitive
instruments failed to confirm the reported
detections~\cite{bib:tysongiffard}. 

In the following years the technology of bar detectors
improved steadily, with the introduction of cryogenic detectors
to reduce thermal noise, cryogenic squid transducers for
more efficient detection of bar excitations, and increasingly
sophisticated analysis techniques~\cite{bib:tysongiffard,bib:igec}.
In the late 1990s, before 1st-generation gravitational wave
interferometers came online, there were five major bar detectors
operating cooperatively in the International Gravitational 
Event Collaboration (IGEC)~\cite{bib:igec}: Allegro~\cite{bib:allegro}
at Louisiana State University; 
Auriga\cite{bib:auriga} at Padua University,
Explorer\cite{bib:explorernautilus} at CERN, 
Nautilus~\cite{bib:explorernautilus} at Frascati Laboratory,
and Niobe~\cite{bib:niobe} at the University of Western Australia.
These bars achieved impressive strain amplitude spectral noise
densities near $10^{-21}/\sqrt{{\rm Hz}}$, but only in narrow  
bands of $\sim$1-30~Hz~\cite{bib:bigbarbandwidths} near their resonant frequencies
(ranging from $\sim$700 Hz to $\sim$900 Hz). Hence waveform
reconstruction for all but very narrowband gravitational wave
sources was not feasible with these detectors. As of 2012,
only the Auriga and Nautilus
detectors are still collecting data,
since the major interferometer detectors LIGO and Virgo have
achieved broadband sensitivities better than the narrowband
sensitivities of the bars. It should be noted, however, that
as of 2012, the LIGO and Virgo detectors are undergoing major
upgrades (discussed in section~\ref{sec:advanceddetectors}),
leaving GEO 600 as the only major gravitational wave interferometer
collecting data routinely for the next several years. Should
a supernova occur in our galaxy during that time, only GEO 600 and
the remaining bar detectors would have a chance of detecting it
in gravitational waves.

\begin{figure}[tb]
\begin{center}
\epsfig{file=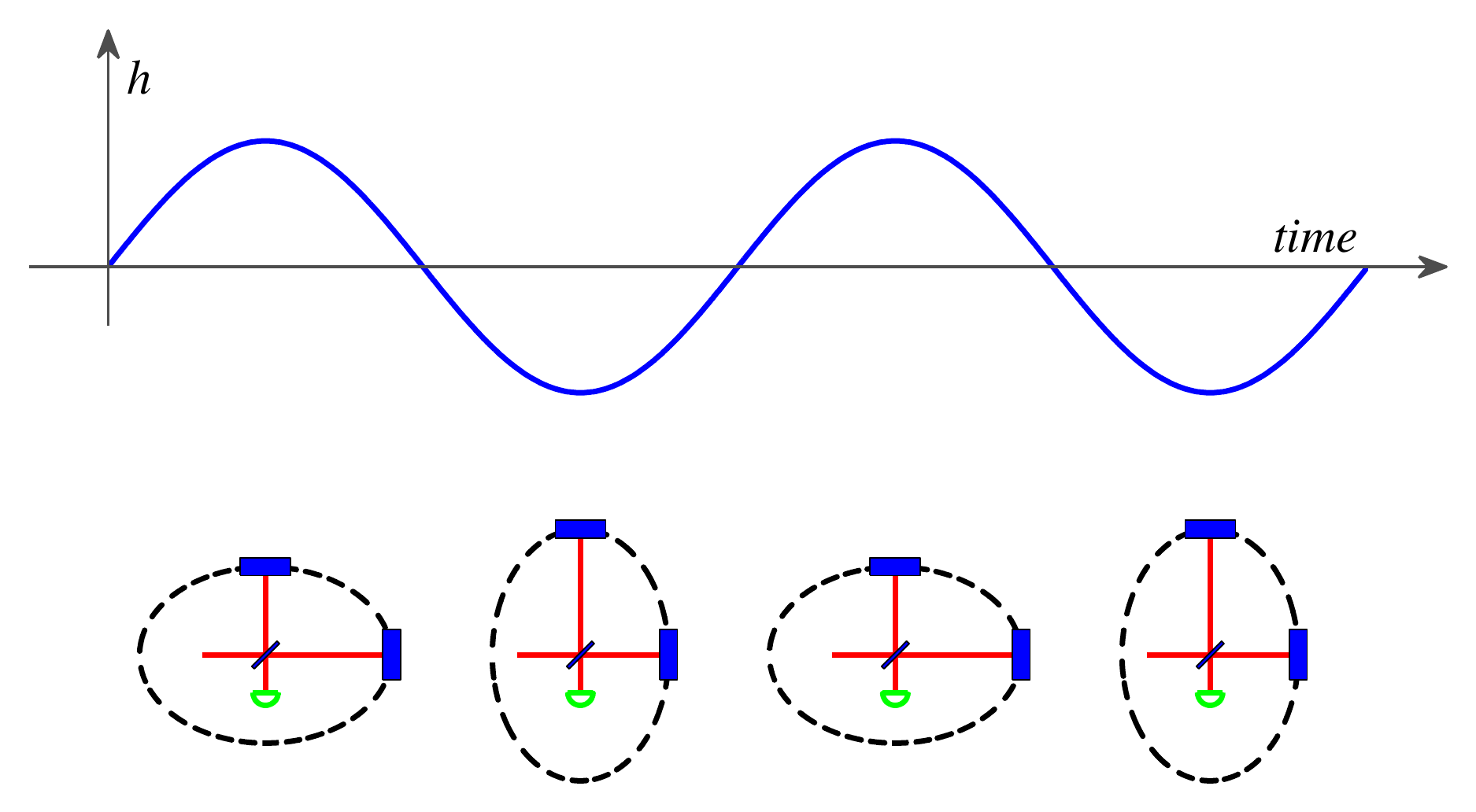,scale=0.5}
\caption{Cartoon illustration of the effect of a gravitational wave on
the arms of a Michelson interferometer, where the readout photodiode is denoted
by the green semi-circle~\cite{bib:s5detectorpaper}.
\label{fig:GWcartoon}}
\end{center}
\end{figure}

Gravitational-wave interferometers take a different approach to
detection from that of resonant bars. Setting aside enhancements to be discussed below,
a simple right-angle Michelson laser interferometer, 
as shown in the cartoon in figure~\ref{fig:GWcartoon}~\cite{bib:s5detectorpaper}
is a natural gravitational-wave detector. For example, a linearly
polarized wave impinging normally on the interferometer with
its polarization axis aligned with the arms will alternately
stretch one arm while contracting the other. One common question
is how this alternation is detected, given that the laser light
is stretched and compressed too. The answer is that the detection
is based on the phase difference between the light returning
from each arm, and that phase difference increases with time,
following the passage of the gravitational wave. The red-shifted
light simply takes longer to complete its round-trip in the
arm than the blue-shifted light. Hence even an idealized, simple
gravitational-wave interferometer has a finite and frequency-dependent
response time. An intuitive elaboration on this concept and
a related gedanken experiment can be found in ref.~\cite{bib:saulsongedanken}.

The basic idea for a gravitational wave interferometer
was first written down by Gertsenshtein and Pustovoit in 1962~\cite{bib:gertpust}.
Weber's group developed this idea further into the first 
gravitational wave interferometer prototype~\cite{bib:moss}
built by Weber graduate Robert Forward at Hughes Aircraft Research Lab~\cite{bib:forward}.
It was early work carried out in parallel by Rai Weiss~\cite{bib:weiss}, however, that
laid the groundwork for present-day gravitational wave interferometers.
As discussed further below, it
became appreciated quickly that laser interferometers had the
potential to surpass bar detectors in sensitivity, and there was
rapid development of ideas and technology. Subsequent improvements 
included (among many others) using Fabry-Perot cavities for the interferometer arms
to increase the time of exposure of the laser light to the
gravitational wave~\cite{bib:fparms}, introduction of a ``recycling''
mirror between the laser and beam-splitter, to increase
effective laser power~\cite{bib:powerrecycling}, and introduction
of another mirror between the beam splitter and photodetector
to allow tuning of the interferometer's frequency 
response~\cite{bib:signalrecycling}.

\subsection{Detector sensitivity and resolution considerations}
\label{sec:detsens}

In keeping with the concepts that led to their invention,
it is most natural to think of a bar detector as an
energy detection device, while an interferometer is more
naturally regarded as a strain amplitude detector. (An 
interesting, alternative perspective from which an interferometer
can be regarded as a non-linear parametric energy transducer
can be found in ref.~\cite{bib:saulsonenergy}.)

Let's consider what sensitivity one might expect from an
ideal bar detector of length $L$, mass $M$, operating at temperature $T$,
and having a resonant frequency (fundamental longitudinal harmonic) $f$
of mechanical quality factor $Q$. From the equipartition theorem,
the average energy of vibration of the mode is $\kb T$, where 
$\kb$ is Boltzmann's constant. For simplicity,
treat the bar's fundamental longitudinal mode of vibration as a
simple harmonic oscillator of spring constant $k$ 
with displacement of one end of the 
bar from its nominal distance of $L/2$ from the center as the 
spring's displacement from equilibrium with
the half of the bar providing mass $M/2$. Then we expect an RMS strain motion of
(from ${1\over2}kx_{\rm RMS}^2 \equiv {1\over2} (M/2)(2\pi f_0)^2 = {1\over2}k_{\rm B}T$):
\begin{equation}
h_{\rm RMS} \quad \sim\quad {x_{\rm rms}\over L/2}\quad \sim \quad {2\over L}\sqrt{{k_{rm B}T\over2\pi^2f_0^2M}}.
\end{equation}
Taking the LSU Allegro bar~\cite{bib:allegro} as an example, for 
which $L$ = 3.0 m, $f_0$ = 907 Hz, $M$ = 2296 kg, and $T$ = 4.2 K,
one obtains $h_{\rm RMS}\sim3\times\sim10^{-17}$.
Naively then, one might think that only gravitational waves
with characteristic amplitude much greater than $10^{-17}$ would
be detectable with such a bar. Fortunately, the fact that 
resonant bars are deliberately designed with high mechanical
quality factors $Q$ allows much better sensitivity than
this naive calculation suggests.
The impulse imparted by a passing wave is dissipated over
a time scale of $\sim Q/ f_0$. Hence by measuring 
over many cycles (but less than $Q$) of the resonance,
one can reduce the effective noise by a factor comparable to $\sqrt{Q}$.

It is interesting to examine relations involving the energy $E_{\rm dep}$ deposited
in the bar by the passing gravitational wave. From refs.~\cite{bib:saulsontext} and~\cite{bib:amaldi},
one has for the characteristic strain amplitude $h_c$ of a burst wave of
duration of characteristic time $\tau_{\rm burst}$:
\begin{equation}
h_c \quad \approx\quad {\sqrt{15}\over2} {L\over\tau_{\rm burst}v_s^2}\sqrt{{E_{\rm dep}\over M}},
\end{equation}
where $v_s$ is the speed of sound in the bar. For a characteristic
amplitude of $10^{-19}$, one obtains for the Allegro bar $E_{\rm dep}\sim4\times10^{-28}$ J.
For reference, for a gravitational wave of frequency 900 Hz,
this corresponds to a loss of about 700 gravitons, each of
energy $3.7\times10^{-12}$ eV.

The above calculations ignore non-fundamental but important additional
sources of noise, such as in the readout electronics or from the
terrestrial environment, including magnetic fields. 
Measuring the deposited gravitational wave energy is non-trivial.
The original piezo-electric transducers at the waists of the original
bar detectors evolved into transducers at the ends of the detectors,
where vibration amplitude is maximum. Using a transducer with an intrinsic
resonant frequency very near that of the bar leads to a coupled oscillator
with two normal modes and a beat frequency that defines the time scale
for the energy of the resonant bar to leak into the transducer.
This amplification trick~\cite{bib:paik} permits more efficient detection
readout.

There is a nominal quantum limit, however, to the performance of
a bar detector (as there is for an interferometer, as discussed below).
The readout of the energy in the fundamental harmonic is limited
by the quantum noise of the system at that frequency.
In summary~\cite{bib:saulsontext},
the nominal quantum limit on strain sensitivity on an ideal bar 
is 
\begin{equation}
h_{\rm RMS} \quad \approx \quad {1\over L}\sqrt{{\hbar\over2\pi f_0M}} \quad \approx \quad {\rm few}\>\times 10^{-21},
\end{equation}
where the numerical value is for a bar of Allegro's dimensions and mass.

This nominal quantum limit need not be truly
fundamental. By exploiting quantum ``squeezing'' (sacrificing phase
information for amplitude information), one can, in principle, do
somewhat better~\cite{bib:braginsky,bib:cavesetal}. But squeezing is 
notoriously delicate, in practice, offering little hope of improvement
by orders of magnitude.

It is amusing to compare the energy loss of a gravitational wave
impinging on a bar detector with that of high-energy neutrinos, which
are famous for their penetration. For example, 1-GeV
muon neutrinos traveling along the axis of a 3-meter long aluminum bar
have a probability of interacting of about $3\times10^{-12}$.
In comparison, a monochromatic (1 kHz), planar linearly polarized
gravitational wave of amplitude $h_+=10^{-19}$ has an energy flux $\mathcal{F}$
through the bar of [see equation~(\ref{eqn:energyflux})] of 1.6 kW/m$^2$,
while the energy deposition rate for a resonant bar with quality factor 
$Q=10^6$ is approximately $1.2\times10^{-22}$ W on resonance~\cite{bib:saulsonenergy},
giving a fractional energy loss of
O(10$^{-25}$), making the ``elusive neutrino'' seem relatively easy
to stop. Saulson~\cite{bib:saulsonenergy} computes
effective ``cross sections'' of resonant bars 
and finds, for example,  $\sigma_{\rm bar}/L_{\rm bar}^2\sim10^{-22}$.
Such a tiny value can be thought of as a measure of the weakness of
the gravitational interaction, or alternatively,
as a measure of the impedance mismatch between matter and 
extremely stiff space-time~\cite{bib:blairimpedance}.

Let's turn now to the expected sensitivity of interferometers.
For concreteness, consider a gravitational wave burst
with a duration of 1 ms and characteristic frequency in
the detector's sensitive band. In order to obtain a ``5-$\sigma$
detection,'' the intensity of the light at the photodetector
must change by at least what is required to 
be seen over shot noise (photon count statistical fluctuations).
For a 10-W ($\lambda$ = 1064 nm) laser impinging on the photodetector (after
recombination at the beam splitter), one has a relative statistical
fluctuation of $1/\sqrt{{(5.3\times10^{19}\,{\rm s}^{-1})(10^{-3}\>{\rm s})}}$ $\sim$ $4\times10^{-9}$.
Assuming a simple Michelson interferometer of the same size
as LIGO (4-km arms) with the beam splitter positioned to give 
a nominal light intensity at half its maximum (constructive
interference of the returning beams), one has a gain factor
of
\begin{equation}
{1\over I}{dI\over dh}\quad = \quad 8\pi {L\over\lambda} \quad \approx \quad 10^{11}.
\end{equation}
Hence to obtain 5-$\sigma$ detection, one needs a strain amplitude
of O($2\times10^{-19}$), which even in this relatively simple configuration,
already gives impressive broadband sensitivity. As discussed in
more detail below, the LIGO and Virgo interferometers have 
achieved significant improvement over this
sensitivity by using Fabry-Perot cavities in the arms and 
a recycling mirror to increase effective light power.
Other improvements, such as elaborate laser intensity and frequency 
stabilization, along with a heterodyne RF readout scheme that allows the interferometer
to operate with near-destructive interference at the photodiode, 
mitigate non-fundamental noise sources that would otherwise
invalidate the above simple model.

As with bars, there is a nominal quantum limit. Naively, one could
improve sensitivity arbitrarily by increasing laser power, 
to reduce shot noise ($\propto1/\sqrt{N_{\rm phot}}$),
but at some intensity, radiation pressure fluctuations ($\propto\sqrt{N_{\rm phot}}$)
become limiting. 
Note that the effects of radiation pressure are reduced by increasing
the masses of the mirrors. For example, Advanced LIGO mirrors will be
40 kg, much heavier than the 11-kg mirrors used for Initial LIGO, in order
to cope with the increased laser power ($\sim$180 W vs $\sim$10 W).
One could imagine increasing mirror mass with laser power indefinitely,
but sustaining the high optical quality and high mechanical quality factor
becomes more challenging. In addition, it becomes more difficult to
prevent internal vibrational modes from contaminating the detection band.

Once again, as with bars, the standard quantum limit can be evaded
via squeezing~\cite{bib:caves}, but with the opposite intent. In interferometer
squeezing, one sacrifices intensity sensitivity to achieve lower phase
noise, using an optical parametric amplifier at the output beam
of the interferometer. A useful way to think about the quantum
fluctuations is that when the interferometer is operated near a point
of destructive interference, vacuum fluctuations in the quantum field
``leak'' back into the interferometer. In some sense, squeezing the 
interferometer is actually squeezing the vacuum state with which
it interacts. Squeezing has been demonstrated not only in tabletop
experiments~\cite{bib:squeezingexperiments}, but also on two 
large-scale gravitational wave interferometers
(GEO 600~\cite{bib:geosqueezing} and LIGO~\cite{bib:h1squeezing}).
In the future squeezing may be used to go beyond design sensitivities
for Advanced LIGO and Virgo, or in the event that technical obstacles
arise at full laser power for those interferometers, squeezing
offers an alternative to reach design sensitivity at lower laser powers.

\begin{figure}[tb]
\begin{center}
\epsfig{file=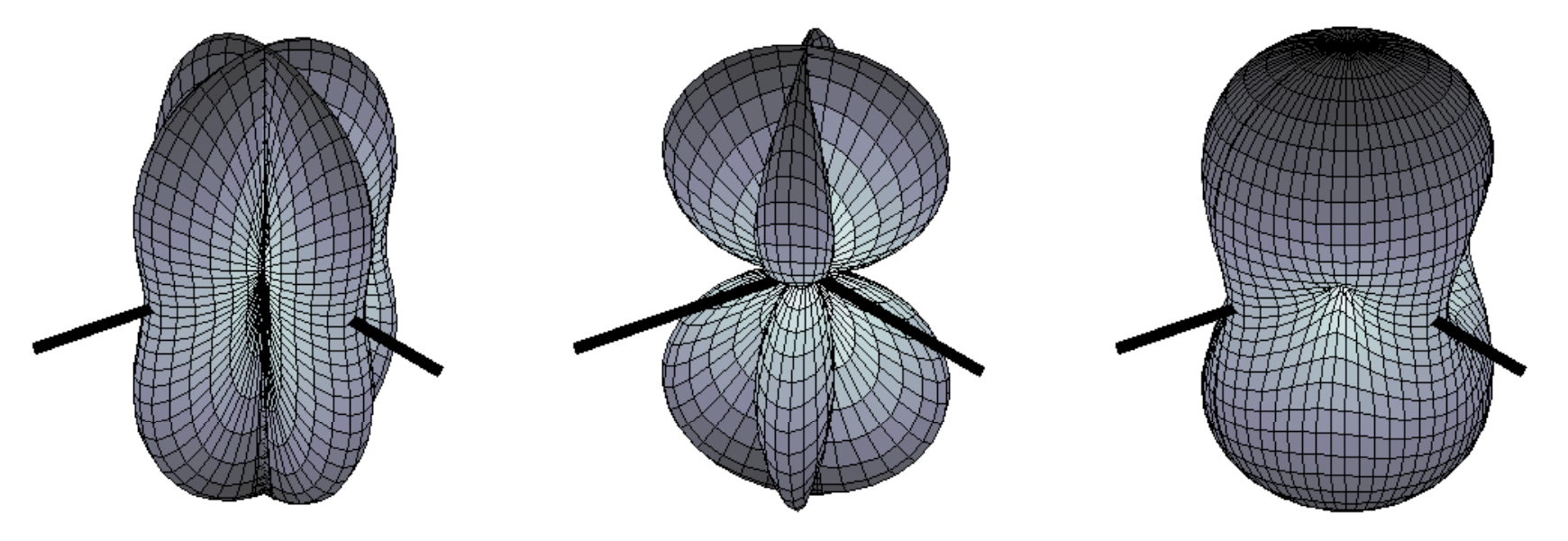,scale=0.65}
\caption{Antenna response pattern for a Michelson interferometer in
the long-wavelength approximation. The interferometer beamsplitter is located at
the center of each pattern, and the thick black lines indicate the orientation of the
interferometer arms. The distance from a point of the plot surface to the center of
the pattern is a measure of the gravitational wave sensitivity in this direction. The
pattern on the left is for $+$ polarization, the middle pattern is for $\times$ polarization, and
the right-most one is for unpolarized waves~\cite{bib:s5detectorpaper}.
\label{fig:PeanutAll}}
\end{center}
\end{figure}

Both bars and interferometers are better thought of as antennae
than as telescopes, because their sizes 
are small compared to the wavelengths they are meant to detect.
For example, a bar detector of length 3 m with a resonant frequency
of 900 Hz has L/$\lambda\sim10^{-5}$, while even the LIGO detectors when
searching at 4 kHz have  L/$\lambda$ of only about 0.05.  
These small ratios imply broad antenna lobes. 
Figure~\ref{fig:PeanutAll}~\cite{bib:s5detectorpaper}
shows the antenna lobes for $+$, $\times$ linear polarizations
and circular polarizations \vs\ incident direction for a
Michelson interferometer in the long-wavelength limit. As a result,
a single interferometer observing a transient event has very poor
directionality. 

One can do substantially better by triangulating
detections via multiple detectors. For a given SNR, consistency of
timing between each detector in a pair leads to an allowed annulus
on the sky with angular thickness $\propto$ 1/SNR. 
Combining each allowed pair in a network of three or 
more detectors favors intersections of these annular rings.
In principle, requiring amplitude consistency of a putative
sky location and the known relative orientations of the
detectors resolves resulting ambiguities from multiple intersections,
but polarization effects complicate that resolution, since detected
amplitudes depend on the typically unknown orientation of the 
gravitational wave source. 

A notable exception to the $L/\lambda\ll1$ rule of thumb is
detection of a long-lived continuous-wave source, where the Earth's
orbit around the solar system barycenter,
gives a single detector over the course of a year an effective aperture radius comparable
to the distance from the Earth to the Sun. For a 
nearly monochromatic, continuous-wave
source at 1 kHz, \eg, from a millisecond pulsar,
Rayleigh's criterion gives an angular resolution of approximately:
\begin{equation}
\Theta \quad \approx\quad {3\times 10^5\> {\rm m}\over 3\times10^{11}\> {\rm m}}\quad =\quad 10^{-6}\> {\rm radians}\>\> (0.2\> {\rm arcsec})
\end{equation}

\subsection{First-generation interferometers}
\label{sec:1stgeneration}

\begin{figure}
\begin{center}
\hbox{
\includegraphics[width=3in]{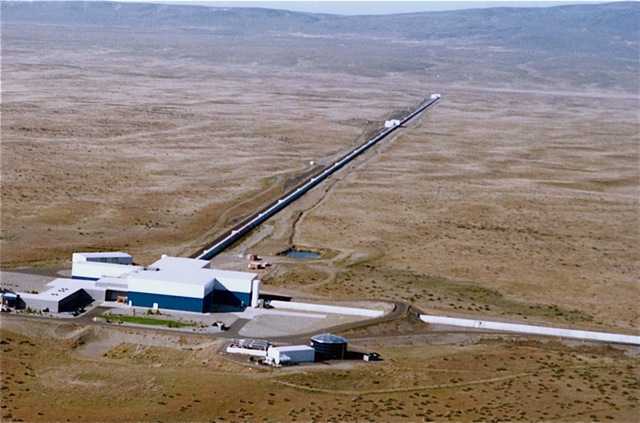}
\includegraphics[width=3in]{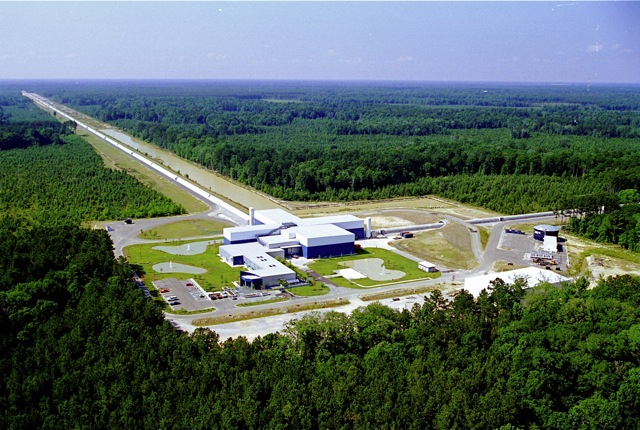}
}
\caption{Aerial photographs of the LIGO observatories at Hanford,
  Washington (left) and Livingston, Louisiana (right)~\cite{bib:s5detectorpaper}. 
  The lasers and optics are contained in the white and blue buildings. From 
  the large corner building, evacuated beam tubes extend at right
  angles for 4 km in each direction (the full length of only one
  of the arms is seen in each photo); the tubes are covered by the 
  arched, concrete enclosures seen here. Credit: LIGO Laboratory.
\label{fig:observatories}}
\end{center}
\end{figure}

The ``first generation'' of ground-based gravitational wave interferometers
is not well defined, since many of the prototype interferometers used
to demonstrate new technology developments were also used in prototype
gravitational wave searches, some of which led to journal publications.
Saulson~\cite{bib:saulsontext} provides a nice summary of these experiments
and searches. Prototype interferometers were built and operated around
the world in Australia, Europe, Japan and the United States.
These prototypes led eventually to the building of six major interferometers:
TAMA (300-m arms) near Tokyo~\cite{bib:tama},
GEO 600 (600-m arms) near Hannover~\cite{bib:geo},
Virgo (3000-m arms) near Pisa~\cite{bib:virgo}, and
LIGO (two with 4000-m arms and one with 2000-m arms ) in the states of
Washington and Louisiana~\cite{bib:abramowici,bib:barishweiss,bib:s1detectorpaper,bib:s5detectorpaper,bib:s6detectorpaper}.
Figure~\ref{fig:observatories} shows aerial views of the two LIGO sites in 
Hanford, Washington and Livingston, Louisiana.
The major interferometers share many design characteristics, but also
display significant differences. 

In the following, the design of the
4-km LIGO interferometers will be described in some detail, followed by
only a brief summary of differences between LIGO and the other major detectors.
As a result, the bibliography is LIGO-centric and makes no attempt to
document all of the critical technical developments leading to the other
major interferometers. A thorough documentation of all important developments
leading to LIGO, however, is also beyond the scope of this review.

We begin with a more detailed description than given above of how a power-recycled 
Fabry-Perot Michelson interferometer works.

\subsubsection{Fabry-Perot cavities}
\label{sec:fpcavities}

\begin{figure}[tb]
\begin{center}
\epsfig{file=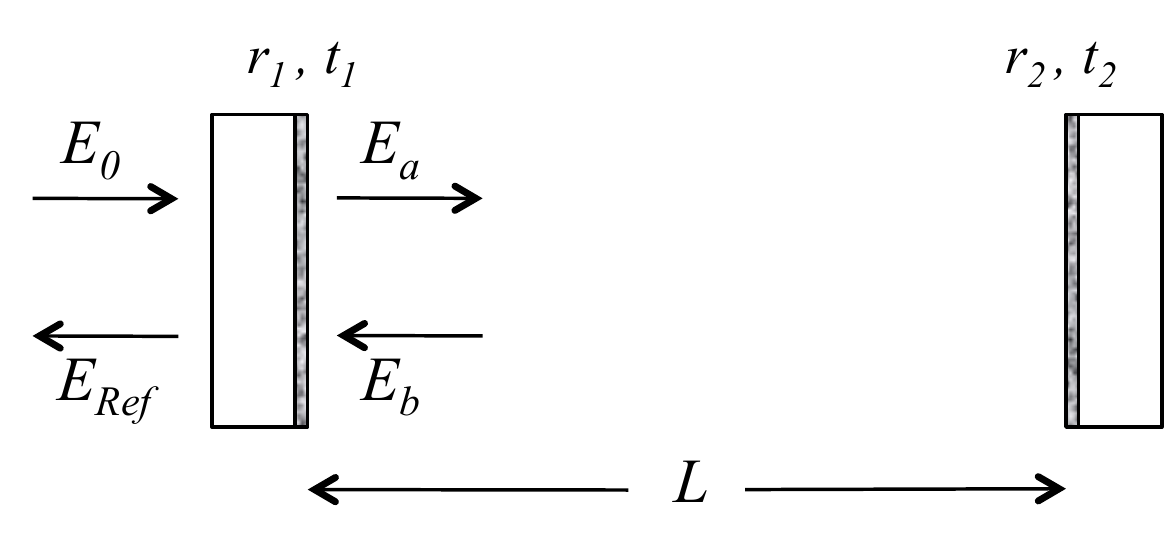,scale=0.7}
\caption{Schematic diagram of a flat-flat Fabry-Perot cavity with 
reflective coatings on the intra-cavity mirror surfaces. The 
$E_i$ labels and arrows refer to (signed) electric field amplitudes
of waves traveling in the directions of the arrow. $E_0$ denotes
the incident field on the cavity, and $E_{\rm Ref}$ denotes the
net reflected field. $r_i$ and $t_i$
denote the amplitude reflection and transmission coefficients of
the coated surfaces.
\label{fig:planarmirrors}}
\end{center}
\end{figure}

It is helpful to start by reviewing the essential principles of a Fabry-Perot cavity.
First, consider a cavity formed by two flat, parallel mirrors,
as shown in figure~\ref{fig:planarmirrors}
with a polarized plane electromagnetic wave of wavelength $\lambda$ 
incident from the left,
where the intra-cavity surfaces have amplitude reflectivity 
coefficients $r_1$ and $r_2$ and where the extra-cavity surfaces
are taken, for simplicity, to have perfect anti-reflective
coatings. For input laser power of electric field 
amplitude $E_0$ and in the steady state,
after start-up transients have settled down, one has the 
following relations among the electric field amplitudes of 
the light entering, leaving and residing in the cavity at the
two mirror surfaces: 
\begin{eqnarray}
E_a & = & t_1E_0 - r_1E_b \nonumber \\ 
E_b & = & -r_2\,e^{i\phi}E_a \nonumber \\
E_{\rm Ref} & = & r_1E_0+t_1E_b
\end{eqnarray}
where $E_a$ refers to the rightward-moving wave at the 1st mirror,
$E_b$ refers to the leftward moving wave at the 1st mirror,
$E_{\rm Ref}$ refers to the wave reflected from the 1st mirror, and
$\phi = 4\,\pi L/\lambda$ is the length-dependent phase shift due
to propagation from the 1st mirror to the 2nd mirror and back.
The sign convention chosen here is to take $r_1$ and $r_2$ both
positive. Solving these steady-state relations leads to
\begin{eqnarray}
E_a & = & {t_1\over1-r_1r_2e^{i\phi}}\,E_0, \nonumber \\
E_b & = &-{t_1r_2e^{i\phi}\over1-r_1r_2e^{i\phi}}\,E_0, \nonumber \\
E_{\rm Ref} & = & {r_1-r_2e^{i\phi}\over1-r_1r_2e^{i\phi}}\,E_0.
\end{eqnarray}
The cavity resonates when $\phi = 2\pi N$ for an integer $N$.
If, as is typically the case for Fabry-Perot arm cavities used
in gravitational wave interferometers, the reflectivity of
mirror 2 is much closer to unity than that of mirror 1,
then on resonance:
\begin{eqnarray}
E_a & \approx & {t_1\over1-r_1}E_0, \nonumber \\
E_b & \approx & -{t_1\over1-r_1}E_0, \nonumber \\
E_{\rm Ref} & \approx & -E_0.
\end{eqnarray}
In practice, there are small losses in the cavity, in the coatings
and transmission through mirror 2 that lead to small corrections
to these relations. Note that in the above lossless approximation,
energy conservation requires $|E_{\rm Ref}| = |E_0|$. 

Imagine that a single Fabry-Perot is on resonance, but
that a gravitational wave passes, leading to a momentary
increase of the cavity's length by an amount $\Delta L$.
Then the change in $E_{\rm Ref}$ is governed by the
derivative:
\begin{equation}
\left[{dE_{\rm Ref}\over dL}\right]_{\phi=2\pi N} \quad = \quad-i\,{(1-r_1^2)\,r_2\over(1-r_1r_2)^2}\,{2\pi\over\lambda}\,E_0 
\end{equation}
Again, taking the case $r_2\rightarrow1$ and $\delta r_1\equiv 1-r_1\ll 1$,
\begin{equation}
\left[{dE_{\rm Ref}\over dL}\right]_{\phi=2\pi N} \quad \approx \quad -i\,{2\over\delta r_1}\,{2\pi\over\lambda}\,E_0,
\end{equation}
which implies a large amplification in phase sensitivity for small $\delta r_1$.
One figure of merit is the cavity {\it finesse} $F \approx {\pi\sqrt{r_1}/(1-r_1)}$. 
In principle, a single cavity can therefore act as a gravitational wave detector,
but one gains in sensitivity by simultaneously monitoring another, identical arm oriented
at a right angle, not merely because of the potential to double the signal
strength from optimum quadrupole orientation, but much more important, mundane noise
sources that affect the input field amplitude $E_0$ and phase cancel in the difference-signal.

The above analysis idealized the laser light as a plane wave and used planar mirrors.
But high-quality lasers produce Gaussian beams with curved wavefronts.
Fabry-Perot cavities are normally designed with at least one curved mirror (usually concave as seen from
the cavity), such that a Gaussian beam resonates with a spherical wavefront at the
mirror(s) with radii of wavefront curvature at those locations to match those of the mirror(s). 
For a cavity of length $L$ with two concave mirrors of radii $R_1$ and $R_2$, it is useful~\cite{bib:siegman}
to define mirror $g$ factors $g_i \equiv 1 - L/R_i$, from which one can derive
the beam's characteristic intensity radii at the mirrors ($w_1$ and $w_2$) and 
the beam's ``waist'' (minimum radius):
\begin{equation}
w_1^2 = {L\lambda\over\pi}\sqrt{{g_2\over g_1(1-g_1g_2)}},\qquad
w_2^2 = {L\lambda\over\pi}\sqrt{{g_1\over g_2(1-g_1g_2)}}
\end{equation}
and
\begin{equation}
w_0^2 = {L\lambda\over\pi}\sqrt{{g_1g_2(1-g_1g_2)\over(g_1+g_2-2\,g_1g_2)^2}},\qquad
\end{equation}
Although the radii of curvature of the initial LIGO mirrors vary slightly,
round numbers for the four different 4-km arm cavities 
are $R_1 \approx 14,000$ m (input mirror near the beam splitter) and 
$R_2\approx 7,300$ m (end mirror), leading to beam radii at the
input mirrors or about 3.6 cm and 4.5 cm, respectively~\cite{bib:s5detectorpaper}, 
with an inferred beam waist of 3.5 cm about 1 km from the input mirror.
Ensuring that the mirror aperture is much larger than the maximum beam radius
is an important design constraint.

It should be noted that an infinite number of Gaussian-modulated waveforms
can resonate in a cavity (\eg, with Hermite-Gaussian or Laguerre-Gaussian envelopes), 
but their resonant lengths differ slightly because of differing Gouy phases~\cite{bib:siegman}.
As a result, a servo-locked cavity resonating in the fundamental mode (normally desired)
will not generally simultaneously resonate in higher order modes that could
introduce confusion into the servo error signal.
Matching the waist size and location of the input laser beam to the cavity is delicate,
with mismatches leading to degraded resonant power and reduced phase sensitivity in
$E_{\rm Ref}$.

Another important consideration is the finite time response of a Fabry-Perot cavity to length
changes, ignored in the above steady-state analysis. There is a characteristic time
scale $\tau\approx L\,/\,\pi c$ for the information of a disturbance to ``leak'' into the reflected
light $E_{\rm REF}$ ($\sim$1 ms for LIGO 4-km interferometers). For a fixed arm length $L$, the time constant $\tau$ increases with the cavity finesse.
Hence, while increasing finesse increases phase sensitivity at DC, it leads to more
rapid onset of amplitude loss with higher frequencies, 
as measured by the {\it cavity pole} $f_{\rm Pole} = 1/(4\pi\tau)$ ($\sim$85 Hz for LIGO 4-km interferometers). 

How does one maintain a Fabry-Perot cavity on resonance, especially when the mirrors are
suspended as free pendula, as discussed below? One needs a negative-feedback servo control system with an
error signal proportional to the deviation of the cavity from resonance and with an
actuation mechanism to bring the cavity back to resonance by forcing the error signal
to zero (to a level consistent with a necessarily finite gain). For gravitational-wave
interferometers, the servo control system is based on Pound-Drever-Hall (PDH) locking~\cite{bib:pdh}.
In this scheme, the laser light is phase-modulated at a radio frequency $f_{\rm mod}$
and a photodetector viewing the reflected light is demodulated at that frequency.
To see why this method is effective, consider the Bessel function expansion of a 
phase-modulated field:
\begin{eqnarray}
\label{eqn:bessel}
E_0 \,e^{i[\omega t + \Gamma\cos(\Omega_{\rm mod}t)]} & = & E_0\>\bigl[J_0(\Gamma)e^{i\omega t}
\>+\>iJ_1(\Gamma)e^{i(\omega+\Omega_{\rm mod})t} 
\>+\>iJ_1(\Gamma)e^{i(\omega-\Omega_{\rm mod})t} \nonumber \\
& & -\>J_2(\Gamma)e^{i(\omega+2\Omega_{\rm mod})t} 
\>-\>J_2(\Gamma)e^{i(\omega-2\Omega_{\rm mod})t}\> +\> ... .\bigr]
\end{eqnarray}
The field can be treated as a carrier with sidebands at integer harmonics of the
modulation frequency, where for a moderate modulation depth ($<$ 1 radian), the
strengths of the higher-order harmonics fall off rapidly.  Note, however, that
the time-averaged intensity of a purely phase modulated beam is monochromatic:
\begin{equation}
|E_0 e^{i\omega + \Gamma\cos(\Omega_{\rm mod}t))t}|^2\quad = \quad E_0^2,
\end{equation}
that is, no sidebands are apparent (imagine measuring power over many cycles of $\omega$,
but over only a fraction of a cycle of $\Omega_{\rm mod}$, where 
typically $\Omega_{\rm mod}/\omega < 10^{-6}$). An exercise for the reader
is verification that explicit summation of the intensity contributions
from equation~(\ref{eqn:bessel}) at $\Omega_{\rm mod}t$ and $2\,\Omega_{\rm mod}t$
cancel to zero. The key to the PDH scheme~\cite{bib:pdhblack}
is that the carrier and sidebands have different resonant characteristics in a
Fabry-Perot cavity. For example, the carrier might resonate, while
the fundamental sidebands reflect promptly with negligible leakage into
the cavity. In that case (for $\Omega_{\rm mod} \gg 2\pi 
f_{\rm FSR}$, where $f_{\rm FSR} = c\,/\,2L$ is the free spectral range 
of the cavity) the beat between the reflected carrier and sideband
will cancel for carrier resonance, but will have a non-zero residual
beat when the carrier is off resonance. For small deviations from resonance, the
strength at the beat frequency is proportional to the deviation in
cavity length from its resonance value. Hence an error signal for
locking the cavity frequency to the laser frequency can be derived.
One can just as well, treat the error signal as a measure of the
laser frequency's deviation from what resonates in the cavity,
and feed back to the laser frequency to lock the servo, \eg,
by actuating on a piezoelectric controller on one mirror of the
Fabry-Perot lasing cavity.

\subsubsection{Power-recycled Michelson interferometry and Initial LIGO}
\label{sec:iligo}

The initial LIGO detector was a set of three power-recycled Michelson
interferometers with the parameters given in 
table~\ref{tab:ligoparams}~\cite{bib:s5detectorpaper}.
Power recycled interferometry is explained in this section,
along with noise considerations. The LIGO laser source was a diode-pumped,
Nd:YAG master oscillator and power amplifier system, and emitted
10 W in a single mode at 1064 nm~\cite{bib:savage}
The beam passed through an ultra-high vacuum system (10$^{-8}$-10$^{-9}$ Torr)
to reduce phase fluctuations from light scattering off residual gas~\cite{bib:scattering}
and to ensure acoustical isolation. The 4-km stainless steel tubes of 1.2-m diameter
were baked at 160$^\circ$ for 20 days to remove hydrogen.

The mirrors defining the interferometer were fused-silica substrates
with multilayer dielectric coatings having extremely low scatter, low absorption
and high optical quality
These mirrors
were suspended as pendula with a natural oscillation frequency of
$\sim$0.76 Hz, designed to respond as essentially free ``test masses'' to gravitational
waves while being isolated from ground motion by the $\sim1/\omega^2$ filtering
of the pendulum. The suspension came from a single loop of steel wire
around each mirror's waist. Mirrors were controlled by electromagnetic
actuators -- magnets bonded to the optics and influenced by currents
in nearby coils mounted on the support structure. For further isolation
from ground motion, the pendulum support structures were mounted on
four stages of mass-spring isolation stacks~\cite{bib:giaime}.

In addition to locking the arms on resonance, in order to obtain
exquisite sensitivity to distance changes between the arm mirrors,
one must also ``lock'' the relative position of the beam splitter
with respect to the arm input mirrors so as to establish a well
defined interference condition at the output photodetector.
Although one might naively choose the interference to be
halfway between fully destructive and fully constructive,
in order to maximize the derivative of intensity with
respect to relative phase of the light returning from
the arms, it pays instead to choose an interference operating
point that is at or near fully destructive (a null condition)~\cite{bib:saulsontext}.

The initial LIGO interferometers chose a null operating point for all but the
final science data runs, where a small offset was introduced,
as discussed below. The primary advantage of destructive
interference is reduction of effective noise. In principle, the disturbance
from a gravitational wave produces a non-zero light intensity
where there was previously only dark current. 
One might worry that the increase in intensity would
lead to a phase ambiguity, but the PDH
signals used to control the interferometer arms provided
a neat solution. By introducing a deliberate (Schnupp)
asymmetry~\cite{bib:schnupp} (355 mm for LIGO) in the distance between the
beam splitter and the arm input mirrors, one obtains non-cancelling
PDH sidebands at the photodetector that stand ready to
beat with any non-cancelling carrier light induced by
a passing gravitational wave, where the phase of the
beat signal reveals the phase of the gravitational wave.

\begin{table}
\begin{center}
\begin{tabular}{lccc}
   &  {\bf H1}  &  {\bf L1}  &  {\bf H2} \\
\hline\hline
Laser type and wavelength & \multicolumn{3}{c}{Nd:YAG, $\lambda = 1064$~nm} \\
Arm cavity finesse & \multicolumn{3}{c}{220} \\
Arm length & 3995 m & 3995 m & 2009 m \\
Arm cavity storage time, $\tau_s$ & 0.95 ms &  0.95 ms & 0.475 ms \\
Input power at recycling mirror & 4.5 W & 4.5 W & 2.0 W \\
Power Recycling gain  & 60 & 45 & 70 \\
Arm cavity stored power  & 20 kW & 15 kW & 10 kW \\
Test mass size \& mass  &  \multicolumn{3}{c}{$\phi\,25\,{\rm cm}\times10\,{\rm cm}$, 
   10.7 kg} \\
Beam radius ($1/e^2$ power) ITM/ETM & 3.6\,cm\,/\,4.5\,cm & 3.9\,cm\,/\,4.5\,cm &
  3.3\,cm\,/\,3.5\,cm \\
Test mass pendulum frequency & \multicolumn{3}{c}{0.76 Hz} \\
\hline\hline
\end{tabular}
\caption{Parameters of the LIGO interferometers. H1 and H2 refer to
the interferometers at Hanford, Washington, and L1 is the
interferometer at Livingston Parish, Louisiana~\cite{bib:s5detectorpaper}.} 
\label{tab:ligoparams}
\end{center}
\end{table}

In the optical configuration described so far there are
three longitudinal degrees of freedom that must be controlled:
the distances between the pairs of mirrors forming the
Fabry-Perot arms and the difference in distance between the
beam splitter and the two input mirrors. Now we add one
more primary mirror to control, namely the recycling mirror
located between the laser and the beam splitter. 
By ensuring that the average optical path length between
the recycling mirror and the arm input mirrors is a half
integer of laser wavelengths, one achieves resonance of
the light in what is called the recycling cavity\cite{bib:powerrecycling,bib:recyclingcontrol}.
Hence the light returning toward the laser that would 
have otherwise been discarded by the Faraday isolation optics
(to prevent interference with the laser itself), is
recycled back into the main interferometer. This recycling
effectively increases the laser power in the entire
interferometer and thereby decreases shot noise due to
limited photon statistics. The addition of this mirror
increases the number of primary longitudinal servo-controlled
degrees of freedom to four. 

As one might expect, simultaneously controlling these
four degrees of freedom is a technical challenge for free-swinging
pendulum mirrors subject to environmental disturbances.
The challenge is increased by several factors: the small time
window in each swing during which the PDH error signal is effective (having high gain),
the power transients associated with individual arm locks,
and the overall change in sign of the differential arm signal
as both arms resonate, requiring the servo feedback to
reverse sign in tens of milliseconds. A technique~\cite{bib:evans}
based on allowing the servo to ``coast'' through that
delicate transition period provides a robust solution,
albeit one that still relies upon stochastic swinging of
mirrors to bring the degrees of freedom under simultaneous
control.

In addition to controlling the longitudinal degrees of freedom,
one must also address alignment of the mirrors, and to a lesser
degree, transverse displacement. Wobbling of the mirrors
modulates interferometer gain, leading to non-linear
noise. Hence each angular degree of freedom is also servo-controlled.
There are two distinct stages of angular control. The first,
which works well for initial control authority, is based on
shining an auxiliary laser on each mirror at a non-normal
angle and observing the transverse displacement of the
reflected beam. This ``optical lever'' method is straightforward
and does not require the longitudinal degrees of freedom
to be locked, but it comes with the risk of introducing
extra noise due to the auxiliary laser and due to any ground motion
of the external photodiode.
The second angular control method is known
as ``wave front sensing'' and uses the PDH sidebands to
sense the misalignment of optical cavity degrees of
freedom~\cite{bib:wfs}. This second method can only be used
when the longitudinal degrees of freedom are locked, at which
point it ``takes over'' from the optical lever (at least at
lower frequencies) after fully
recycled lock is achieved. Because the method is
based on spatial asymmetries in carrier-sideband phase differences caused by cavity
misalignment, one can obtain a DC alignment signal without
relying on the external reference points that introduce additional
positional noise.

There are also degrees of freedom to control
for the input laser light before it impinges on the power
recycling mirror. The laser itself, of course, has
servo-controlled mirrors defining the lasing cavities.
In addition, a triangular ``mode cleaner'' cavity between it and the
recycling cavity provides a means to transmit only a
Gaussian wavefront to the main interferometer.
The intrinsic frequency stability of the laser is
increased by locking the laser to the interferometer arms. It should be
noted that tidal effects from the Moon and Sun lead 
to compression and stretching of LIGO's arms at
the level of hundreds of microns, well outside the 
dynamic range of the low-noise direct actuation 
on the mirrors (based on sending currents through
voice coils surrounding cylindrical magnets bonded to
the faces of the mirrors near their edges). 
To cope with this predictable and very slowly changing
disturbance, a feed-forward actuation is applied via
piezoelectric transducers to the positions of the
vacuum chambers supporting the mirrors and to the
frequency of the laser via a temperature controlled
reference cavity with respect to which the laser frequency
is offset ($\sim$10 pm), via an acousto-optic modulator.

\begin{figure}
\begin{center}
\includegraphics[width=13cm]{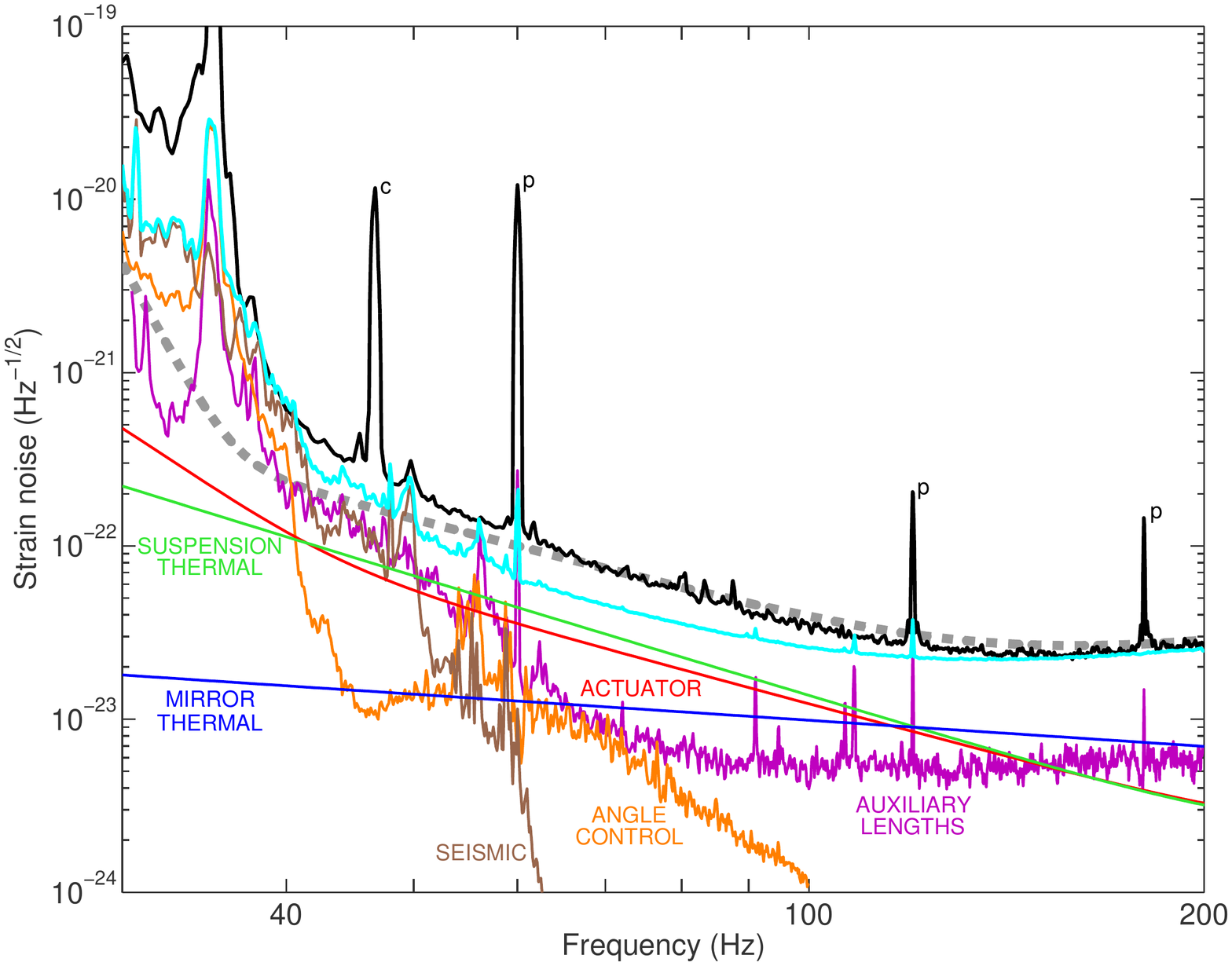}
\includegraphics[width=13cm]{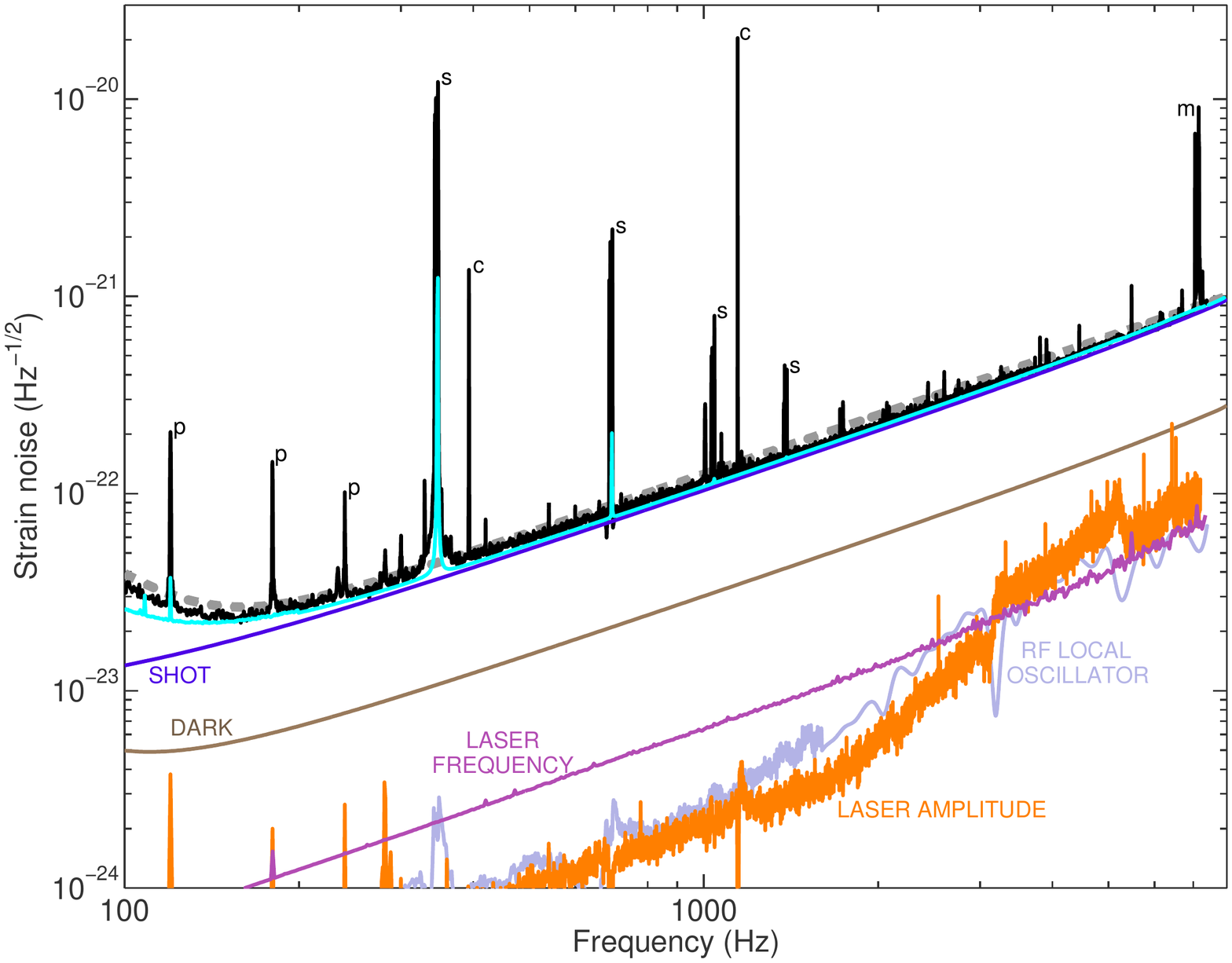}
\caption{Primary known contributors to the H1 detector noise
  spectrum~\cite{bib:s5detectorpaper}. The upper panel shows the displacement noise components,
  while the lower panel shows sensing noises (note the different
  frequency scales). In both panels, the black curve is the measured
  strain noise (same spectrum as in Fig.~\ref{fig:s5strainspectra}), the
  dashed gray curve is the design goal, and the cyan curve is the
  root-square-sum of all known contributors (both sensing and
  displacement noises). The labeled component curves are summarized
  in the text and described in more detail in ref.~\cite{bib:s5detectorpaper}.
  The known noise sources explain the observed noise very
  well at frequencies above 150~Hz, and to within a factor of 2 in the
  40\,--\,100~Hz band. Spectral peaks are identified as follows: c,
  calibration line; p, power line harmonic; s, suspension wire
  vibrational mode; m, mirror (test mass) vibrational
  mode. \label{fig:noisebudget}}
\end{center}
\end{figure}

It is beyond the scope of this article to discuss
interferometer noise sources in detail~\cite{bib:s5detectorpaper}.
But it is useful to summarize the primary known contributing sources.
Figure~\ref{fig:noisebudget} shows an example of a LIGO
noise budget graph for the Hanford 4-km (H1) interferometer.
At low frequencies ($<\sim$45 Hz), the noise is dominated by seismic ground
motion, despite the strong  isolation provided by the multiple
stages of passive oscillators. At high frequencies ($>\sim$100 Hz), sensing shot noise dominates.
At intermediate frequencies important known contributions come
from noise coupled to two auxiliary servoed degrees of freedom, which can
be thought of as the positions of the beam splitter and recycling mirrors;
from thermal noise in the suspension wires; from noise coupled to
mirror alignment fluctuations (due to residual beam non-centering); 
and from current noise in the actuation electronics, which 
must support demanding dynamic range requirements~\cite{bib:s5detectorpaper}.
Less important noise contributions come from thermal noise in the mirrors;
from dark-current noise in the photodiodes; from laser frequency noise;
from laser amplitude noise; and from phase noise in the RF oscillator used for
PDH locking and for the heterodyne readout. Some of the noise curves shown
in the budget are based on models, while others can be determined from
measured transfer functions.

At the minimum of the noise curve one expects (by design) the most
important noise source to be from the thermal noise of the suspension wires, where
there is $\kb T$ of vibrational energy over the entire band.
One strives for high mechanical Q's for the wires so that
the bulk of the energy is contained in a narrow band around
the wire resonant frequencies (``violin modes'' -- $\sim$350 Hz and harmonics). 
{\it A priori} predictions of suspension thermal noise are
challenging, depending on detailed modeling of the dissipative
losses in the wire and in the contact points with the mirrors
and supports. 

There are also electromagnetic environmental 
effects from ambient power mains magnetic fields, despite the careful anti-alignments
of magnets used in actuation, as suggested from the 60-Hz
harmonics seen in figure~\ref{fig:noisebudget}.

In addition to these mostly well understood noise sources,
there were suspected additional technical sources of noise
to account for the difference between measured and expected sensitivity seen
in figure~\ref{fig:noisebudget}. 
In particular, it is likely that the gap in agreement between $\sim$40-120 Hz
was mainly due to non-linear upconversion of low-frequency
noise. Upconversion can arise, for example, from mirror wobble, 
modulated beam apertures, and modulated beam backscattering~\cite{bib:backscattering}.
Scattering from surfaces attached rigidly to the ground is strongly suppressed
by elaborate serrated-edge baffling along the length of beam pipe 
and in other strategic locations, but even tiny 
scattering contributions can be deleterious when the surface is 
moving relative to the mirrors~\cite{bib:backscattering}.

More important, however, it is now believed that a substantial contribution to initial
LIGO detector upconversion was Barkhausen noise from interactions between
magnetic material used in the mirror actuation system and the voice actuation
coils. The stochastic flipping of magnet domains creates a fluctuating
force contribution. Another likely contributor at times was electrostatic
charging of mirrors, which can lead to fluctuating forces on the mirrors
as charges move to reduce local charge density~\cite{bib:s5detectorpaper}.

Other sources of noise come from imperfection in the optical configurations,
including small asymmetries between the effective reflectivities and losses
in the interferometer arms and higher-order Gaussian modes (carrier and sideband).
A particularly troublesome noise source was instability of the recycling
cavity with respect to sidebands. The recycling cavity was nearly flat-flat,
leading to intrinsic instability (``walk off'') of the beam. Since the sidebands
resonated in the cavity, but not in the arms, their strengths were especially susceptible to 
misalignment and wobble of the input mirrors, recycling mirror and beam splitter.

It should be noted that passive isolation did not suffice to
enable 24-hour operations at LIGO Livingston Observatory in
Louisiana. The observatory is surrounded by a pine forest used
by loggers. The sawing and removal of trees generated
excess seismic noise in the few-Hz band, which the passive 
mass-and-spring stacks did little to mitigate and, in some cases,
amplified. To cope with this nearly constant weekday
environmental disturbance, an active feed-forward system~\cite{bib:hepi}
based on hydraulic actuation exerted on vacuum chambers, driven by
signals from seismometers, geophones and accelerometers was
installed and commissioned, using technology originally developed
as part of Advanced LIGO research and development.

Another technical issue arose from higher-than-expected thermal absorption
in the input mirrors of each arm, causing thermal lensing and
degrading the matching of beam shapes into the arm Fabry-Perot
cavities~\cite{bib:tcs1}. To cope with this degradation, a thermal compensation system
was developed, based on shining a CO$_2$ laser on the input mirrors
so as to compensate the thermal lensing (to lowest order)~\cite{bib:tcs2}.

\begin{figure}
\begin{center}
\includegraphics[width=11cm]{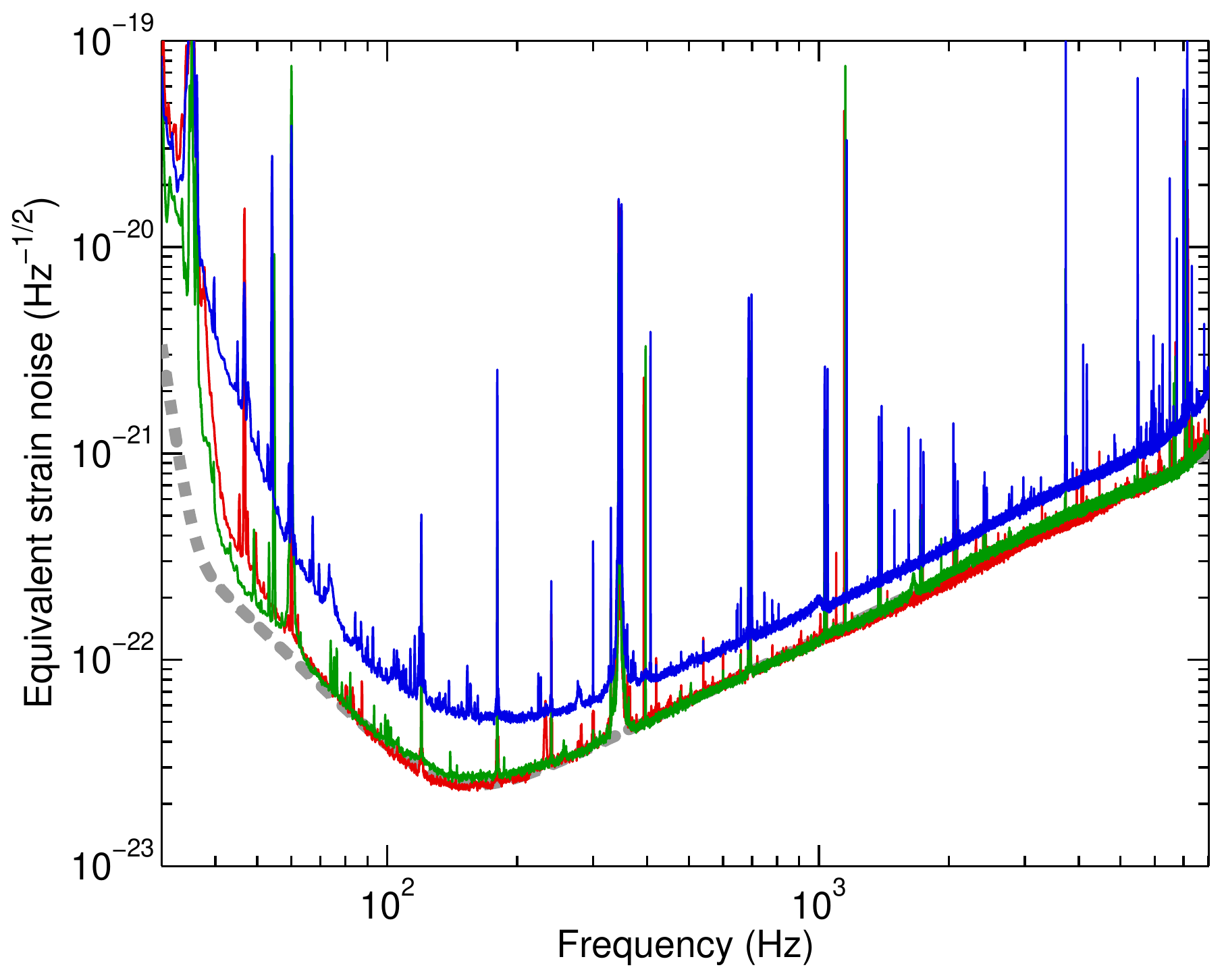}
\caption{Strain sensitivities (amplitude spectral densities)
  of the LIGO interferometers during
  the S5 data run~\cite{bib:s5detectorpaper}.
  and Virgo interferometer~\cite{bib:s5detectorpaper}.
  Shown are typical high sensitivity spectra for each of the three
  interferometers (red: H1; blue: H2; green: L1), along with the
  design goal for the 4-km detectors (dashed gray).
\label{fig:s5strainspectra}
}
\end{center}
\end{figure}

Commissioning of the LIGO interferometers required several years, as the sensitivities
of the instruments approached their designs. The official design requirement~\cite{bib:srd}
was to reach a band-limited RMS strain in a 100-Hz band as low as 10$^{-21}$. In addition,
a more optimistic aspirational target curve was produced. 
Figure~\ref{fig:s5strainspectra}~\cite{bib:s5detectorpaper}
shows typical sensitivities of the three interferometers in the S5 data run 
(November 2005 - September 2007), along with the target curve.
As seen, the final S5 sensitivities of the two 4-km interferometers did indeed reach
the target curve over a broad band and easily achieved the design band-limited strain
in the best (non-contiguous because of 60-Hz harmonics) 100-Hz band. These measured
sensitivities depend in part on a model of the interferometer response to
gravitational waves~\cite{bib:iforesponse}, but primarily upon stimulus-response
calibration. Three distinct methods have been used for calibrating LIGO
interferometers~\cite{bib:calibs5paper}: 1) calibration of voice coil actuators via fringe counting
in a simpler, unlocked Michelson interferometer configuration~\cite{bib:calibvoicecoil};
2) frequency modulation of the laser in a 1-arm configuration~\cite{bib:calibfreqmod}
in which frequency modulation is mapped to length modulation of the cavities;
and 3) calibration from photon radiation pressure using an auxiliary laser 
in the full interferometer configuration~\cite{bib:calibpcal}.

\subsubsection{Enhanced LIGO}
\label{sec:eligo}

Following the S5 run, the LIGO interferometers underwent an ``enhancement'' to
improve strain sensitivity by a factor of two in the shot-noise regime. This upgrade
was based on increasing laser power from 6 W to more than 20 W, but a simple increase
in power would have led to unacceptably high noise from higher-order-modes 
light impinging on the photodetector. To avoid this problem, an ``output mode cleaner''
(4-mirror bow-tie configuration) was installed between the beam splitter and the photodetector, to ensure that higher-order
modes (carrier {\it and} sidebands) were filtered out, 
leaving only a clean Gaussian measure of interferometer light~\cite{bib:s6detectorpaper,bib:dcreadout}.
This cleaning method also filters out the PDH sideband light, preventing those sidebands
from being used in the fully null experiment described above. 

\begin{figure}[tp]
\vskip -0.4cm
\begin{center}
\includegraphics[width=13cm]{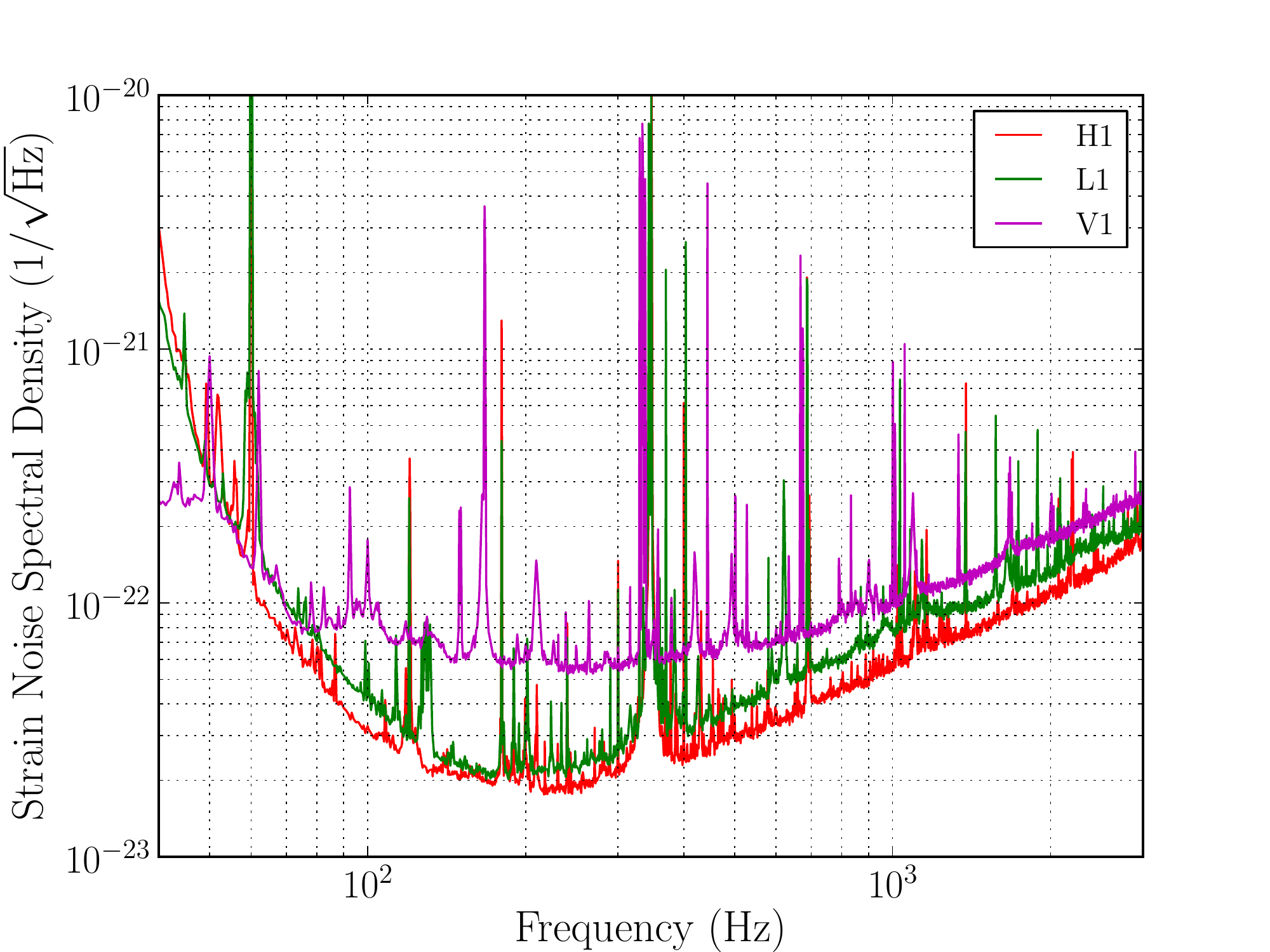}
\caption{Typical detector strain amplitude spectral densities for the LIGO S6
and Virgo VSR2/3 runs. From lowest to highest at $10^2$ Hz, the curves are for
the LIGO interferometers (H1 and L1) and Virgo interferometer~\cite{bib:cbcsearchs6}.
\label{fig:s6sensitivity}}
\end{center}
\end{figure}

Instead, a small, deliberate offset ($\sim$10 pm)
was introduced into the differential arm servo so that a gravitational wave disturbance
would lead to a change in intensity of photodetector light (``DC Readout'')~\cite{bib:dcreadout}. Although
this technique would seem to lead to the worry of intensity fluctuations in the laser
mimicking a gravitational wave, aggressive gain in the laser intensity stabilization servo
allowed operation in this mode~\cite{bib:dcreadout}. The Enhanced LIGO upgrades were applied
to the Hanford and Livingston 4-km interferometers (H1 and L1) from fall 2007 to summer 2009,
with commencement of the sixth science run (S6) in July 2009 and completion in October 2010.
By the end of the S6 run, the sensitivities of the interferometers had reached the curves shown
in figure~\ref{fig:s6sensitivity}, with an approximate factor of two improvement in 
instantaneous sensitivity above $\sim$300 Hz, as expected, given the higher laser power.
It had been hoped that noise at lower frequencies would also be reduced after replacement
of the primary actuation magnets, using material with smaller Barkhausen noise~\cite{bib:barkhausen}
(NdFeB replaced with SmCo), but upconversion noise remained,
and it was later hypothesized (but not conclusively established) that magnetized metal
components slightly further from the mirrors were the source of the noise~\cite{bib:barkhausenrevised}.

\subsubsection{Virgo interferometer}
\label{sec:virgo}

The Virgo interferometer has a quite similar design to that of LIGO and comparable
performance. 
The primary differences are in the arm lengths (3 km \vs\ 4 km), laser power (17 W vs 10 W)
and in the seismic isolation. While not as sensitive as LIGO in the most sensitive
band near 150 Hz, Virgo is substantially more sensitive at frequencies below 40 Hz
because of aggressive seismic isolation. Virgo's mirrors are suspended as 5-stage pendula
supported by a 3-legged inverted pendulum~\cite{bib:superattenuator}, a system
known as the {\it superattenuator}.
This extreme seismic isolation 
permitted Virgo to probe gravitational waves down to $\sim$10-20 Hz, in contrast to 
LIGO's $\sim$40-50 Hz. 
This lower reach offers the potential to detect
low-frequency spinning neutron stars like Vela that are inaccessible to LIGO.
Figure~\ref{fig:s6sensitivity} shows the sensitivity achieved by the Virgo
interferometer during the VSR2/3 runs.

\subsubsection{GEO 600 interferometer}
\label{sec:geo}

The GEO 600
interferometer~\cite{bib:geo} has served not only as an observatory
keeping watch on the nearby galaxy when the LIGO and Virgo interferometers have
been down (and serving as a potential confirmation instrument in the event of a very 
loud event candidate), it has also served as a testbed for Advanced LIGO technology.
With 600-meter, folded (non-Fabry-Perot) arms and a 12-W input laser and built
on a relatively small budget, GEO 600 cannot match the sensitivity of the LIGO or Virgo 
interferometers, but it has pioneered several innovations
to be used in Advanced LIGO: multiple-pendulum suspensions, signal recycling,
rod-laser amplification, and squeezing. As of 2012, GEO 600 is operating at high duty
factor in ``AstroWatch'' mode, primarily in case of a nearby galactic supernova, as the
LIGO and Virgo detectors undergo major upgrades.

\subsubsection{TAMA interferometer}
\label{sec:tama}

The 300-meter TAMA
interferometer~\cite{bib:tama} in Japan was similar to the LIGO
detectors (power recycled Michelson interferometer), but with much shorter arms
and comparable laser power, its sensitivity was limited. Nonetheless, it 
operated at comparable sensitivity to LIGO in LIGO's early runs, and joint data analysis
was carried out on S2 data~\cite{bib:cbctama,bib:bursttama}. The Japanese collaboration
that built TAMA is now building the 2nd-generation KAGRA detector discussed briefly below.

\subsection{Second-generation interferometers}
\label{sec:advanceddetectors}

The LIGO and Virgo detectors are now undergoing major upgrades
to become Advanced LIGO~\cite{bib:advligo} and Advanced Virgo~\cite{bib:advvirgo}. 
These upgrades are expected to improve their broadband strain
sensitivities by an order of magnitude, thereby increasing their
effective ranges by the same amount. Since the volume of accessible
space grows as the cube of the range, one can expect the advanced
detectors to probe roughly 1000 times more volume and therefore
have expected transient event rates O(1000) times higher than for the 1st-generation detectors.

\begin{figure}[tb]
\begin{center}
\epsfig{file=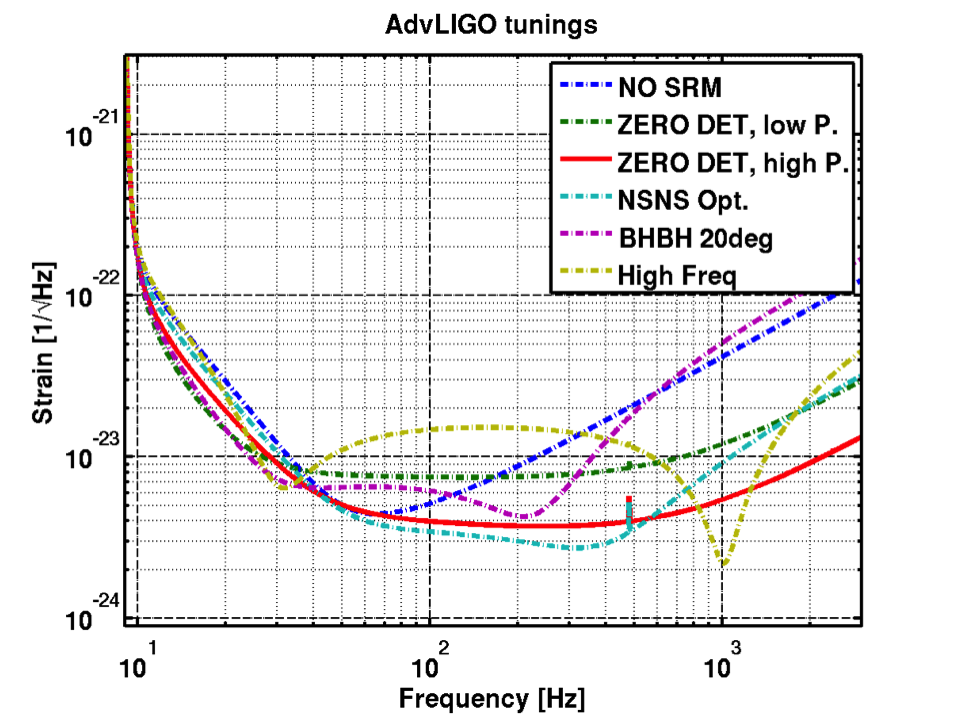,scale=0.8}
\caption{Projected Advanced LIGO strain amplitude spectral noise densities
for several different possible configurations~\cite{bib:advligo}. The
curve labeled ``ZERO DET, high P' corresponds to the nominal Advanced
LIGO high-power broadband operation. Sensitivity curves are also shown
for a lower power and for sample interferometer tunings that favor 
certain frequency bands.
\label{fig:advligosens}}
\end{center}
\end{figure}

The key improvements
for Advanced LIGO are 1) increased laser power ($\sim$10 W $\rightarrow$ $\sim$180 W) 
with rod-laser amplification developed by GEO collaborators~\cite{bib:advligolaser}, to
improve shot noise at high frequencies;
2) quadruple-pendulum suspensions (also pioneered by GEO) to lower the seismic wall to
just above $\sim$10 Hz~\cite{bib:quadsuspension};
3) high-mechanical-Q silica-fiber suspension to reduce suspension thermal
noise; 
4) more massive, higher-mechanical-Q test-mass mirrors
to reduce thermal noise from the mirrors and mitigate the increased
radiation pressure noise from the higher laser power; 
5) active, feed-forward in-vacuum active isolation of optical tables,
using accelerometer, seismometer and geophone sensing (supplemental to
hydraulic pre-isolation discussed above which was used already for the
Livingston interferometer); and
6) the addition of a signal recycling mirror between the beam splitter and
photodetector~\cite{bib:advligo}.
The addition of a signal-recycling mirror with its adjustable relative
position with respect to the beam splitter will give Advanced LIGO
some flexibility in its frequency-dependent sensitivity.
Figure~\ref{fig:advligosens} shows example design curves for different
laser powers and optical configurations.

In the following, the eventual broadband, highest-power configuration 
labeled ``ZERO DET, high P'' in figure~\ref{fig:advligosens}
will be assumed (non-zero tunings allow shaping of spectral sensitivity
to favor certain bands) . With this improvement in sensitivity,
the average NS-NS inspiral range (20 Mpc achieved in S6) should reach 
200 Mpc, while the NS-BH for 10-$\msolar$ black holes should
reach 410 Mpc, and the corresponding BH-BH range should reach 980 Mpc.
The resulting increases in expected detected coalescence rates will be presented below in 
section~\ref{sec:summary}.

As of mid 2012, two LIGO interferometers (H1 at Hanford and L1 at Livingston)
are under construction. In the original plan a second 4-km interferometer (H2)
was to be built at Hanford, but it was appreciated that moving H2
to another location on the globe would reap great scientific dividends from
improved triangulation of transient sources~\cite{bib:bettertriangulation}.
One possibility pursued aggressively was placing the third interferometer
in Australia at the Gingin site~\cite{bib:gingin}, but the Australian
government declined to provide the funding necessary for civil construction,
including buildings, vacuum tube and enclosure, along with vacuum chambers for
the optics.
Placing the third interferometer in India was proposed in parallel~\cite{bib:indigo}, 
and currently is being pursued seriously enough that 
installation of the H2 interferometer at Hanford has been suspended
and preparations made for transport to India. If all
goes well, the Hanford and Livingston interferometers will begin operations
in 2015, with the third LIGO interferometer commissioned and operated by 
the Indian Initiative in Gravitational-wave Observations (IndIGO) by 2020 
(although projections for IndIGO
are understandably uncertain at this early stage).

While the detailed design parameters for Advanced Virgo are not 
completely settled, there exists a baseline design~\cite{bib:advvirgo}.
Its general outlines include: higher laser power and improved thermal
noise in suspensions and test masses. Note that the aggressive passive
isolation used already in Virgo means that no major changes are expected
in order to match the current, already-impressive low-frequency seismic
wall of 10-20 Hz. Ultimate Advanced Virgo sensitivity is expected to be
comparable to that of Advanced LIGO.

In parallel, a primarily Japanese collaboration is proceeding to build
an underground 3-km interferometer 
(KAGRA -- KAmioka GRAvitational wave telescope, formerly
known as LCGT- Large Scale Cryogenic Gravitational Wave Telescope)~\cite{bib:lcgt}
in a set of new tunnels in the Kamiokande
mountain near the famous Super-Kamiokande neutrino detector. 
A 100-m prototype interferometer, CLIO (Cryogenic Laser Interferometer Observatory)
has operated successfully in a shorter Kamiokande tunnel~\cite{bib:clio}.
Placing the interferometer underground dramatically suppresses
noise due to ambient seismic disturbances. The rigid rock in the
base of the Kamiokande mountain suffers much smaller displacements
due to seismic waves than surface soil. Hence the terrestrial gravity
gradients due to motion of rock and soil are much reduced.
It is hoped that a first version of KAGRA will be operational for
a short run in 2015, followed by a major upgrade to cryogenic mirrors
(at 20 K) to reduce thermal noise due to suspensions and internal modes
and resumption of operations in 2018. 
In its final configuration KAGRA is expected to have sensitivity
comparable to that of Advanced LIGO.

\subsection{Third-generation ground-based interferometers}
\label{sec:3rdgeneration}

With construction of second-generation interferometers well under way,
the gravitational wave community has started looking ahead to
third-generation underground detectors, for which KAGRA will provide
a path-finding demonstration. A European consortium is in the
conceptual design stages of a 10-km cryogenic underground
trio of triangular interferometers called Einstein Telescope~\cite{bib:et}, which would use
a 500-W laser and aggressive squeezing, yielding a design sensitivity an order of magnitude
better than the 2nd-generation advanced detectors now under construction.
With such capability, the era of {\it precision} gravitational wave astronomy
and cosmology is expected to open. Large statistics for detections
and immense reaches ($\sim$Gpc) would allow new distributional analyses
and cosmological probes. LIGO scientists too are starting to consider
a 3rd-generation cryogenic detector, with a possible location in the 
proposed DUSEL facility~\cite{bib:dusel}.

\subsection{Space-based interferometers}
\label{sec:spaceifos}

Building interferometers underground offers the prospect of probing 
frequencies down to $\sim$1 Hz, but to reach astrophysically interesting
sensitivities at much lower frequencies will likely require placing
interferometers in space.
An ambitious and long-studied proposed joint NASA-ESA project called LISA 
(Laser Interferometer Space Antenna) envisioned
a triangular configuration (roughly equilateral with sides of $6\times10^6$ km)
of three satellites. These spacecraft were to comprise a double quasi-interferometer
system, where each satellite would send a laser to the other two and
receive another laser from each, where each laser would be phase locked 
to the other two, yielding a total of six phase-locked lasers. As discussed
above, there are many low-frequency gravitational wave sources expected
to be detectable with LISA, and the proposed project has received
very favorable review by a number of American and European
scientific panels. Nonetheless the project has been turned down by NASA.
Subsequently, NASA and ESA have solicited separate and significantly descoped 
new proposals. The funding prospects for these new proposals are quite uncertain,
with ESA having recently passed over a descoped version of LISA called NGO
(New Gravitational-wave Observer) in favor of a mission to Jupiter. A launch
before 2020 of any space-based gravitational wave interferometer seems unlikely
at this point.

\subsection{Pulsar timing arrays}
\label{sec:pta}

An entirely different effort is under way in the radio astronomy
community to detect stochastic gravitational waves by way of 
precise pulsar timing. Very-low-frequency waves ($\sim$ several nHz)
in the vicinity of the Earth could lead to a quadrupolar pattern
in the timing residuals from a large number of pulsars observed
at different directions on the sky~\cite{bib:ptaidea}. 

An informative recent review of this approach can be found in ref.~\cite{bib:ptareview}.
Only a summary of salient issues is presented here. 
Galactic millisecond stars provide extremely regular clocks (after
correcting for tiny, measurable spindowns). By measuring a single
pulsar over many years, one could, in principle detect the presence
of a very low-frequency stochastic background of gravitational waves
affecting the space between the Earth and the pulsar. In practice, however,
with only a single pulsar's timing residuals, it is easier to set upper limits 
than to establish detection. 

Hence pulsar astronomers have mounted several systematic
multi-year efforts to monitor the timing residuals of many millisecond pulsars
over the sky, to allow a cross-correlation determination of a $\sim$several-nHz
component near the Earth that affects all of the timing measurements.
Three collaborations have
formed in recent years to carry out the precise observations
required: 1) The Parkes Pulsar Timing Array (PPTA -- Australia)~\cite{bib:ppta}, 2) the
European Pulsar Timing Array (EPTA -- U.K., France, Netherlands, Italy)~\cite{bib:epta},
and 3) the North American NanoHertz Observatory for Gravitational
Waves (NANOGrav -- U.S.A. and Canada)~\cite{bib:nanograv}. In addition, these three separate
efforts have agreed to collaborate on joint analysis to improve sensitivity,
forming an International Pulsar Timing Array~\cite{bib:ipta}. 
This work builds upon previous searches for evidence of gravitational waves
from pulsar timing residuals, including searches for both stochastic 
radiation~\cite{bib:ptaresult1,bib:ptaresult2,bib:ptaresult3,bib:ptaresult4} 
and continuous radiation~\cite{bib:ptaresult5} from a postulated nearby supermassive binary
black hole system in 3C66B~\cite{bib:proposedsbbh}.
The two most recent stochastic background searches from EPTA~\cite{bib:ptaresult3} 
and NANOGrav~\cite{bib:ptaresult4} achieve limits on a stochastic background
characteristic strain amplitude in the several-nHz band of O(several $\times$ 10$^{-15}$).

By continued monitoring of known millisecond pulsars and finding still other stars
with small timing residual, these consortia hope to improve upon current array
sensitivity and achieve detection of a stochastic gravitational wave background.
The background from supermassive black hole binaries with masses of 10$^{9-10}\msolar$ 
at redshifts $z\sim1$ may lie only a factor of a few below current detection
sensitivity~\cite{bib:sesana}.
Important to this effort are identifying and mitigating systematic
uncertainties in pulsar timing,
of which some are purely instrumental
(\eg, radio observatory clock synchronization), some are terrestrial 
(\eg, ionosphere effects), and some are astrophysical (\eg, plasma fluctuations
in the intervening interstellar medium, variable pulsar torque, and
magnetospheric motions of emission regions).
A recent assessment~\cite{bib:cordesshannon} concludes that prospects 
for substantial improvement in current array sensitivity depend primarily on the
nature of these astrophysical sources of noise. If dominant noise sources are
red and therefore resemble the expected astrophysical background from binary SMBHs,
a gravitational wave signal will be harder to establish than 
if the astrophysical timing noise is predominantly white. 
In that scenario, to establish firm detection and characterize the gravitational wave background
requires 50-100 stable millisecond pulsars, a significant increase
over the current total among the three major pulsar timing arrays.

In a more favorable scenario (with respect to noise or signal), however,
pulsar timing arrays could well make the first direct detection of gravitational waves,
before the advanced ground-based interferometers reach sufficient sensitivity.

\section{Gravitational Wave Searches with LIGO and Virgo}
\label{sec:gwsearches}

\subsection{Overview of gravitational wave data analysis}
\label{sec:searchoverview}

In the following, 
a sampling of results to date from searches in the LIGO S5-S6 and Virgo VSR2 data will be presented.
Searches in data from the early runs (S1-S4, VSR1)  will be discussed only briefly.
Analyses of LIGO and Virgo data are carried out by joint working groups focusing on four distinct source types:
1) compact binary coalescences (CBC), 2) unmodeled bursts, 3) continuous waves (CW), and 4) stochastic background,
in keeping with the waveform categories discussed above in section~\ref{sec:overviewsources}.
Results from searches for these four source types will be discussed in turn below. It should be kept in mind,
however, that these types represent archetypal extremes and that some sources fall between these extremes and
can be attacked with complementary methods arising from approaches developed for the extremes.

\begin{figure}[tb]
\begin{center}
\epsfig{file=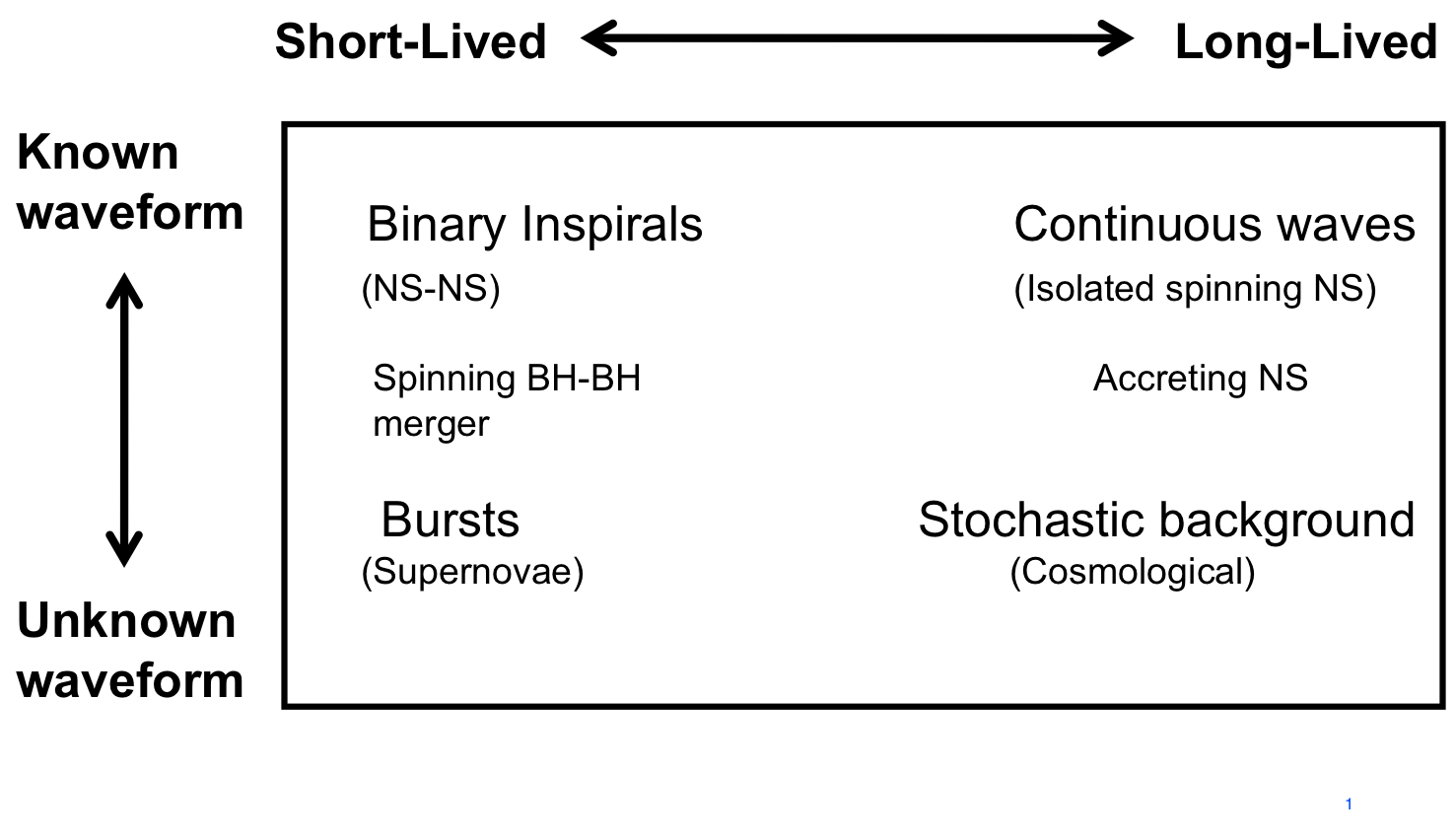,scale=0.5}
\caption{Schematic diagram illustrating the gravitational waveform categories
that affect search strategies. As one moves from left to right, waveforms increase
in duration, while as one moves from top to bottom, {\it a prior} waveform 
definition decreases.
\label{fig:boxdiagram}}
\end{center}
\end{figure}

Figure~\ref{fig:boxdiagram} shows a schematic outline of the way in which LIGO and Virgo searches can be broken down. As one moves from left to right on the diagram, waveforms increase in duration,
while as one moves from top to bottom, {\it a priori} waveform definition decreases. Populating the upper left corner is
the extreme of an inspiraling compact binary system of two neutron stars in the regime where corrections
to Newtonian orbits can be calculated with great confidence. Populating the upper right corner are
isolated, known, non-glitching spinning neutron stars with smooth rotational spindown and measured 
orientation parameters. Populating the lower left corner of the diagram are supernovae, rapid bursts of gravitational
radiation for which phase evolution cannot be confidently predicted, and for which it is challenging to make
even coarse spectral predictions. At the bottom right one finds a stochastic, cosmological background of 
radiation for which phase evolution is random, but with a spectrum stationary in time.
Between these extremes can live sources on the left such as the merger phases of a BH-BH coalescence.
On the right one finds, for example, an accreting neutron star
in a low-mass X-ray binary system where fluctuations in the accretion 
process lead to unpredictable wandering phase.

Uncertainties in the phase evolution of a gravitational wave affect not only the algorithm used for
a search, but also affect the attainable sensitivity. If one knows the precise evolution of the
phase, one can apply matched-filter techniques~\cite{bib:matchedfilter} to optimize 
sensitivity (\eg, highest detection efficiency for a fixed false alarm rate).
But as uncertainty increases, one must search over a larger volume of parameter space,
in which case the signal-to-noise ratio (SNR) required to achieve a fixed false alarm rate
necessarily increases. Roughly speaking, the SNR threshold rises only logarithmically with
the number of distinct search templates, but for many searches, the number of templates
searched can be greater than 10$^{13}$, leading to a significant degradation in intrinsic
achievable sensitivity. In many cases, however, the requirement of coincident detection
in two or more interferometers or the requirement of signal coherence among interferometers
permits a reduction of signal strain threshold (see section~\ref{sec:cbccoherence}).

Requiring coincidence between an interferometer signal and an external astrophysical signal is another
powerful way to reduce the search space, in this case the search over start times. For
example, a gamma ray burst trigger permits defining a search window of only a few minutes duration
(to allow for uncertainties in the production mechanism for the gamma rays created
in an outgoing jet). The same principle applies to the Fourier domain, where a search
for gravitational waves from a known pulsar can require precise agreement between electromagnetic
periodicity and gravitational wave periodicity.

It should be noted that an alternative tradeoff between sensitivity and parameter volume
searched can be struck by applying generic algorithms that are robust against phase 
evolution uncertainty. That is, instead of searching a large number of precisely
defined templates, one uses a coarser basis for which intrinsic SNR is limited.
As discussed below, specific examples of the tradeoff between sensitivity and parameter space 
searches include searches for late stages of distant compact binary neutron star coalescences and
searches for unknown spinning neutron stars in our own galaxy.

Published and ongoing LIGO and Virgo searches are based on as series of data runs,
starting with LIGO science run S1 in August 2002 and ending with the Virgo VSR4 run
in summer 2011. Table~\ref{tab:dataruns} lists all of these runs and their durations.
While results from searches in the early runs will be discussed briefly below, the
primary focus will be on the results from the LIGO S5 \&\ S6 runs,
and the Virgo VSR2 and VSR3 runs.

\begin{table}
\begin{center}
\begin{tabular}{lcc}
Data Run & Period  &  Duration (days) \\
\hline\hline
LIGO S1 & August-September 2002 & 17 \\ 
LIGO S2 & February-April 2003 & 59 \\
LIGO S3 & October 2003 - January 2004 & 70 \\
LIGO S4 & February-March 2005 & 30 \\
LIGO S5 & November 2005 - September 2007 & 696 \\
LIGO S6 & July 2009 - October 2010 & 470 \\
\hline
Virgo VSR1 & May-September 2007 & 136 \\
Virgo VSR2 & July 2009 - January 2010 & 187 \\
Virgo VSR3 & August-October 2010 & 71 \\
Virgo VSR4 & June-September 2011 & 110 \\
\hline\hline
\end{tabular}
\caption{Data runs taken with the LIGO and Virgo detectors from 2002 to 2011.
The LIGO H1 (4 km), H2 (2 km) and L1 (4 km) interferometers were operated
in data runs S1-S5, while only H1 and L1 were operated in the S6 run.}
\label{tab:dataruns}
\end{center}
\end{table}

\subsection{Searching for coalescences}
\label{sec:cbcsearches}

A variety of searches for coalescences have been carried out in LIGO and Virgo data,
with increasing sophistication, sensitivity and coverage of source parameter space.
Here we summarize the search methods used and the (so-far negative) results
of those searches.

\subsubsection{Expected inspiral waveforms}
\label{sec:cbcwaveforms}

As discussed above, the coalescence of two compact massive objects (neutron stars and
black holes) into a single final black hole 
can be divided into three reasonably distinct stages: inspiral, merger and
ringdown. Let's first address the problem of detecting the inspiral
stage, for which analytic expressions can be derived in perturbative
expansions. Equations~(\ref{eqn:fvstimeequalmass}) and~(\ref{eqn:hvstimeequalmass})
describe the frequency and amplitude evolution of a circular binary of
two equal-mass stars. Generalizing this quasi-Newtonian model to include
unequal stellar masses $M_1$ and $M_2$ leads to 
\begin{equation}
\label{eqn:fvstimeunequalmass}
\fgw \quad = \quad {1\over8\,\pi}
       \left[1\cdot5^3\right]^{1\over8}
       \left[{c^3\over G\mchirp}\right]^{5/8}
       {1\over(\tcoal-t)^{3\over8}}.
\end{equation}
and
\begin{equation}
\label{eqn:hvstimeunequalmass}
h_0(t) \quad = \quad {1\over r}
         \left[{5\,G^5\mchirp^5\over c^{11}}\right]^{1\over4}
         {1\over(\tcoal-t)^{1\over4}},
\end{equation}
where $\mchirp\equiv (M_1M_2)^{3\over5}\,/\,(M_1+M_2)^{1\over5}$ is known as the {\it chirp mass}.
Note that {\it both} the frequency and amplitude evolution depend on the chirp mass;
at this level of approximation, one cannot separately determine the stellar
masses $M_1$ and $M_2$.

As the radius of the orbit approaches zero, the above expression breaks down.
The stellar velocities become relativistic, and the Newtonian framework used
to derive the relation between $\omega$ and the orbit size no longer applies.
In the later stages of the inspiral, further post-Newtonian approximations are 
necessary. A review of various approaches can be found in~\cite{bib:sathyaschutz}.
As one example, the following expression gives an expansion to seventh order
in the dimensionless parameter $\taubar \equiv \left[5G(M_1+M_2)\,/\, c^3\nu(\tcoal-t)\right]^{1\over8}$ for
the orbital phase of the system~\cite{bib:sathyaschutz}: 
\begin{eqnarray}
\label{eqn:phasevstime}
\phi(t) & = & -{1\over\nu\taubar^5}\Biggl\{1+\left({3715\over8064}+{55\over96}\nu\right)\taubar^2 - {3\pi\over4}\taubar^3
+\left({9275495\over14450688}+{284875\over258048}\nu+{1855\over2048}\nu^2\right)\taubar^4 \nonumber \\
& & + \left(-{38645\over172032}+{65\over2048}\nu\right)\pi\taubar^5\ln(\taubar)+
\Bigl[{831032450749357\over57682522275840}-{53\over40}\pi^2-{107\over56}(\gamma+\ln(2\taubar)) \nonumber \\
& & + \left(-{126510089885\over4161798144}+{2255\over2048}\pi^2\right)\nu + {154565\over1835008}\nu^2
-{1179625\over1769472}\nu^3\Bigr]\taubar^6 \nonumber \\
& & + \left({188516689\over173408256}+{488825\over516096}\nu-{141769\over516096}\nu^2\right)\pi\taubar^7\Biggr\},
\end{eqnarray}
where $\gamma$ is Euler's constant, where $\nu \equiv M_1M_2/(M_1+M_2)^2$ is known as the symmetric mass ratio 
(reduced mass / total mass: $0<\nu\le{1\over4}$) and where this expansion is
classified as Post-Post-Newtonian of order 3.5 in the relativistic velocity parameter $(v/c)^2$.
The expression for $\fgw$ in equation~(\ref{eqn:hvstimeunequalmass})
can be obtained from the first term of ${1\over\pi}d\phi/dt$. This perturbative expression
has terms depending on $M_1$ and $M_2$ in different ways, allowing them (in principle) to be
determined separately from measured waveforms, in contrast to the lowest-order term
[equation~(\ref{eqn:hvstimeunequalmass})] where they appear only in the chirp mass combination.
In searches carried out to date, it has been assumed that by the time the binary systems
have decayed to where their gravitational wave frequencies lie in the terrestrial band,
their orbits have circularized enough (because of gravitational wave emission!) that
corrections to gravitational waveforms due to non-zero eccentricity can be neglected.
For elliptical orbits, gravitational radiation leads to a more rapid decrease in eccentricity
than in semi-major axis~\cite{bib:peters,bib:jktext}. It has been pointed out, however,
that stellar-mass black holes can be captured in highly elliptical orbits by galactic
nuclei, leading possibly to periodic bursts of gravitational radiation, preceding 
coalescence~\cite{bib:periodicbursts}.
In the case of NS-NS systems, it has also been assumed that spin effects can be neglected.
Corrections due to black hole spin will be discussed briefly below.

\subsubsection{Search algorithm for a coalescence event}
\label{sec:cbcalgorithm}

Now we turn to {\it how} one might measure such a waveform and derive the associated
astrophysical parameters. Unlike the events recorded at a high-energy collider, triggered
by, for example, high calorimeter energy or the presence of hits in an external muon
detector, gravitational wave data comes as a steady stream of digitized data 
(a time series, commonly known as ``one damned thing after another''~\cite{bib:fisher}).
Although details vary among different interferometers, there is typically a primary
gravitational strain channel sampled at 10 kHz or more, accompanied by hundreds
or thousands of auxiliary channels that monitor the state of the interferometer
and physical environment, to allow assessment of the credibility of any detection
candidate in the primary channel.

A naive way to search in the data for a waveform, such as that consistent
with equation~(\ref{eqn:phasevstime}), would be to compute a $\chi^2$ 
or likelihood statistic
based on the match between the time series data and the sum of 
random noise plus putative waveform,
while stepping in time by one data sample each iteration and searching over
a parameter set that describes the waveform. Although this approach could,
in principle, succeed for very strong gravitational wave signals, it is 
non-optimum because it fails to exploit the spectral characters of the signal
and of the noise. Gravitational wave interferometers have highly non-white
sensitivities, as shown, for example, in figure~\ref{fig:s5strainspectra}.
To increase SNR, one should emphasize (weight) signal content near the
frequencies of best sensitivity ($\sim$100-200 Hz for LIGO \&\ Virgo) and
de-emphasize (de-weight) signal content at other frequencies~\cite{bib:matchedfilter}. 

Hence Fourier analysis is more natural than time-domain analysis for
most waveform searches. It is beyond the scope of this article to
review detection theory in time series data, but a few key concepts
merit description. For more thorough treatments in the context of
gravitational wave data analysis, see~\cite{bib:jktext} or~\cite{bib:jolienwarren}.
For simplicity, continuous data and continuous
Fourier transforms will be used for illustration; the
same conceptual framework applies in the use of discrete Fourier
transforms to finite, discretely sampled data. 
Assume the strain data stream $x(t)$ is a sum of a
deterministic signal $h(t)$ and
Gaussian (but not white) noise $n(t)$:
\begin{equation}
x(t) \quad = \quad h(t) + n(t)
\end{equation}
Define the Fourier transform of the data stream:
\begin{equation}
\tilde x(f) \quad \equiv \quad \int_{-\infty}^{\infty}\,dt\,e^{-2\pi\,ift}\,x(t)
\end{equation}
and the noise power spectral density:
\begin{equation}
S_n(f)\quad = \quad |\tilde n(f)|^2,
\end{equation}
where it is assumed $S_n(f)$ can be estimated from off-source
data (\eg, from neighboring time intervals when searching for a
transient signals or from neighboring frequency bins when
searching for a long-lived, narrowband signal).

One can then define a weighted measure of signal strength \vs\
time from 
\begin{equation}
\label{eqn:zoft}
z(t) \quad = \quad 2\int_{-\infty}^{\infty}{\tilde h(f)x^*(f)\over S_n(f)}\>e^{2\,\pi ift}\,df\,.
\end{equation}
Matched filtering theory~\cite{bib:matchedfilter} then 
leads to an SNR given by
\begin{equation}
\rho(t)\quad = \quad {|z(t)|\over\sigma},
\end{equation}
where the variance parameter:
\begin{equation}
\label{eqn:sigmadef}
\sigma^2 \quad = \quad 2\int_{-\infty}^{\infty}{|\tilde h(f)|^2\over S_n(f)}\,df
\end{equation}
is the matched filter output due to detector noise. The detection
statistic $\rho$ weights most strongly those spectral bands where the signal sought is expected
to be strong and the detector noise low.

One can also coherently combine the data from multiple interferometers,
taking into account the expected time differences among signals reaching
the different detectors for an assumed direction of the source on the sky.
For an all-sky search for an unknown transient, the source location,
\eg, right ascension and declination, add two more parameters to the
search space, whereas the direction would be known {\it a priori} for, say, a GRB-triggered
search. For example, in the first reported LIGO search for inspirals~\cite{bib:cbcsearchs1},
a coherent detection statistic was used: 
\begin{equation}
\rho^2_{\rm coh}(t) = {\rm max}_{{\rm over}\>\tdelay}\left[{|z_{{\rm L}1}(t)+z_{{\rm H}1}(t+\tdelay)|^2\over\sigma_{{\rm L}1}^2+\sigma_{{\rm H}1}^2}\right],
\end{equation}
where $\sigma_{{\rm L}1}^2$ and $z_{{\rm H}1}^2$ are defined separately for the LIGO Livingston (L1) and Hanford (H1) 4-km
interferometers from equation~(\ref{eqn:sigmadef}). This statistic implicitly assumes that
the L1 and H1 interferometers have the same antenna pattern sensitivity to a given
source, which is a good approximation for most source directions, since the interferometer
arms were designed to align as much as possible, given their 3000-km separation and the curvature
of the Earth. More generally, one must correct for the antenna pattern differences, which
depend not only upon the assumed source location, but also the source orientation. Those
corrections are especially important when combining data from interferometers distributed
widely on the globe, such as in joint LIGO-Virgo analysis, as discussed in more detail in
section~\ref{sec:burstsearches}.

For nearly every gravitational wave search carried out to date,
there are unknown parameters, \eg, chirp mass in an inspiral or
orientation of a rotating star, that affect the phase evolution of
the putative source, thereby affecting the integral in equation~(\ref{eqn:zoft}). 
As a result, one normally searches over a bank of template waveforms
with fine enough stepping in parameter space to maintain 
satisfactory efficiency, often characterized by an SNR mismatch
parameter that is kept below a certain allowed maximum value, e.g., 3\%.

An additional consideration, specific to inspiral searches, is the range
of validity in time of the assumed waveform. Equation~(\ref{eqn:phasevstime}),
for example, cannot be assumed to hold all the way until coalescence
at $t=\tcoal$ ($\taubar\rightarrow\infty$). Traditionally, such expressions
have been used only up until the ``innermost stable circular orbit,''
(ISCO)~\cite{bib:hartletext} inside of which the two stars may be said to plunge together in the
merger phase to form a single black hole. This merger phase should provide
additional signal for detection, but it was long assumed that
higher-order corrections beyond the reach of post-Newtonian calculations
made it unwise to search explicitly for an assumed analytic waveform model
with a coherent continuation of the matched filter used up until the ISCO.

With recent breakthroughs in numerical relativity, however, it has become
appreciated that the merger phase can sensibly treated as a smooth continuation
of the inspiral phase, using effective one-body theory (EOB)~\cite{bib:eob},
with relatively mild departures until quite late in the merger phase~\cite{bib:sathyaschutz}.
The coordinated NINJA effort~\cite{bib:ninja} mentioned in section~\ref{sec:cbcsources} intends
to bring these promising developments to fruition through creation and evaluation
of templates for carrying out searches and for parameter estimation in the event
of detection~\cite{bib:parameterestimation}. 

The final phase of coalescence is the ringdown of the black hole, which is
formed with large distortion and is expected to ``shake off'' that deviation
from axisymmetry about its spin axis through emission of gravitational waves~\cite{bib:sathyaschutz}.
These waves can be considered as emission from vibrational quasi-normal modes
(QNM) of the star~\cite{bib:press,bib:kokkotasschmidt}. 
The frequencies and damping times of these modes are determined
uniquely by the black hole's mass and angular momentum. The initial amplitudes
of those modes, however, are governed by the initial conditions forming the hole.
Comparison of measured mode radiation with numerical relativity calculations constrained
(to some precision) by the initial conditions determined from the inspiral phase should
allow interesting tests of general relativity in a highly non-perturbative, strong-field
regime.

While initial stellar spin effects are expected to be small for 
neutron stars~\cite{bib:manchester,bib:apostetal},
they are expected to induce measurable distortions of waveforms for rapidly
spinning black holes~\cite{bib:apostetal,bib:blackholespinimportance,bib:spinningalgorithm}. 
Nonetheless, searches carried out using spinless black-hole templates retain significant sensitivity
to high-spin systems~\cite{bib:vandenbroeck}. 

\subsubsection{Coping with data artifacts}
\label{sec:cbcartifacts}

Before summarizing the results of searches for coalescences,
it is worth noting that in analysis of gravitational wave interferometer data
there are many non-ideal considerations that affect search strategies.
The necessity to use spectral weighting
of matched-filter templates for the non-white noise spectra has
already been discussed. That technique comes from signal processing theory
in which it is assumed the data is stationary (or at least can be treated as so
over time scales long compared to the waveforms being sought) and Gaussian, \eg, 
the real and imaginary coefficients in a given frequency bin of a Fourier transform are 
distributed normally with zero mean and well defined (but frequency-dependent) 
variance.

In reality, data taken with interferometers at the frontier of strain sensitivity 
rarely display ideal characteristics. The optics and laser control servos
are tuned to minimize stationary noise, but that tuning can lead to non-robustness
against changes in the environment or even in the instrument. An important environmental example is
ground motion, which can lead to non-linear upconversion of low-frequency noise
into higher-frequency noise in a variety of ways, including modulations 
of cavity power via wobbling Fabry-Perot mirrors, inducement of Barkhausen noise 
through interactions of actuation currents and magnets, or shaking of surfaces
that scatter laser light back into the interferometer.
In some cases such non-Gaussian noise sources are understood and their
effects can be modeled, but many artifacts in the time domain (and to a lesser degree
in the spectral domain) are not understood at all and must be dealt with in an
{\it ad hoc}, phenomenological manner. When a ``glitch'' is detected simultaneously
in the gravitational wave strain channel and an auxiliary channel {\it and} there
is no plausible mechanism for a gravitational wave to create the auxiliary-channel glitch,
then one can sometimes veto the short interval of strain data affected in one's
search. Ideally, a physical mechanism should explain how the glitch in the auxiliary
channel affects the strain channel, but sometimes the association is established
only statistically from a collection of similar glitches~\cite{bib:s5glitchstudies}.
Considerable effort has gone into carrying out data quality studies, to allow vetoing
that is effective and safe (against false dismissal of signals), as discussed in
references~\cite{bib:s6dataquality}, and used in all LIGO and Virgo
searches published to date.

Despite this data quality work, however, the data remains contaminated with
non-Gaussian artifacts that can interfere with searches. Hence one must adopt
search algorithms that are robust against such artifacts. As mentioned in 
section~\ref{sec:searchoverview}, coincidence requirements among two or 
more detectors can allow search thresholds to be lowered in Gaussian noise,
but perhaps more important, such coincidence is especially effective in
coping with non-Gaussian glitches, which can occasionally produce apparent signals
of enormous SNR in a single detector, relative to the ambient near-Gaussian
noise characterizing most of the data. If, for example, the background glitch rate and duration 
of glitches in detector 1 are $R_1$ and $\delta t_1$ (with $R_1\delta t_1\ll1$), and detector 2 has
corresponding values $R_2$ and $\delta t_2$, then the approximate rate of
accidental coincidence is $R_{\rm coinc}\sim R_1R_2(\delta t_1+\delta t_2)$, where
the coefficient depends on the details of the coincidence definition. The coincidence
rate can be very low. For example for $\delta t_1=\delta t_2 =$ 10 ms and $R_1 = R_2$ = once per hour,
one obtains $R_c\sim1.5\times10^{-9}$ s$^{-1}$ or about once every 20 years.

How does one accurately estimate such a rate for the complicated coincidence criteria applied
in actual searches, which can use single-interferometer events of fuzzily defined
window lengths and matching criteria based on the similarity of waveform shapes, comparableness
of amplitudes, \etc? The technique used most often to date has been
time lag (or ``time slide'') background estimation. In this method, one artificially shifts
one interferometer's data by a  set of time strides that are longer than
the duration of signals being sought and measuring the artificial coincidence rate for
each time stride. For example, one might shift the data of interferometer 2
by $t_{{\rm lag}_n} = n\delta t$, with $n = -N, -(N-1), ... , -1, +1, ... , N-1, N$, giving
$2N$ different rate measures. Thus one has a ``black box'' measure of not only the
average rate of accidental coincidence, but also its variance and other measures of
statistical distribution. Specifically, one can directly estimate the false alarm rate
for different single-interferometer SNR thresholds or even for criteria depending on
measurements in more than interferometer. 

There are two important implicit assumptions
in this method: 1) there is no preference in the background (non-gravitational-wave) accidental rate
for a time lag of zero; and 2) true gravitational wave signals are so rare that their
contribution to the background estimate through random coincidence can be neglected.
The first assumption is assumed to hold when detectors are far enough apart on the Earth's
surface that there are no appreciable common and undetected environmental disturbances consistent
with the light-travel-time between the detectors. For the LIGO Hanford and Livingston
detectors, a hypothetical example of such a common disturbance would be a lightning
strike in Kansas creating an electromagnetic glitch that affects the electronics or
magnets of interferometers at both observatories. In fact, the effects of distant
lightning have been measured and found negligible. In addition, magnetometers
at each site monitor the environment continuously. A more insidious potential source
of correlated transients are the nearly identical controls and data acquisition systems used
at these two observatories, which are both clocked by synchronized GPS receivers.
The precise synchronization is essential to accurate triangulation of astrophysical sources
on the sky, but imperfections in the electronics can lead to false coincidences. 

The second assumption of a weak gravitational contribution to the background estimate 
seems reasonable, but in fact, a loud gravitational wave signal
in one detector can appear in time-lag coincidence with an ordinary, weaker glitch in
the other detector. While the significance of this lag-coincidence event may be appreciably
weaker than that of the true zero-lag coincidence of the gravitational wave signal,
the significance can also be appreciably stronger than that of all lag-coincidences that
involve no true gravitational-wave triggers. Hence the distribution of combined significance
of the estimated background can be quite distorted, leading to an extraordinarily rare
occurrence being ranked as only moderately rare. One solution to this problem of signal
contaminating its own background estimate is to exclude single-interferometer ``foreground''
triggers of a given zero-lag candidate from the set of time-lagged triggers used to estimate its
own background. That solution is a clean one -- if the candidate is indeed a true gravitational
wave signal, but it has been argued~\cite{bib:cbcsearchs6} that such exclusion could itself
lead to a bias that overestimates the importance of a coincident candidate, the apparent significance of which
stems primarily from one single-detector instrumental glitch. This dilemma in background
estimation is not an academic one; a deliberately``blind'' injection into the LIGO and Virgo
detectors in 2010 revealed its importance, as discussed below.

\subsubsection{Coincidence or coherence?}
\label{sec:cbccoherence}

A question that comes up frequently when analyzing data from two or more detectors
is whether one should use a combined detection statistic from all detectors,
\eg, $Z_{\rm combined}>Z^*$ where $Z^*$ is a single threshold, 
or require separate detection statistics for each detector 
to exceed separate thresholds, \eg, $Z_1>Z_1^*$ {\it and} $Z_2>Z_2^*$ {\it and} \etc.
This issue arises in both transient searches and continuous-wave searches.
There are many ways to combine data and many ways to apply individual criteria, 
making a general quantitative answer difficult, even when making optimistic assumptions about the 
Gaussianity of the background noise.
But as a general rule for Gaussian data, combined detection statistics (especially when combined coherently
to exploit phase coherence present in a signal in all detectors and absent in the 
background noise) is statistically more powerful than requiring that individual
detection statistics satisfy individual criteria. To achieve the same false alarm
rate in coincidence as in combination typically sacrifices signal detection efficiency
for well behaved data.

Nonetheless, separate detection thresholds are frequently used in gravitational
wave searches. Why? One technical reason is that the computational cost of 
pursuing candidate outliers can be reduced with negligible loss in efficiency by applying 
low individual thresholds as an initial step. But there are searches for which relatively high
individual thresholds are applied, despite an appreciable efficiency loss.
The usual reason for this choice is to cope with non-Gaussian detector artifacts.
Nominal false alarm probabilities for a combined detection statistic can 
skyrocket if even one detector misbehaves. 

One can, of course, impose additional
consistency requirements to avoid accepting triggers created by the coincidence
of a large glitch in one detector with a Gaussian excursion in the other, but
such criteria tend inevitably to evolve {\it de facto} into the kind of single-detector
threshold requirements one tried to avoid via the combined statistic.
It should be pointed out that even within a single detector, similar considerations
apply. For example, in LIGO and Virgo searches for coalescences to date, there
has been a waveform consistency requirement in the form of an additional $\chi^2$ 
statistic~\cite{bib:cbcsearchs1,bib:cbcchisquare} that essentially requires the SNR from different
bands of the detector to be consistent, thereby suppressing triggers due to a
single spectral artifact.

One can hope that the elaborate seismic isolation systems of the advanced detectors
now being built will dramatically reduce the glitches tied directly or indirectly to
ground motion, allowing more effective use in the future of truly coherent multi-interferometer
detection statistics. But once again, these interferometers will be pushing the
frontier of technology and may well be subject to unexpected, non-Gaussian disturbances, 
at least in the initial years.

\subsubsection{Results of all-sky searches for coalescence}
\label{sec:cbcresults}

Let's turn now to results from all-sky searches to date for coalescences in interferometer
data. Triggered searches for coalescences from short gamma ray bursts will be discussed
in section~\ref{sec:burstsearches}. First -- there have been no detections. But the improvements over the last two
decades in detector sensitivity have been dramatic and have been accompanied by substantial 
improvements in algorithms (including those to cope with non-Gaussian data).
As we approach the advanced detector era, prospects for discovery look very promising,
as discussed below in section~\ref{sec:summary}.

All LIGO and Virgo searches for coalescences to date owe much to a pioneering
analysis~\cite{bib:allen40meter} of 25 hours of data taken in 1994 with the Caltech
40-meter interferometer prototype. Although the detector sensitivity fell far
short of what LIGO achieved later and the observation span was short, the exercise
proved valuable in developing matched-filter approaches to inspiral searches and
in learning how to carry out analysis in glitchy, non-Gaussian data. The 40-meter's
sensitivity was good enough during the data run to detect NS-NS inspirals in most, but not 
all of our galaxy. In the end, a 95\%\ CL upper limit was set on the galactic NS-NS coalescence rate
of 0.5 per hour, about eight orders of magnitude higher than the ``realistic'' rates
quoted in section~\ref{sec:cbcsources}. A search~\cite{bib:tamacbcsearch} in 6 hours of early TAMA 300
data from 1999 yielded a similar limit on coalescence rate of 0.59 events per hour in the galaxy.

A long series of searches for NS-NS inspirals have been
carried out in LIGO and Virgo 
data~\cite{bib:cbcsearchs1,bib:cbcsearchs2,bib:cbcsearchbbhs2,bib:cbcsearchs3s4,bib:cbcsearchs5a,
bib:cbcsearchs5b,bib:cbcsearchs5c,bib:cbcsearchs6}, 
using the FindChirp algorithm~\cite{bib:findchirp} and systematic template banks~\cite{bib:cbctemplatebank},
starting with the LIGO S1 data, for which event
rates limits obtained were O(10$^{+2}$ year$^{-1}\cdot$MWEG$^{-1}$) and culminating in the joint search of LIGO S6 and Virgo VSR2/VSR3 data,
for which resulting event rate limits were O(10$^{-2}\cdot$year$^{-1}\cdot$MWEG$^{-1}$). In addition, there have been searches for
NS-BH and BH-BH coalescences, sometimes reported in separate 
publications~\cite{bib:cbcsearchprimordials2,bib:cbcsearchspinnings3,bib:cbcsearchringdowns4,bib:cbcsearchimrs5}
with divisions based on the assumed mass range of the binary mass components. In particular,
searches for high-mass black hole systems
require special care -- the terminal inspiral frequency is low enough that signals tend to 
accumulate less SNR in band than a NS-NS inspiral of the same amplitude when entering the band,
and the lower-frequency bands ($<\sim$200 Hz) are typically more non-Gaussian than higher-frequency
bands, leading to more frequent false triggers.

\begin{figure}[htb]
\begin{center}
\vskip -0.5cm
\hspace*{-2mm}
\hbox{
\includegraphics[width=3.75in]{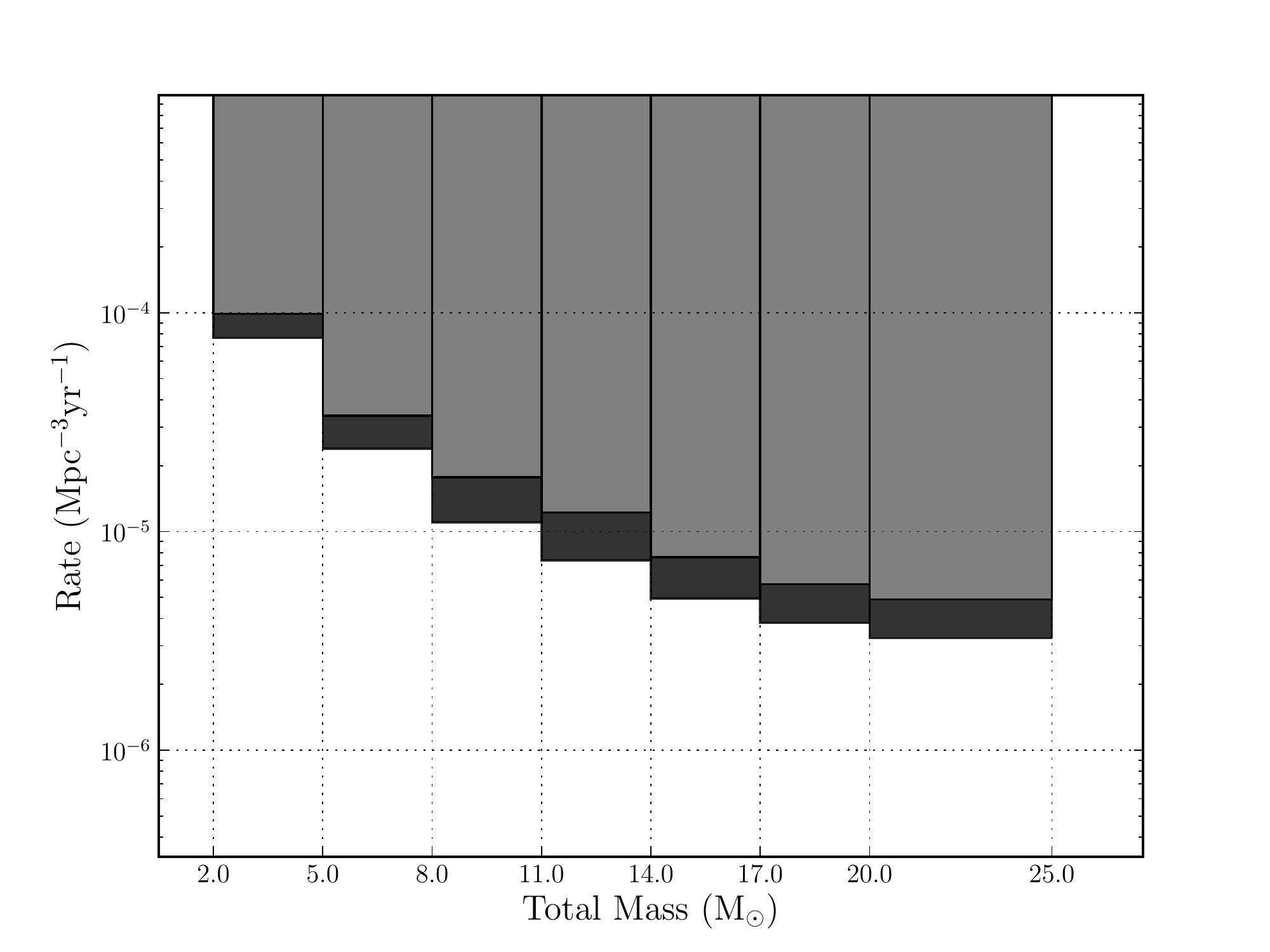}
\includegraphics[width=3.75in]{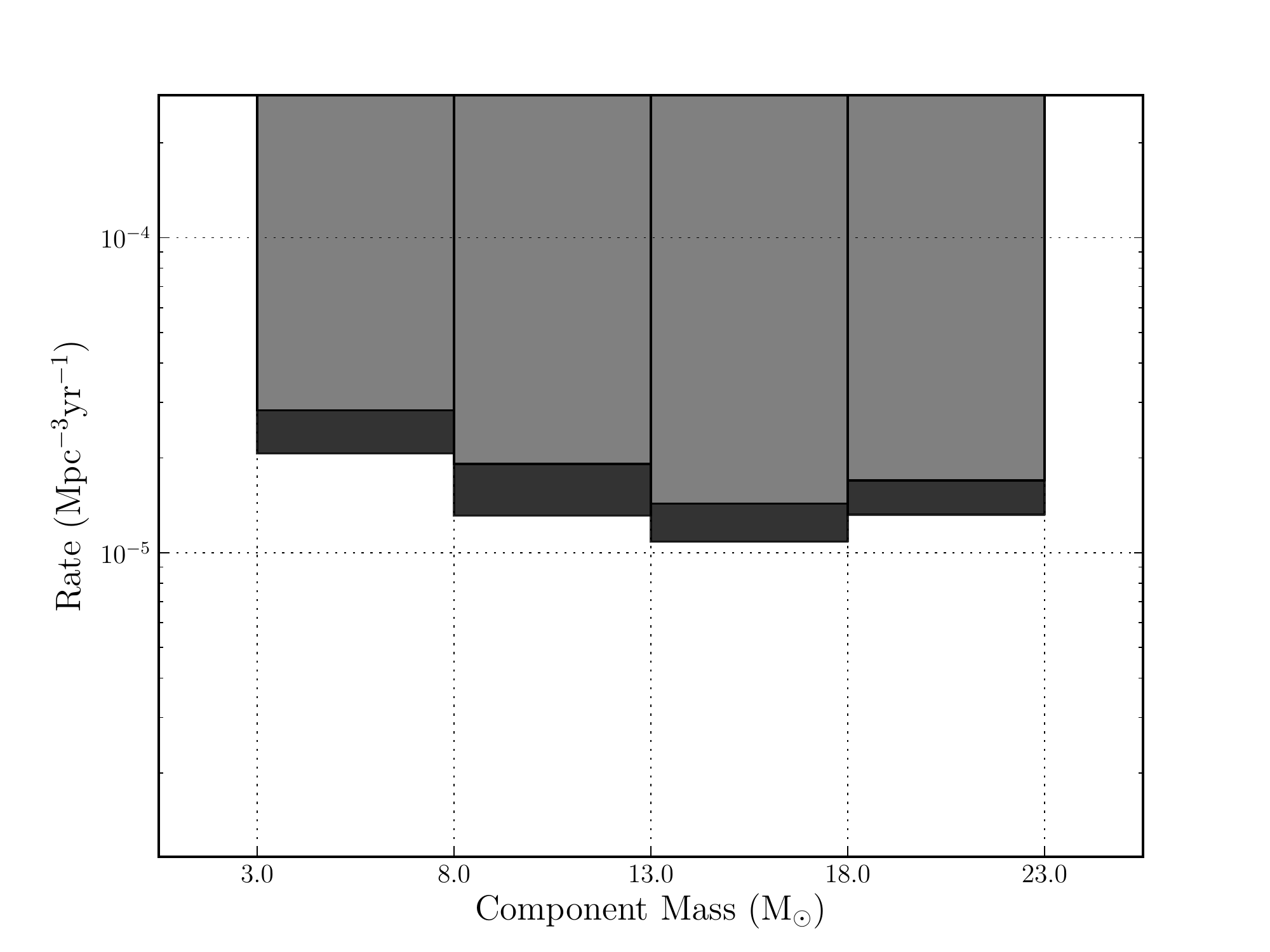}
}
\caption{
Marginalized upper limits on coalescence rates as a function of mass, based on searches in
LIGO S6 and Virgo VSR2-3 data~\cite{bib:cbcsearchs6}. The left plot
shows the limit as a function of total system mass $M$, using a distribution uniform in 
$m_1$ for a given $M$. The right plot shows the limit as a function of an assumed black hole
mass, with the companion neutron star mass restricted to the range 1-3 $\msolar$. The
light bars indicate upper limits from previous searches. The dark bars
indicate the combined upper limits including the results of the S6 / VSR2-3 search.}
\label{fig:cbclimitss6ab}
\end{center}
\end{figure}

\begin{figure}[ht]
\vskip -0.5cm
\hspace*{-2mm}
\begin{center}
\includegraphics[width=13cm]{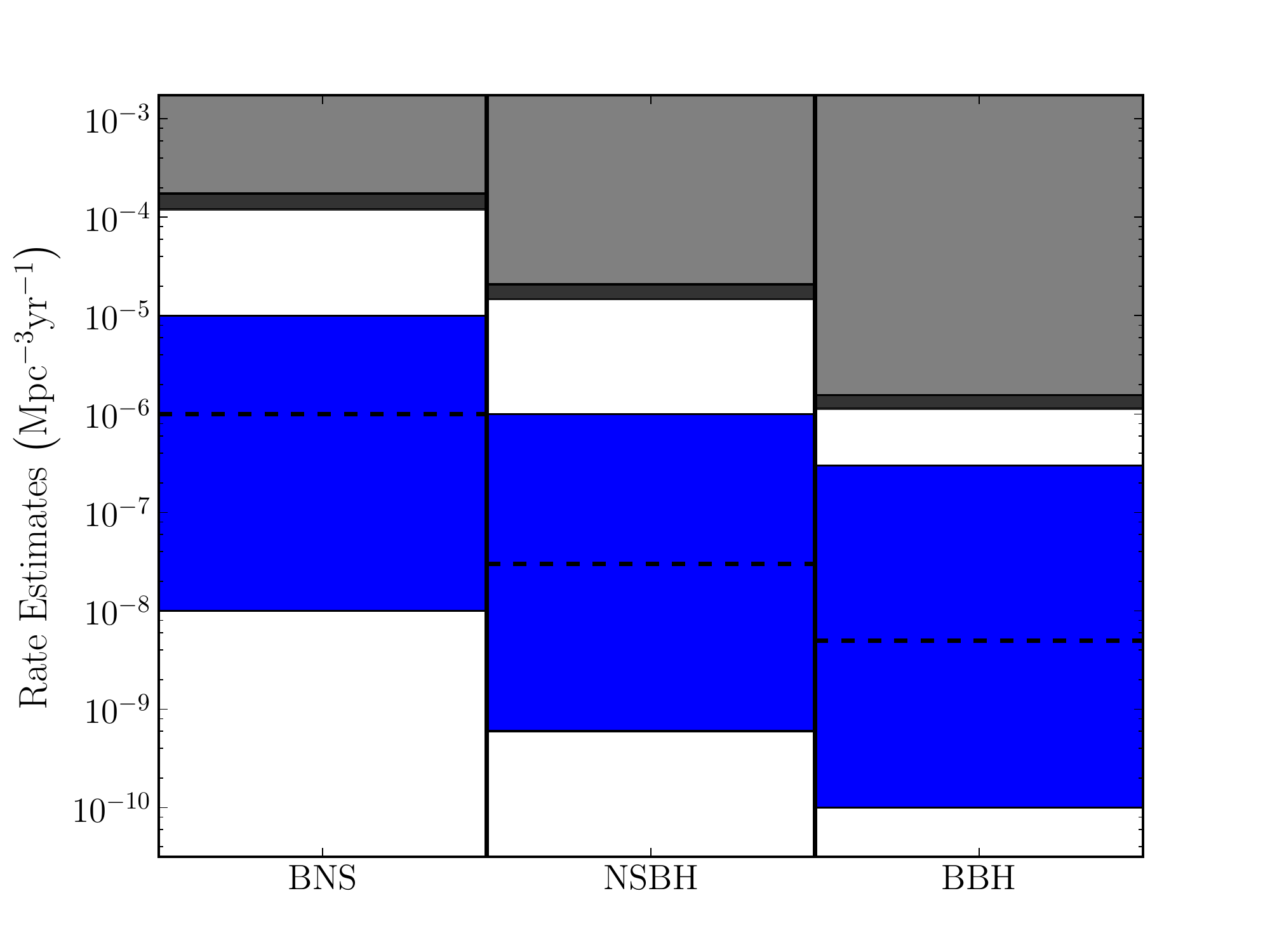}
\caption{Comparison of coalescence upper limit rates for NS-NS, NS-BH and
BH-BH systems~\cite{bib:cbcsearchs6}.  The light gray regions display the upper limits obtained in
the S5-VSR1 analysis; dark gray regions show the upper limits obtained in the
S6-VSR2-3 analysis, using the S5-VSR1 limits as priors. The lower (blue) regions
show the spread in the astrophysically predicted rates, with the dashed-black
lines showing the ``realistic'' estimates \cite{bib:cbcratespaper}.}
\label{fig:cbclimitss6c}
\end{center}
\end{figure}

Figure~\ref{fig:cbclimitss6ab} shows the most recent LIGO and Virgo limits~\cite{bib:cbcsearchs6} on coalescence event
rates (in units of Mpc$^{-3}$yr$^{-1}$) for NS-NS, NS-BH and BH-BH systems for a total binary
system mass up to 25 $\msolar$ as a function of total mass $M_{\rm total}$, where the individual component masses
are allowed to vary uniformly for each $M_{\rm total}$. The limits on rate per volume decrease dramatically 
with increasing total mass, since the higher masses can be seen to larger distances. These limits are
generic in applying to spinless neutron stars and/or black holes. Figure~\ref{fig:cbclimitss6ab} also shows
the resulting marginalized NS-BH limits when one component mass is restricted to the range 1-3 $\msolar$.
Figure~\ref{fig:cbclimitss6c} shows a comparison~\cite{bib:cbcsearchs6} of the S6/VSR2/VSR3 limits
obtained on NS-NS, NS-BH and BH-BH coalescence rates to the expected rates summarized in ref.~\cite{bib:cbcratespaper},
assuming a black hole mass of 10 $\msolar$. Limits remain 2-3 orders of magnitude above realistic
expectations and about ten times higher than optimistic rate estimates. Table~\ref{tab:cbclimits} shows
specific sample numerical rate limits from ref.~\cite{bib:cbcsearchs6} for nominal neutron star and
black hole masses of 1.35 $\msolar$ and 5.0 $\msolar$, respectively.

\begin{table}
\begin{center}
\begin{tabular}{l|ccc}\hline\hline
System & NS-NS & NS-BH & BH-BH \\
\hline
Component masses ($\msolar$) & 1.35 / 1.35 & 1.35 / 5.0 & 5.0 / 5.0 \\
$D_{\rm horizon}$ (Mpc) & 40 & 80 & 90 \\
Non-spinning upper limit (Mpc$^{-3}$yr$^{-1}$) & $1.3\times10^{-4}$ & $3.1\times10^{-5}$ & $6.4\times10^{-6}$ \\
Spinning upper limit (Mpc$^{-3}$yr$^{-1}$) & -- & $3.6\times10^{-5}$ & $7.4\times10^{-6}$ \\
\hline\hline
\end{tabular}
\caption{Upper limits on NS-NS, NS-BH and BH-BH coalescence rates,
assuming canonical mass distributions~\cite{bib:cbcsearchs6}. $D_{\rm horizon}$ is the
horizon distance averaged over the time of the search.
The sensitive distance average over all sky locations and
binary orientations is $D_{\rm avg}\approx D_{\rm horizon}/2.6$\cite{bib:finnchernoff}. 
The first set of upper limits are those
obtained for binaries with non-spinning components. The second set of upper limits are
produced using black holes with a spin uniformly distributed between zero and the maximal value of $GM_{\rm BH}^2/c$.}
\label{tab:cbclimits}
\end{center}
\end{table}

\subsubsection{The Big Dog}
\label{sec:cbcbigdog}
In the analysis leading to these results a strong coincidence signal candidate was found
in the LIGO H1 and L1 data, with evidence of a weaker signal seen simultaneously in the
Virgo data. Reconstruction of the most likely locations on the sky included a region in
the constellation Canis Major, leading to the informal dubbing of the event as the
``Big Dog''. It was known in advance that ``blind'' signal injections might be made 
into the LIGO and Virgo data by a small team sworn to silence until an eventual
``opening of the envelope.'' While the signal was strong enough to convince nearly
everyone relatively quickly that it was either a genuine astrophysical event or a
blind injection, there were a number of non-trivial issues to address: the evidence
for a signal in Virgo data was much weaker, and its statistical significance 
sensitive to how the trigger was combined with data from H1 and L1. There was a
substantial glitch nine seconds prior to the trigger in L1, which led to a worry of
a coincidental artifact. Parameter estimation studies found better agreement of 
the measured waveform with a lower order phase model than with the most advanced phase
model available. Despite these small inconsistencies and mildly nagging worries,
LIGO and Virgo scientists wrote a polished journal article and voted to submit it
for publication if the envelope proved to be empty. 

In the end, of course, the
envelope was {\it not} empty. The signal was indeed an injected coalescence,
and some of the puzzles seen in parameter reconstruction turned out to arise from slightly
out-of-date software used in the blind injection. 

\begin{figure}[t]
\vskip 0.1cm
\hspace*{-1.5mm}
\begin{center}
\includegraphics[width=13cm]{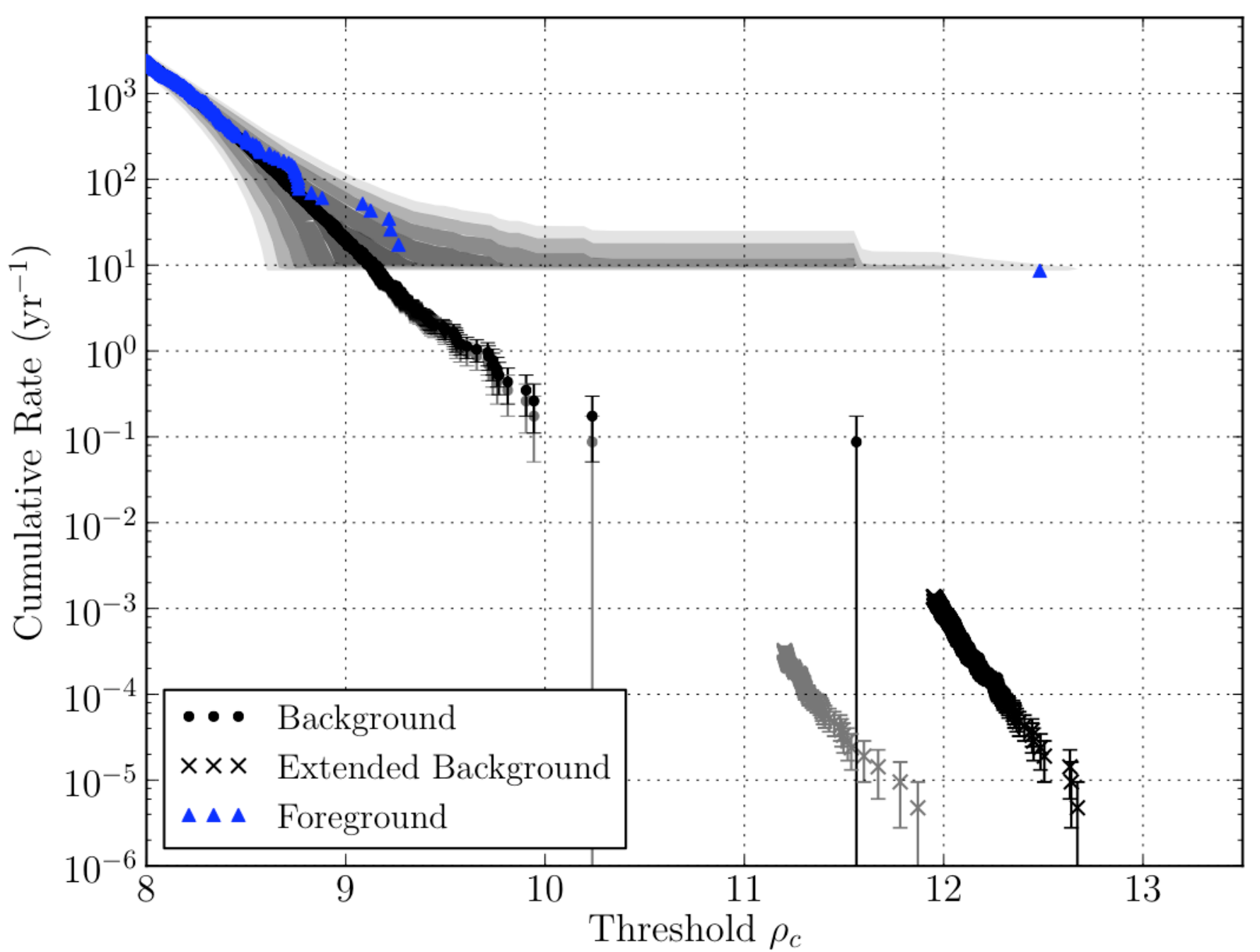}
\caption{The cumulative rate of events in the S6 / VSR2-3 coalescence
search with chirp mass $3.48 \le
\mathcal{M}/\msolar < 7.40$ coincident in the H1 and L1 detectors, seen in
four months of data around the September 16, 2010 blind injection, as a function of
the threshold ranking statistic $\rho_c$~\cite{bib:cbcsearchs6}.  The blue triangles show
true coincident events (foreground).  Black dots show the background estimated from 
an initial set of 100
time-shifts. Black crosses show the extended background estimation from
all possible 5-second shifts on this data restricted, for computational
reasons, to only the tail of loudest events.  The gray dots and crosses
show the corresponding background estimates when 8 seconds of data
around the time of the candidate are excluded (foreground-less).  Gray shaded contours
show the 1-5$\sigma$ (dark to light) consistency of coincident events
with the estimated background including the extended background
estimate, for the events and analysis time shown, including the
candidate time.}
\label{fig:cbcfarplot}
\end{center}
\end{figure}

As mentioned above, the detection
of this event uncovered an important issue in background estimation, namely
whether or not to include the foreground triggers of a coincident candidate 
in time-lagged estimates of its own background. 
In this case the foreground triggers were those outliers seen
in true coincidence in the H1 and L1 interferometers that led to the
combined Big Dog candidate.
The difference in estimated 
false alarm rates turned out to be substantial. 
Combining the (louder) H1 foreground trigger with background (time-lagged) L1 triggers
led to a conservatively estimated false alarm
rate of 1 in 7,000 years. It was not feasible to estimate accurately the
false alarm rate when excluding the H1 foreground trigger
(\ie, using only H1 background triggers) simply because no lagged-coincidence
candidates were found to give as high a combined significance as the Big Dog candidate. Figure~\ref{fig:cbcfarplot}
shows the background estimates with and without including the two H1 and L1 foreground
triggers. An extrapolation of the foreground-less background estimate suggests
(but does not establish!) a false alarm rate about two orders of magnitude lower
than 1 in 7,000 years.

The Big Dog exercise was an extreme one in the effort and time devoted to the analysis
and writing of results. But it proved valuable in preparing the collaborations
for not only the technical difficulties in signal detection in multiple 
interferometers of differing sensitivities and with different idiosyncrasies,
but also for confronting the ambiguities and philosophical issues that
arise in establishing a true discovery. A professional sociologist who has
been ``embedded'' in the LIGO Scientific Collaboration for many years has documented the
Big Dog in a forthcoming volume~\cite{bib:collins1}, that continues a recording of the ongoing
process toward eventual gravitational wave detection~\cite{bib:collins2}.

Finally, more specialized searches for coalescences 
have included a search~\cite{bib:cbcsearchprimordials2} for low-mass ($<1\msolar$) primordial black holes
in LIGO S2 data; a search~\cite{bib:spinningalgorithm,bib:cbcsearchspinnings3} for binaries with spinning stars
in LIGO S3 data; a search~\cite{bib:ringdownalgorithm,bib:cbcsearchringdowns4} for black hole ring-downs following
coalescence in LIGO S4 data; and a search~\cite{bib:cbcsearchimrs5} for modeled combined waveforms
corresponding to the inspiral, merger and ringdown phases of coalescence in LIGO S5 data.
In addition, a search~\cite{bib:burstimbh} was carried out in LIGO S5 and Virgo VSR1 data
for intermediate black hole binaries ($M_{\rm total}\approx100-450\msolar$)
using a polarization-optimized version of the Coherent Waveburst program discussed
below in section~\ref{sec:burstallsky}.

\subsection{Searching for bursts}
\label{sec:burstsearches}

Searches for unmodeled bursts date from the birth of the gravitational
wave detection field with Weber's early bar measurements. 
See Saulson~\cite{bib:saulsontext} for a discussion of early searches,
including some claims of detection.
Until the LIGO interferometers
approached design sensitivity, the most sensitive burst searches were carried
out with bar detectors, focused on $\sim$1-Hz bands near 900 Hz. A notable recent example
was the joint IGEC analysis of data taken in coincidence among the Allegro,
Auriga, Explorer and Nautilus detectors in 2005, leading to upper limits on burst
rates of $\sim$10/year for burst spectral amplitudes of $\sim2\times10^{-21}$ Hz$^{-1}$,
with a threshold corresponding to a false alarm rate of $\sim$1/century~\cite{bib:igec}.

Searches for unmodeled bursts have been carried out in LIGO and Virgo data for
both untriggered sources anywhere on the sky and triggered transient sources of
known sky location and time, in particular, gamma ray transients detected by satellites.
These searches attempt to be ``eyes wide open'' in the sense that a variety of
different and unpredictable waveforms could be detected with reasonable efficiency.
Several approaches have been tried, with the community now converging
on coherent multi-detector algorithms that strongly favor consistency
among the data sets, but allow for a broad family of such waveforms.
Because one cannot safely apply only matched-filter techniques in these generic
searches, one is more subject to false alarms from instrumental artifacts
than, for example, in NS-NS coalescence searches. Since the LIGO instruments
are intended to behave identically, they tend to be subject to the same
classes of glitch phenomena, making work in data quality studies and vetoing
all the more critical. In addition to using external triggers to focus
on interesting intervals of data to scrutinize, LIGO and Virgo scientists
have also begun collaboration with other astronomical collaborations
(electromagnetic and neutrino) to
provide prompt triggers based on results of the all-sky searches, to
allow independent confirmation of transient detection and to permit
better understanding of source astrophysics. The effort to use external
electromagnetic and neutrino detections to trigger follow-up gravitational
wave searches and \vv\ is part of what has come to be known
as multi-messenger astronomy.

\subsubsection{All-sky burst searches}
\label{sec:burstallsky}

The first published burst search~\cite{bib:glasgowmpiburstsearch} using gravitational wave interferometers
was based on a 62-hour coincident data run taken in 1989 with interferometer
prototypes at the University of Glasgow and the Max Planck Institute for
Quantum Optics. The algorithm was based on coincidence of triggers from
each detector, where triggers were based on a generic ``boxcar'' broadband
filter in the Fourier domain. The  upper limits on ambient burst
signals were astrophysically uninteresting ($h\sim10^{-16}$), but the
coincidence experiment and broadband analysis were useful forerunners of 
the observational programs carried out in recent years by the major
interferometer collaborations. 

The first published burst search in LIGO (S1) data~\cite{bib:burstsearchs1} was also based on coincidence
of triggers in two or more interferometers, where single-interferometer triggers were generated by
two distinct algorithms. The first, called SLOPE~\cite{bib:slope}, used a
time-domain filter that amounted to a differentiator and hence triggered on 
slope changes inconsistent with ambient Gaussian data. The second, called TFCLUSTER~\cite{bib:sylvestre},
used clustering of excess power in the time-frequency domain (spectrogram).
As for the first LIGO coalescence search, the burst search used time lag analysis to
estimate background. Limits on astrophysical strain were placed on
a family of {\it ad hoc} phenomenological waveforms with the waveform
strengths parametrized in terms of the ``root sum square of h'', $\hrss$, defined by
\begin{equation}
(\hrss)^2 \quad = \quad \int [|h_+(t)|^2+|h_\times(t)|^2]\, dt,
\end{equation}
where $h_+(t)$ and $h_\times(t)$ are the ``$+$'' and ``$\times$'' quadrupolar polarizations
of the detectable strain waveforms. This figure of merit is a proxy for the energy
content of a transient wave. Qualitatively, waveforms of similar spectral power 
content tend to be detected with comparable sensitivities. When evaluating the
performance of a burst algorithm, it helps to measure the efficiency of
detection \vs\ $\hrss$ for different families  of waveforms, with different waveforms
within a single family governed by one or parameters.

For example, the first LIGO burst search used two families of waveforms to
evaluate performance. The first family was that of Gaussian bursts defined
by
\begin{equation}
h(t)\quad =\quad h_0\,e^{−{(t-t_0)^2\over\tauτ^2}} ,
\end{equation}
where $t_0$ is a time of peak amplitude $h_0$, and $\tau$ is a parameter
defining the waveform's characteristic duration.
The second family was that of sine-Gaussian bursts defined by
\begin{equation}
h(t) \quad = \quad h_0\, \sin (2\pi f_0t)\,e^{−{(t-t_0)^2\over\tau^2}},
\end{equation}
where $f_0$ is an additional parameter describing the characteristic
central frequency of the waveform. These waveforms correspond to Gaussians
centered on $f_0$ in the Fourier domain with a quality factor
$Q= 2\pi\tau f_0$. Although $Q$ was fixed to $2\sqrt2\pi$ in the first
LIGO search, it was treated as a free parameter in later searches.

\begin{figure*}[!htbp]
\begin{center}
 \includegraphics[width=1.01\textwidth]{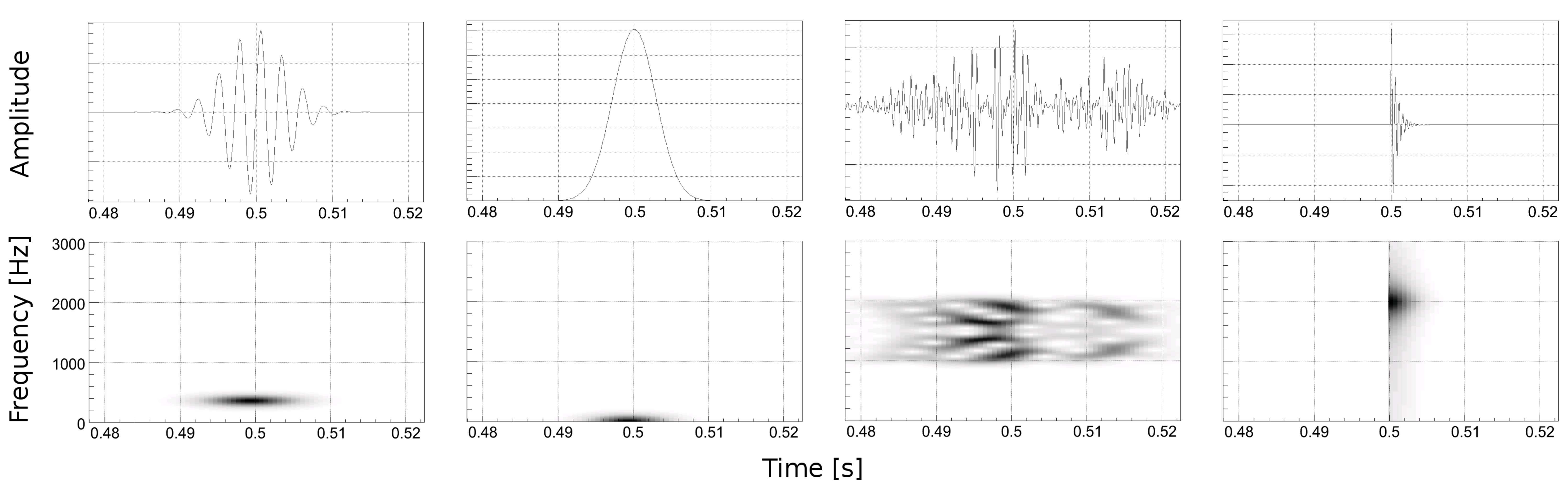}
\caption{Representative waveforms injected into data for simulation studies for the
S6 / VSR2-3 run~\cite{bib:burstsearchs6}. 
The top row is the time domain and the bottom row is a time-frequency domain representation of the waveform. 
From left to right: a 361 Hz $Q=9$ sine-Gaussian, 
a $\tau=4.0$~ms Gaussian waveform, 
a white noise burst with a bandwidth of 1000--2000 Hz and characteristic duration of $\tau=20$~ms, 
and a ringdown waveform with a frequency of 2000 Hz and $\tau=1$~ms. 
}
\label{fig:burstwaveforms}
\end{center}
\end{figure*}

The most recent all-sky burst search of LIGO and Virgo data 
(S6 \&\ VSR2-3)~\cite{bib:burstsearchs6} used additional waveform
families for evaluating performance. One family resembles
the waveform expected from the ringdown following black hole
formation ($t>t_0$):
\begin{eqnarray}
h_+(t) & = & h_0\, e^{-{(t-t_0)\over\tau}}\, \sin(2\pi f_0 t) \nonumber \\
h_\times(t) & = & h_0\, e^{-{(t-t_0)\over\tau}}\, \cos(2\pi f_0 t).
\end{eqnarray}
Another family describes ``white noise bursts''  with
uncorrelated, band-limited $h_+(t)$ and $h_\times(t)$ polarization components
and Gaussian envelopes in time. 
Figure~\ref{fig:burstwaveforms} shows a sample of waveforms from
these families~\cite{bib:burstsearchs6} in both the
time domain and time-frequency domain (spectrogram).
In addition, a small family of numerical relativity-derived waveforms~\cite{bib:baiotti}
of neutron star collapse to a black hole
were also used to evaluate performance, again, via detector efficiency
\vs\ $\hrss$.

Over the last decade the burst search algorithms have steadily
improved, as measured by detection efficiency \vs\ $\hrss$,
localization of sources on the sky and
robustness against noise artifacts. LIGO's second all-sky burst
search introduced wavelet decomposition (Waveburst) of the data as an alternative
to a strictly Fourier treatment~\cite{bib:waveburst}. The deliberate admixture of time
and frequency content embodied in wavelets is well suited to the
problem of recovering transient waveforms in data with a smoothly
varying band sensitivity. 
There were additional improvements in pre-search data conditioning,
to remove stationary and quasi-stationary spectral lines, along
with more comprehensive veto studies and waveform consistency
tests based on cross-correlation~\cite{bib:rstatistic}.
Similar searches using the Waveburst program were carried out in the data of the next three LIGO
data runs (S2-S4)~\cite{bib:burstsearchs2,bib:burstsearchs3,bib:burstsearchs4}.

By the time of the 5th science run (S5), the Waveburst algorithm had
evolved to a fully coherent version~\cite{bib:cwb},
which carried out an explicit search over possible time delays among interferometers, corresponding
to putative source sky locations and implemented a maximum likelihood method,
including regulators to disfavor sky locations and polarizations for
a given set of data that favor unlikely relative antenna patterns.
The regulation was motivated by single-detector glitches that could
be construed as true gravitational wave signals coming from a particular
source location and orientation that leads to a small antenna
pattern sensitivity for all but one detector. 
For example, an intrinsically strong source lying near the bisector of a
LIGO interferometer's arms would appear weak to that interferometers, but
possibly strong to the Virgo detector.
While physically possible,
such events are rare, and suppressing them leads to a small (and measurable)
reduction in efficiency while greatly mitigating false alarms due
to instrumental artifacts.
In addition, two alternative algorithmic approaches had evolved
in parallel for use on the S5 data: BlockNormal, described in~\cite{bib:blocknormal} 
and 
Q-Pipeline, described in~\cite{bib:qpipeline}.

All three of these search programs were applied in parallel to 
the first year of S5 data (LIGO only) and upper limits
reported in ref.~\cite{bib:burstsearchs5a} for a variety
of waveform families and parameters. For the 2nd year of the
run, including five months of coincidence with the Virgo VSR1 run,
a burst search  was carried out~\cite{bib:burstsearchs5b} using three different algorithms:
Coherent Waveburst, $\Omega$-Pipeline~\cite{bib:qpipeline} (improved
version of the Q-Pipeline) and Exponential Gaussian Correlator~\cite{bib:egc},
which is based on a matched-filter search using a family of templates similar to sine-Gaussians.

\begin{figure}
\begin{center}
\includegraphics*[width=0.8\textwidth]{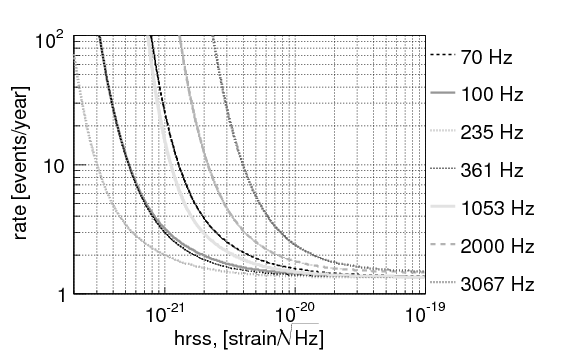}
\caption{Upper limits at 90\% confidence on the rate of gravitational-wave bursts at Earth as a function
of $\hrss$ signal amplitude for selected sine-Gaussian waveforms with $Q=2\sqrt2\pi$. The results include all the 
LIGO and LIGO--Virgo observations since November 2005.}
\label{fig:burstlimitss6}
\end{center}
\end{figure}

The most recently published all-sky burst search (from the LIGO S6 and Virgo VSR2 \&\ VSR3 runs, which
strongly overlapped in time), used the Coherent Waveburst program and
reached the most sensitive limits to date on all-sky gravitational wave bursts. 
Figure~\ref{fig:burstlimitss6} shows 90\%\ confidence upper limits on
burst rate \vs\ $\hrss$ for a family of sine-Gaussian waveforms of different
central frequencies, but with a common quality factor $Q=2\sqrt2\pi$. 
The best sensitivity is achieved among these waveforms for a central
frequency of 235 Hz, reflecting the best average detector spectral behavior,
including non-Gaussian artifacts.

Other all-sky burst searches in the last decade have included joint searches
with the TAMA detector~\cite{bib:bursttama} (in addition to a joint coalescence
search~\cite{bib:cbctama}), with the AURIGA detector~\cite{bib:burstauriga} and
with the GEO 600 detector~\cite{bib:burstgeo}. In addition, there have been
special searches for high-frequency bursts~\cite{bib:bursthighfreq} (central
frequency in the range  $1-6$ kHz in S5 data and for cosmic string cusp bursts~\cite{bib:burstcosmicstring}
(see section~\ref{sec:burstsources}) in S4 data.

\subsubsection{Triggered burst searches}
\label{sec:bursttriggered}

Triggered burst searches permit somewhat improved sensitivity compared to all-sky
searches. With an electromagnetic (or neutrino) trigger, one has not only a relatively
small time window to search, but also a direction on the sky with a known antenna
pattern sensitivity for each detector for any given assumed polarization. This
knowledge allows significant reduction of the ``trials factor'' for the search and
hence a significant reduction in the SNR threshold required for a fixed false alarm rate.

The first triggered burst search carried out with major interferometers was
for a coincidence between GRB 030329 (March 2003 coincided with the LIGO S2 
data run) and excess power in the Hanford H1 (4-km) and H2 (2-km) interferometers
which were operational during the burst's electromagnetic detection. The search
was based on excess power in the cross-correlation~\cite{bib:finnromanomohanty} between the H1 and H2 data
in a three-minute time window, starting two minutes before the detection of the
gamma rays by the HETE-2 satellite~\cite{bib:hete2}. This relatively large time
window was meant to be conservative, to reflect potential uncertainties in
gamma ray burst emission models. Naively, one would expect the gravitational
wave burst's arrival at the Earth to precede that of the gamma rays slightly, as they
are likely secondary particles~\cite{bib:fryeretal}. Although this GRB was
one of the closest with an established redshift from an optical counterpart~\cite{bib:opticalafterglow},
it was nonetheless approximately 800 Mpc distant, making its detection
extremely unlikely, especially with the interferometers of S2 sensitivity~\cite{bib:grb030329search}.
No signal was seen, and upper limits on $\hrss$ as low as $\sim6\times10^{-21}$ Hz$^{-{1\over2}}$
for sine-Gaussian waveforms were placed.

Subsequent searches for gravitational wave bursts coincident with 
detected gamma ray bursts have detected no 
signals~\cite{bib:grbsearchs2s3s4,bib:grbsearchs5,bib:grbsearchs5cbc,bib:grbsearchs6}, as expected,
given detector sensitivities to date. Recent searches~\cite{bib:grbsearchs5,bib:grbsearchs6} for GW bursts coincident with
gamma ray bursts have been carried out with the X-Pipeline~\cite{bib:xpipeline}, which uses coherent
network measures of coherent and incoherent strain energy detected, along with automated threshold
tuning based on measured (non-stationary) noise characteristics. In addition,
explicit searches for compact binary coalescences coincident with the GRBs have been carried out~\cite{bib:grbsearchs5cbc,
bib:grbsearchs6}

An especially interesting gamma ray burst was
GRB 070201, a short hard gamma ray burst with a reconstructed position
consistent with M31 (Andromeda). The absence of any plausible signal
in LIGO data~\cite{bib:grb070201search} at the time of the burst and the nearness of M31 make it very
unlikely that the GRB was a binary merger in M31. More likely it was an
SGR giant flare in M31. Another interesting event coincident with LIGO
data taking was GRB051103 which had a triangulated position consistent with
the relatively nearby M81 galaxy ($\sim$ 3.6 Mpc). Again, no evidence was seen~\cite{bib:grb051103search}
of a gravitational wave burst, strongly suggesting this burst was another
giant flare, making it the most distant flare detected.
A much closer giant flare~\cite{bib:sgr1806} in December 2004 of SGR 1806-20 prompted
a search in LIGO data~\cite{bib:sgr1806search} from the H1 and H2 detectors, which were operating in
an astrowatch mode at the time. No signal was seen, allowing upper limits on the gravitational energy
emitted in the flare to be placed of O(10$^{-8}\msolar$c$^2$), comparable to the
electromagnetic energy emitted by the galactic neutron 
star (see also ref.~\cite{bib:barsearchSGR1806} for a search in Auriga data).
Other searches for soft gamma ray repeaters have been carried out in
S5 data~\cite{bib:sgrsearchs5}, and a special search~\cite{bib:sgr1900search} 
was carried out for the 2006 storm of
SGR 1900+14~\cite{bib:sgr1900storm}, using stacking of LIGO data for individual bursts in the storm.
Additional electromagnetic triggers from six magnetars were used as triggers
for gravitational wave searches in LIGO, Virgo and GEO 600 data 
from November 2006 to June 2009~\cite{bib:burstmagnetarsearch}.
Similarly, a search~\cite{bib:burstvelasearch} was carried out in LIGO data for
a gravitational wave burst coincident with a glitch in the Vela pulsar in August 2006.
Finally, a search~\cite{bib:burstantares} was carried out in LIGO and Virgo data for gravitational wave
bursts coincident with high energy triggers in the Antares underwater neutrino detector~\cite{bib:antares}.

\subsubsection{Using gravitational wave candidates as triggers for electromagnetic follow-up}
\label{sec:burstemfollowup}

For the LIGO S6 and Virgo VSR2-3 runs, there was a concerted effort
to provide prompt gravitational wave triggers for use in follow-up by
electromagnetic telescopes across a broad electromagnetic spectrum from radio waves
to gamma rays. For example, an agreement was reached among the LIGO, Virgo and Swift satellite~\cite{bib:swift}
collaborations to follow up interesting gravitational wave triggers in the LIGO 
and Virgo data with Swift observations using the Swift Target of Opportunity 
program. In addition to this target-of-opportunity arrangement with Swift, 
agreements were made between LIGO/Virgo and a large number of ground-based
astronomical telescopes for
rapid follow-up of low-latency gravitational-wave triggers.  
In the optical band, agreements were reached with 
the Liverpool Telescope~\cite{bib:liverpool}, 
the Palomar Transient Factory~\cite{bib:ptf},
Pi of the Sky~\cite{bib:piofthesky}
Quest~\cite{bib:quest}, 
ROTSE III~\cite{bib:rotse},
SkyMapper~\cite{bib:skymapper}, 
TAROT~\cite{bib:tarot},  
and Zadko~\cite{bib:zadko}. 
In the radio band, an agreement
was reached with LOFAR~\cite{bib:lofar}.

An important goal for S6/VSR3 was to reach
latencies between gravitational wave reception and external release of triggers
of O(hours) or better. That required an aggressive effort not only to collect
data from the LIGO, Virgo and GEO 600 detectors (four sites) in near real-time and
carry out computationally demanding analysis algorithms, but also to gather information
on data artifacts rapidly so that a scientist on call could decide whether or not
a signal candidate merited follow-up by electromagnetic telescopes, in which case
an alert was sent to astronomical partners. 

A critical consideration in sending out an alert is the directional resolution of
the gravitational wave candidate. To lowest order, that directionality comes
simply from triangulation based on time delays among the interferometers.
As discussed in section~\ref{sec:detsens}, a physically allowed 
time delay and uncertainty between a pair of non-colocated interferometers defines 
a annular ring on the sky. The intersections of those annuli for all possible detector pairs
define preferred regions to search. To next order, one can use the antenna
patterns associated with the different interferometer orientations to favor
one or more intersections over others. The formalism for such triangulation
was worked out in some detail by Gursel \&\ Tinto in 1989~\cite{bib:gurseltinto}
and has received intensive study in recent years with the active collaboration among
major interferometers worldwide and with the prospect of moving one of the
advanced LIGO interferometers to Australia or India (see section~\ref{sec:advanceddetectors} above).
A number of algorithmic approaches
have been tried, including that of Coherent Waveburst~\cite{bib:cwb}, described above,
in addition to $\Omega$-Pipeline~\cite{bib:qpipeline}.
Achieved spatial resolutions vary with signal SNR and location,
but a rough estimate for low-SNR signals that could have been seen in the S6/VSR2-3 run
was as much as tens of square degrees. Because an area that large is not well matched to
the fields of view of partner telescopes, ranked tiles on the sky were provided to assist in
prioritizing observing time, where ranking could depend on not only SNR-based likelihood,
but also on the existence in a tile of a nearby galaxy.

The software and network infrastructure for low-latency detection among the multiple interferometers
became available in December 2009, allowing for a short first observing run December 18, 2009 to
January 8, 2010, coinciding with the end of the Virgo VSR2 run. The relatively
poor directionality possible with only the two LIGO detectors made release of
triggers unattractive during the S6 period between VSR2 and VSR3. The second 
observing run for external trigger release took place from September 2, 2010
to October 20, 2010, coinciding with the end of the S6 run. Two trigger pipelines
were used, one based on the Coherent Waveburst program, the other based on
a low-latency coalescence detection baseline called Multi Band Templated Analysis (MBTA)~\cite{bib:mbta}.
Two candidates were detected by the Waveburst program, one on January 7, 2010
and one on September 16, 2010, which was the Big Dog blind injection
discussed above in section~\ref{sec:cbcbigdog}.. Three candidates were detected with the MBTA program
and judged to have enough scientific potential to release triggers to partners. One of
the MBTA triggers arrived during a test period and was not sent to partners. 
Another trigger
on October 6, 2010 had a sky location too close to the Sun to permit follow-up. 
The remaining trigger, which occurred on September 19, 2010, was released to partners.
Results of its image analysis will appear in a forthcoming publication.
Typical delays between apparent signal arrival and the decision on trigger release 
during the run varied from about 20 to 40 minutes. Not all partner telescopes (distributed
worldwide) were positioned to observe favored sky locations for all triggers (or to observe 
at all if in daylight). 
The infrastructure used in the MBTA-based pipeline is described in detail in ref.~\cite{bib:emfollowup},
and the results of the Swift follow-up exercise described in ref.~\cite{bib:burstsearchswift}.
Further publications based on the follow-up exercise with ground-based telescopes are planned.

It was recognized that the chances of successful detection of a gravitational wave
transient in the S6/VSR2-3 runs {\it and} its successful electromagnetic follow-up
were remote, but mounting of the low-latency infrastructure and
establishing the prototype communication protocols with astronomical partners are
expected to pay off in the upcoming advanced detector era, when chances
of success will be far higher and the number of participating partners also higher.

\subsection{Searching for continuous waves}
\label{sec:cwsearches}

As discussed in section~\ref{sec:cwsources}, continuous-wave (CW) gravitational 
radiation detectable by ground-based detectors is expected only
from rapidly spinning neutron stars in our galaxy. Search strategies
for CW radiation vary dramatically with the \apriori\ knowledge one
has about the source. It is useful to classify CW searches into three
broad categories~\cite{bib:dawhitepaper2011}: 1) {\it targeted} searches in which
the star's position and rotation frequency are known, \ie, known 
radio, X-ray or $\gamma$-ray pulsars; 
2) {\it directed} searches in which the star's position is known, but rotation
frequency is unknown, \eg, a non-pulsating X-ray source at the
center of a supernova remnant; and 3) {\it all-sky} searches for unknown
neutron stars. The parameter space over which to search increases
in large steps as one progresses through these categories. In each
category a star can be isolated or binary. For 2) and 3) any unknown binary
orbital parameters further increase the search volume.

In all cases we expect (and have now verified from unsuccessful searches to date!)
that source strengths are very small. Hence one must integrate data over long
observation times to have any chance of signal detection. How much one knows about
the source governs the nature of that integration. In general, the greater that
knowledge, the more computationally feasible it is to integrate data
coherently (preserving phase information) over long observation times, for reasons
explained below.

In principle, a definitive continuous-wave source detection can be
accomplished with a single gravitational wave detector because the source
remains on, allowing follow-up verification of the signal strength and of
the distinctive Doppler 
modulations of signal frequency due to the Earth's motion (discussed below). 
Hence a relatively large number of CW searches were carried out with both bar
detectors and interferometer prototypes in the decades before the major
1st-generation interferometers began collecting data, 
as summarized in ref.~\cite{bib:cwtargeteds1}.
The most sensitive of the resulting upper limits came from bar detectors
in their narrow bands of sensitivity.
The Explorer detector reported~\cite{bib:cwexplorer1} an upper limit on 
spindownless CW signals from the
galactic center of $2.9\times10^{-24}$ in a 0.06-Hz band near 921 Hz, based on
96 days of observation.
A broader-band ($\sim$1 Hz) upper limit of $2.8\times10^{-23}$ 
was also reported~\cite{bib:cwexplorer2} from the Explorer detector
based on a coherent 2-day search that allowing for stellar spindown.
In addition, searches for spindownless CW waves from the galactic center and from
the pulsar-rich globular cluster 47 Tucanae in two 1-Hz bands near 900 Hz
were carried out in Allegro detector data, yielding upper limits~\cite{bib:cwallegro} of
$8\times10^{-24}$. 
Finally, a narrowband (0.05 Hz) 
search~\cite{bib:cwtama} was carried out with the TAMA interferometer near
935 Hz for continuous waves from the direction of Supernova 1987A, with
an upper limit of $5\times10^{-23}$ reported.

\subsubsection{Targeted CW searches}
\label{sec:cwtargeted}

For known pulsars with measured ephemerides from radio, optical, X-ray or $\gamma$-ray
observations valid over the gravitational wave observation time, one can apply
corrections for phase modulation (or, equivalently, frequency modulation) due
to the motion of the Earth (daily rotation and orbital motion), and in the case of binary
pulsars, for additional source orbital motion. For the earth's motion, one has
a daily relative frequency modulation of $v_{\rm rot}/c\approx10^{-6}$ and a much
larger annular relative frequency modulation of $v_{\rm orb}/c\approx10^{-4}$. 
The pulsar community has developed a powerful and mature software infrastructure for
measuring ephemerides and applying them in measurements, using the TEMPO program~\cite{bib:tempo}.
The same physical corrections for Sun and Earth's motion (and Jupiter's motion),
along with general relativistic effects
including gravitational redshift in the Sun's potential and Shapiro delay for
waves passing near the Sun, have been incorporated into the LIGO and Virgo software
libraries~\cite{bib:lal}. 

Three distinct approaches have been used in targeted searches to date:
1) A time-domain heterodyne method~\cite{bib:dupuiswoan} in which Bayesian posteriors are determined on
the signal parameters that govern absolute phase, amplitude and
amplitude modulations; 2) a matched-filter method in which marginalization
is carried out over unknown orientation parameters (``\fstatistic'')~\cite{bib:jks}; 
and 3) a Fourier-domain determination of a ``carrier'' strength along with the strengths
of two pairs of sidebands created by amplitude modulation from the Earth's sidereal
rotation of each detector's antenna pattern (``5-Vector'' method)~\cite{bib:5vector}.

Method 1 has been used in all LIGO and Virgo publications to date on targeted CW 
searches~\cite{bib:cwtargeteds1,bib:cwtargeteds2,bib:cwtargeteds3s4,
bib:cwtargetedcrab,bib:cwtargeteds5,bib:cwtargetedvela},
while method 2 was used in the first LIGO targeted search~\cite{bib:cwtargeteds1} for the 
rapidly spinning millisecond pulsar J1939+2134 and
(in a different implementation) for the recent targeted search for the young Vela 
pulsar~\cite{bib:cwtargetedvela}, along with method 3.
The heterodyne method will be described here for illustration for targeted searches. 

For a rotating rigid triaxial ellipsoid (model for a neutron star), 
the strain waveform detected by an interferometer can be written as
\begin{equation}
\label{eqn:cwhdefinition}
h(t) \quad=\quad F_+(t,\psi)\,h_0{1+\cos^2(\iota)\over2}\,\cos(\Phi(t)) 
\>+\> F_\times(t,\psi)\,h_0\,\cos(\iota)\,\sin(\Phi(t)),
\end{equation}
where $\iota$ is the angle between the star's spin direction and the propagation
direction $\hat k$ of the waves (pointing toward the Earth), 
where $F_+$ and $F_\times$ are the (real) detector antenna pattern response factors
($-1 \le F_+,F_\times \le 1)$ to the $+$ and $\times$ polarizations. $F_+$ and $F_\times$ 
depend on the orientation of the detector and the source, and on 
the polarization angle $\psi$~\cite{bib:cwtargeteds1}. Here, $\Phi(t)$ is
the phase of the signal.

The phase evolution of the signal can be usefully expanded
as a truncated Taylor series:
\begin{equation}
\label{eqn:phasedefinition}
\Phi(t) \quad = \quad \phi_0 + 2\,\pi\left[f_s(T-T_0) + {1\over2}\dot f_s(T-T_0)^2 + 
{1\over6}\ddot f_s(T-T_0)^3\right],
\end{equation}
where
\begin{equation}
\label{eqn:phaseevolution}
T \quad = \quad t + \delta t \quad = \quad t - {\vec r_d\cdot\hat k\over c}+ \Delta_{E\odot}-\Delta_{S\odot}.
\end{equation}
Here, $T$ is the time of arrival of a signal at the solar system barycenter (SSB),
$\phi_0$ is the phase of the signal at fiducial time $T_0$, $\vec r_d$ is the position
of the detector with respect to the SSB, and $\Delta_{E\odot}$ and $\Delta_{S\odot}$ 
are solar system Einstein and Shapiro time delays, respectively~\cite{bib:taylorssb}.
In this expression $f_s$ is the nominal instantaneous frequency of the gravitational
wave signal [twice the star's rotation frequency for a signal created by a rotating star's
non-zero ellipticity, as in equations~(\ref{eqn:ellipticity}-\ref{eqn:hexpected})].

A complex heterodyne is computed from equation~(\ref{eqn:phasedefinition})
with a unit-amplitude complex function that has a phase evolution equal
but of opposite signal to that expected for the signal from 
equation~(\ref{eqn:phaseevolution}). 
The heterodyne is evaluated and downsampled to measured values $B_k$ at
times $t_k$ (once per minute) for
the span of the observation time, 
allowing comparison with expectation for a signal model:
\begin{equation}
y(t_k;\vec a)\quad  = \quad {1\over4}F_+(t_k;\psi)\,h_0(1+\cos^2(\iota))e^{i2\phi_0} 
\>-\> {i\over2}F_\times(t_k;\psi)\,h_0\cos(\iota)\,e^{i2\phi_0},
\end{equation}
where $\vec a$ represents the set of signal parameters $(h_0,\iota,\psi,\phi_0)$.

The joint Bayesian posterior pdf for these four parameters is
defined by 
\begin{equation}
p(\vec a|\{B_k\}) \quad \propto\quad p(\vec a)\times\exp
\left[-\sum_k{R\{B_k-y(t_k;\vec a)\}^2\over2\sigma^2_{R\{B_k\}}}\right]
\times \exp\left[-\sum_k{I\{B_k-y(t_k;\vec a)\}^2\over2\sigma^2_{I\{B_k\}}}\right],
\end{equation}
where $p(\vec a)$ is the prior on $\vec a$, and $\sigma^2_{R\{B_k\}}$ 
and $\sigma^2_{I\{B_k\}}$ are the variances on the real and imaginary
parts of the $B_k$ values. Results are insensitive to reasonable choices
of the prior distribution for $h_0>0$, while chosen priors on $\iota$ and $\psi$ 
depend on the knowledge (more commonly, the ignorance of)
the star's spin axis direction, and the prior on $\phi_0$ is taken
uniform over $[0,2\,\pi]$. One can derive posterior pdfs on any
single parameter by marginalizing over the other three (and over any
nuisance parameters, such as instrumental noise~\cite{bib:dupuiswoan}):
\begin{equation}
p(h_0|\{B_k\})\quad\propto\quad\int\!\!\int\!\!\int p(\vec a|\{B_k\})\,d\iota\,d\psi\,d\phi_0,
\end{equation}
normalized so that $\int_0^\infty p(h_0|\{B_k\})\,dh_0=1$. The resulting curve
represents the distribution of one's degree of belief in any particular
value of $h_0$, given the model of the pulsar signal, the priors for
the pulsar parameters, and the data. The width of the curve indicates the
range of values consistent with one's state of knowledge. In this framework,
there is a probability of 95\%\ that the true value of $h_0$ lies below
$h_0^{95\%}$, where
\begin{equation}
0.95 \quad = \quad \int_0^{h_0^{95\%}}\!\!p(h_0|\{B_k\})\,dh_0\,.
\end{equation}

\begin{figure}[t!]
\begin{center}
\includegraphics[height=12.cm]{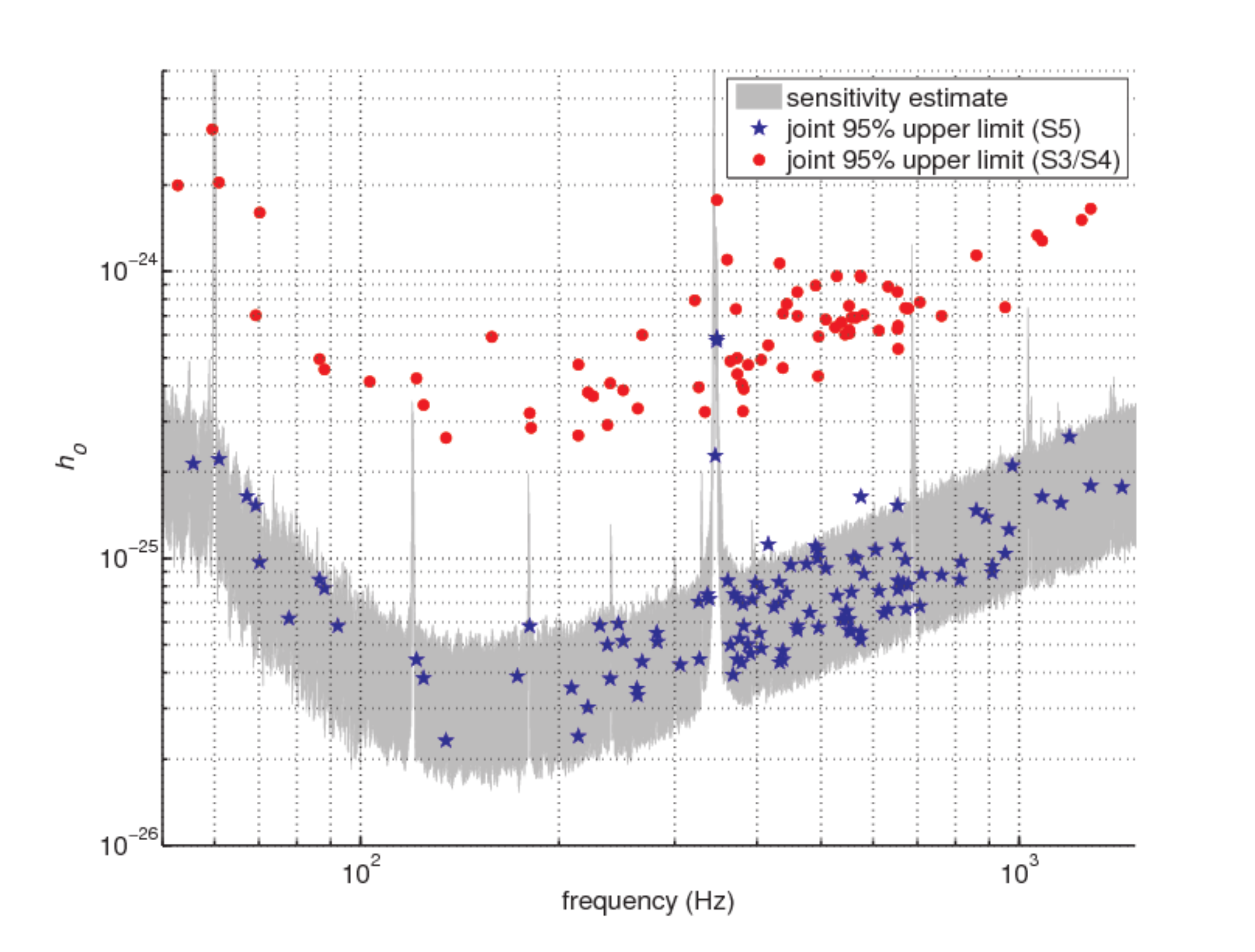}
\caption{Upper limits on $h_0$ for known pulsars from searches in the 
LIGO S5 data~\cite{bib:cwtargeteds5}. The gray band shows the {\it a priori} estimated 
sensitivity range of the search. Also plotted 
are the limits from the S3/S4 search~\cite{bib:cwtargeteds3s4}.
\label{fig:cwtargeteds5}}
\end{center}
\end{figure}

The first application of this method~\cite{bib:cwtargeteds1} 
in LIGO and GEO 600 S1 data (separately to each interferometer) led to upper limits
on $h_0$ of a few times 10$^{-22}$ for J1939+2134 ($f_{\rm rot}$ = 642 Hz).
Comparable upper limits were obtained from the (frequentist) \fstatistic\ method described 
in section~\ref{sec:cwdirected}. Later applications of this method included a variety
of improvements, including coherent treatment of multiple interferometers, marginalization
over noise parameters and a Markov Chain Monte Carlo search method for parameter estimation.
At the same time the number of stars searched in each data run increased, along with
closer partnership with radio and X-ray astronomers who provided ephemerides.
In the S2 data, limits were placed on 28 pulsars, with a lowest strain limit
of $1.7\times10^{-24}$. In the S3 and S4 data (analyzed jointly), limits were
placed on 78 pulsars, with a lowest strain limit of $2.6\times10^{-25}$.
In the S5 data, limits were placed on 116 pulsars, with a lowest strain
limit of $2.3\times10^{-26}$ (J1603-7202). The lowest limit placed on ellipticity
from the S5 search was $7.0\times10^{-8}$(J2124-3358). 
Figure~\ref{fig:cwtargeteds5} shows the resulting upper limits on
$h_0$ for the 116 pulsars searched in LIGO S5 data, along with results
from the previous S4 search.

To date, the direct spindown limit [equation~\ref{eqn:spindownlimit}] has
been beaten for only the Crab and Vela pulsars. 
For the Crab pulsar ($f_{\rm GW}\sim59.5$ Hz), the 95\%\ CL upper limit on $h_0$ 
(based on LIGO S5 data~\cite{bib:cwtargeteds5}) stands at $2.0\times10^{-25}$,
implying that no more than 2\%\ of the star's rotational energy 
loss can be attributed to gravitational wave emission. 
For the Vela pulsar ($f_{\rm GW}\sim22.4$ Hz),
the 95\%\ CL upper limit on $h_0$ 
(based on Virgo VSR2 data~\cite{bib:cwtargetedvela}) stands at $2\times10^{-24}$,
implying that no more than 35\%\ of the star's rotational energy 
loss can be attributed to gravitational wave emission.

\subsubsection{Directed CW searches}
\label{sec:cwdirected}

Unlike targeted searches, where the phase evolution of the signal
is (assumed to be) known precisely enough to permit a coherent integration over
the full observation time, in a directed search one has limited or no
information about the phase evolution of the source, while knowing precisely
the sky location of the star. 
The implied parameter space volume of the search will
then depend sensitively upon the assumed age of the star. For a very young
pulsar, one must search over not only the frequency and first frequency derivative
(spindown), but also over the second and possibly higher derivatives.

To understand the scaling, imagine carrying out a coherent search, where
one wishes to maintain phase coherence over the observation span $T_{\rm obs}$ of no
worse than some error $\Delta\Phi$. From equation~(\ref{eqn:phasedefinition}), one needs
to search over $f_s$ in steps proportional to $1/T_{\rm obs}$,
over $\dot f_s$ in steps proportional to $1/T_{\rm obs}^2$, and
over $\ddot f_s$ in steps proportional to $1/T_{\rm obs}^3$. Hence a search
that requires stepping in $\ddot f_s$ will have a parameter space
volume proportional to $T_{\rm obs}^6$, with search time through the data of
length proportional to $T_{\rm obs}$ that (typically) entails an additional
power of $T_{\rm obs}$. Hence, even when the source direction is precisely
known, the computational cost of a {\it coherent} search over $f_s$, $\dot f_s$ and $\ddot f_s$
grows extremely rapidly with observation time. One can quickly exhaust all available computing capacity
by choosing to search using a $T_{\rm obs}$ value that coincides with a full data
set, \eg, two years. In that case, one may simply choose the largest $T_{\rm obs}$ value
with an acceptable computing cost, or one may choose instead a {\it semi-coherent}
strategy of summing strain powers from many smaller time intervals, as discussed below
in the context of all-sky searches.

Two published directed searches have been carried out to date in LIGO
data. The first~\cite{bib:cwtargetedcrab} was an extremely narrowband search centered on the Crab
pulsar's nominal GW frequency, but allowing for a relative frequency mismatch of O(10$^{-4}$),
in the event that the gravitational wave emitting component of the star spins
at a slightly different frequency from the electromagnetically emitting component,
while constrained by internal torques that tend to enforce co-rotation. Because
of the assumed tight agreement between electromagnetic and gravitational wave
phase evolution, this analysis could scan over a tiny range in $\dot f_s$ and
fix $\ddot f_s$, despite searching over a 6-month observation period during the
LIGO S5 run. A multi-detector implementation of 
the \fstatistic~\cite{bib:multifstatistic} in a search over
$3\times10^7$ templates yielded an upper limit on $h_0$ five times higher
than the corresponding targeted search over the same data, using the
Bayesian heterodyne method described in section~\ref{sec:cwtargeted}.

A second search~\cite{bib:cwcasa}, also based on the \fstatistic\ algorithm,
was carried out for the compact central object (X-ray source) at the center
of the Cassiopeia supernova remnant. 
As discussed in section~\ref{sec:cwsources}, given the $\sim$300-year presumed
age of the star, one can derive a frequency-dependent upper limit on
its strain emission of $1.2\times10^{-24}$, assuming its rotational energy loss 
has been dominated by gravitational wave emission. A coherent search was carried out
in a 12-day period of LIGO S5 data over the band 100-300 Hz, for which it was
expected that the age-based limit could be tested with that data set~\cite{bib:cwcasamethod}.
The resulting upper limits did indeed beat the age-based limit over that
band, reaching a minimum upper limit of $7\times10^{-25}$ at 150 Hz.
That the limits were more than an order of magnitude higher than found
in the full-S5 targeted searches for known pulsars in that band reflected
not only the much shorter observation time used (12 days \vs\ 23 months), 
but also the higher SNR threshold
necessary to apply when searching over $\sim10^{12}$ templates in $f_s$, $\dot f_s$ and
$\ddot f_s$ for a 300-year old star.

Another proposed approach~\cite{bib:xcorrmethod} for directed searches is based on 
cross correlation of data streams, similar to the method used in early GRB
searches (see section~\ref{sec:bursttriggered} above) and very similar to a
method used in directional searches for stochastic gravitational radiation
(see section~\ref{sec:stochsearchesdirected} below) except that it uses finer
frequency binning and includes explicit demodulation of Doppler effects.
Such a method is especially robust against wrong assumptions about phase
evolution and is attractive in searching for a very young object, such as
a hypothetical neutron star remaining from Supernova 1987A~\cite{bib:sn1987a}.

\subsubsection{All-sky CW searches for isolated neutron stars}
\label{sec:cwallsky}

In carrying out all-sky searches for unknown neutron stars, 
the computational considerations grow worse. The corrections for
Doppler modulations and antenna pattern modulation due to the Earth's 
motion must be corrected, as for the targeted and directed searches,
but the corrections are sky dependent, and the spacing of the
demodulation templates is dependent upon the inverse of
the coherence time of the search. Specifically, for a coherence time $T_{\rm coh}$
the required angular resolution is~\cite{bib:cwallskys4}
\begin{equation}
\label{eqn:angres}
\delta\theta \quad \approx \quad {0.5\, {\rm c}\, \delta f\over f\,[v\sin(\theta)]_{\rm max}},
\end{equation}
where $\theta$ is the angle between the detector's velocity relative
to a nominal source direction, where the maximum relative frequency shift 
$[v\sin(\theta)]_{\rm max}/c\approx10^{-4}$, and where $\delta f$
is the size of the frequency bins in the search. For $\delta f=1/T_{\rm coh}$,
one obtains:
\begin{equation}
\delta\theta \quad \approx \quad 9\times 10^{-3}\>{\rm rad}\>\left({30\>{\rm minutes}\over T_{\rm coh}}\right)
\left({300\>{\rm Hz}\over f_s}\right),
\end{equation}
where the nominal $T_{\rm coh}$ = 30 minutes has been used in several all-sky searches to date.
Because the number of required distinct points on the sky scales like $1/(\delta\theta)^2$,
the number of search templates scales like $(T_{\rm coh})^2(f_s)^2$ for a fixed signal frequency $f_s$.
Now consider attempting a search with a coherence time of 1 year for a
signal frequency $f_s=1$ kHz. One obtains $\delta\theta\sim0.3$ $\mu$rad and
a total number of sky points to search of O(10$^{14}$) -- again, for a fixed
frequency. Adding in the degrees of freedom to search over ranges in 
$f_s$, $\dot f_s$ and $\ddot f_s$ makes a fully coherent 1-year all-sky
search utterly impractical, given the Earth's present computing capacity.

As a result, tradeoffs in sensitivity must be made to achieve tractability
in all-sky searches. The simplest tradeoff is to reduce the observation
time to a manageable coherence time, as was done in an 
all-sky search in the 2-month LIGO S2 data (160-730 Hz) based on the \fstatistic\ algorithm,
using a coherence time of 10 hours~\cite{bib:cwfstats2}. 
It can be more attractive, however, to reduce the coherence time still further
to the point where the total observation time is divided into $N=T_{\rm obs}/T_{\rm coh}$,
segments, each of which is analyzed coherently and the results added incoherently
to form a detection statistic. One sacrifices intrinsic sensitivity per 
segment in the hope of compensating (partially) with the 
increased statistics from being able to use more total data. This approach has been used
extensively in all-sky searches. One finds a
strain sensitivity (threshold for detection) that scales as the inverse fourth root 
of $N$~\cite{bib:cwallskys2}. Hence, for a fixed observation time, the sensitivity degrades
as $N^{1\over4}$ as $T_{\rm coh}$ decreases. This degradation is a price one pays
for not preserving phase coherence over the full observation 
time, in order to make the search computationally tractable.

Several semi-coherent algorithmic approaches have been tried, all based on the
``Stack Slide'' algorithm~\cite{bib:stackslide} in which the power from Fourier
transforms over each coherently analyzed segment is stacked on each other after
sliding each transform some number of bins to account for Doppler modulation of
the source frequency.
One algorithm is a direct implementation of this idea called StackSlide~\cite{bib:stackslideimplementation}. Another implementation~\cite{bib:houghmethod} is based on the Hough transform approach~\cite{bib:houghibm}, 
in which for each segment a detection statistic is compared to a threshold and given
a weight of 0 or 1. The sums of those weights are accumulated in parameter space ``maps,''
with high counts warranting follow-up. The Hough approach offers, in principle, somewhat
greater computational efficiency from reducing floating point operations, but its greater
utility lies in its robustness against non-Gaussian artifacts~\cite{bib:cwallskys4}. 
A third implementation, known as PowerFlux~\cite{bib:powerflux},
improves upon the StackSlide method by weighting segments by the inverse variance
of the estimated (usually non-stationary) noise and by searching explicitly over
different assumed polarizations while including the antenna pattern correction factors
in the noise weighting. A comparison of these three algorithms on the S4 data~\cite{bib:cwallskys4}
established a somewhat better detection efficiency for the PowerFlux approach, as one
would expect. (More recent work on an alternative approach to the Hough algorithm
suggests improved efficiency can be achieved~\cite{bib:freqhough}.)

\begin{figure}[htb]
\begin{center}
  \includegraphics[width=6in]{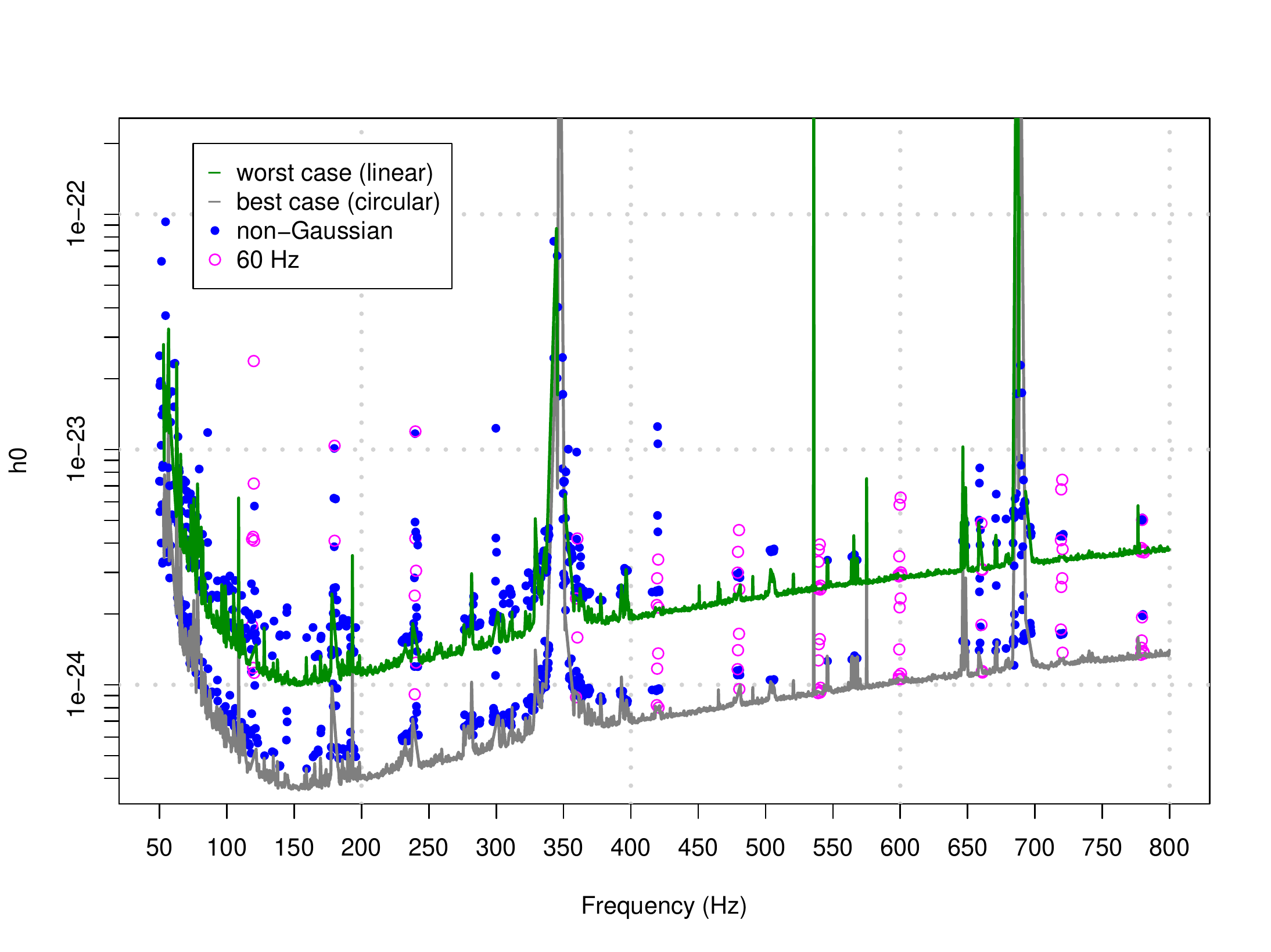}
\vspace{-\abovedisplayskip}
 \caption{All-sky upper limits on unknown sources of continuous waves from the LIGO S5 search~\cite{bib:cwpowerflux2}.
The upper (green) curve shows worst-case upper limits (most unfavorable orientation of
a linearly polarized source) in analyzed 0.25~Hz bands. 
The lower (gray) curve shows upper limits assuming a circularly polarized source. 
The values of solid points (marking non-Gaussian behavior) and 
circles (marking power line harmonics) are not considered reliable. 
They are shown to indicate contaminated bands.}
\label{fig:powerfluxlimitss5}
\end{center}
\end{figure}

The Hough algorithm was used to produce all-sky upper limits in the 200-400 Hz band
of the LIGO S2 data~\cite{bib:cwallskys2}, based on a total of 3800 30-minute segments of data
from the three LIGO interferometers. All three 
of the above semi-coherent methods were used to produce all-sky upper limits in 
the 50-1000 band of the LIGO S4 data~\cite{bib:cwallskys4}. The PowerFlux algorithm was
used to produce all-sky upper limits in the 50-1100 Hz band
of the first eight months of LIGO S5 data~\cite{bib:cwpowerflux1}. The sheer length
of data for the full 23-month S5 run required substantial upgrade of the program which
was then used to produce all-sky upper limits in the 50-800 Hz band of the full data set.
Figure~\ref{fig:powerfluxlimitss5} shows these S5 upper limits based
on a total of more than 80,000 (50\%-overlapped) 
30-minute segments from the H1 and L1 data. In this strict frequentist
analysis the highest 95\%\ CL upper limit obtained from every sky point searched is
shown for the best-case (circular polarization) and worst-case (linear polarization)
assumption for the source orientation. The most recent PowerFlux result~\cite{bib:cwpowerflux2} included a
follow-up procedure of loud candidates, based on a ``loose coherence'' which allows
continuous adjustment of the assumed degree of coherence among Fourier transforms of
successive segments of data~\cite{bib:loosecoherence}. 

These three semi-coherent algorithms compute a detection statistic based on the
strain powers measured in short (30-minute) segments, times over which frequency
modulation effects can be neglected [for a skygrid compatible with equation~(\ref{eqn:angres})]. Another
approach is to use much longer segments and construct a coherent demodulated
power estimate, using the \fstatistic\ algorithm. This approach has been taken
by the Einstein@Home project. Using the same software infrastructure (BOINC)~\cite{bib:boinc}
developed for the Seti@Home project~\cite{bib:seti@home}, Einstein@Home encourages
volunteers to download narrow-band segments of LIGO data and carry out a semi-coherent
\fstatistic\ search over a small patch of sky. Results are automatically returned to 
an Einstein@Home server and recorded, with every set of templates analyzed independently
by host computers owned by at least two different volunteers. LIGO scientists then carry out post-processing
to follow up on promising outliers found. This project has been remarkably successful
in engaging the public ($\sim$225,000 volunteers and $\sim$750,000 host computers to date) 
in forefront science  while making
good use of idle computer cycles to carry out searches that would otherwise exceed
the capacity of available LIGO and Virgo computers. Two searches have been published
to date, one on the LIGO S4 data set~\cite{bib:cwe@hs4} and one on the early 
LIGO S5 data~\cite{bib:cwe@hearlys5}, both of which
achieved comparable sensitivity to other semi-coherent searches while exploring a
significantly larger frequency band (the total computing cost of an all-sky search
is roughly proportional to the cube of the upper frequency bound). Technical constraints
in transferring data to host computers and running on a large variety of computing platforms
have limited achievable sensitivity in previous searches, but new approaches promise
improved performance in future searches~\cite{bib:pletsch,bib:prixshaltev}. In addition, the eventual incorporation of automated
follow-up of outliers on host computers offers the prospect of fully automated
discovery of new continuous wave sources and greatly streamlined post-processing.

It should be noted that the Einstein@Home infrastructure and user base, developed originally
for gravitational wave searches, has now been used successfully to detect pulsars
in radio data from the Arecibo PALFA survey~\cite{bib:e@hradiodetections} and is being
used in searches for gamma-ray pulsars in data from the Fermi
satellite~\cite{bib:fermi9pulsars}. In addition, some of the search techniques developed for gravitational
wave analysis (hierarchical stages of coherent and semi-coherent methods, along with systematic 
search template placement) have been used to 
improve gamma ray search algorithms and have led to new pulsar discoveries~\cite{bib:fermi9pulsars,bib:fermi1pulsar}.
\subsubsection{CW searches for accreting or unknown binary neutron stars}
\label{sec:cwbinary}

For known binary pulsars with measured timing ephemerides,
targeted searches work well, and upper limits have been reported for many 
stars, as described in section~\ref{sec:cwtargeted}.
But searching for known (possibly accreting) binary neutron stars
not exhibiting pulsations  or for entirely unknown binary stars
once again significantly increases the parameter space,
relative to the corresponding isolated star searches,
posing new algorithmic challenges and computing costs. 

Because of its high X-ray flux and the torque-balance
relation for low-mass X-ray binaries [equation~(\ref{eqn:torquebalance})],
Scorpius X-1 is thought to be an especially promising search
target for advanced detectors and has been the subject
of searches in initial LIGO data. From equation~(\ref{eqn:torquebalance}),
one expects a strain amplitude limited by~\cite{bib:cwfstats2}:
\begin{equation}
h \quad \sim \quad (3\times10^{-26})\,\left({540\>{\rm Hz}\over f_{GW}}\right)
\end{equation}
While its rotation frequency
remains unknown, its orbital period is well measured~\cite{bib:scox1period},
which allows some reduction in search space. An early
\fstatistic-based search analyzed
six hours of LIGO S2 data~\cite{bib:cwfstats2}, using an explicit set of
templates in the gravitational wave frequency bands 464--484 Hz and 604--624 Hz
as well as the two relevant binary circular orbit parameters
of the projected semi-major axis (in light-seconds) and an
orbital reference time for the star to cross the ascending node
of the orbit. This pioneering and computationally limited analysis
chose the two restricted search bands based on the drift range of observed
quasi-periodic X-ray oscillations~\cite{bib:cwfstats2}.

More recent searches for Sco X-1 have been carried out using
cross-correlation methods in directional searches for stochastic
radiation, as described below in section~\ref{sec:stochsearchesdirected}.
A potentially more powerful method based on incoherent summing
of orbital sidebands from a coherent search over many orbital
periods has also been proposed~\cite{bib:sidebandmethod}, along
with a frequency-demodulated cross correlation method~\cite{bib:xcorrmethod}.

It should be stated that obtaining more definitive information on
the rotation frequency of Sco X-1 could potentially make the
difference between missing and detecting its gravitational
waves in advanced detector data, by reducing the statistical 
trials factor and thereby the threshold needed to identify
an interesting outlier.

Given the computational difficulty in carrying out a search
over the two unknown orbital parameters of a known binary star
with known period and assumed circular orbit, it should come as no surprise that carrying
out a search over three or more unknown orbital parameters for an
unknown binary star anywhere in the sky is especially challenging.
Two methods have been proposed for carrying out such an all-sky binary search,
which approach the problem from opposite extremes. The first method,
known as TwoSpect~\cite{bib:twospectmethod} carries out a
semi-coherent search over an observation time long compared to
the maximum orbital period considered. Fourier transforms are carried
out over each row (fixed frequency bin) in a $\sim$year-long
spectrogram and the resulting frequency-frequency plot searched
for characteristic harmonic patterns. The second method, known
as Polynomial~\cite{bib:polynomial}, searches coherently using
matched filters over an observation time short compared to the minimum orbital period
considered. A bank of frequency polynomials in time is used for
creating the matched filters, where for a small segment of an orbit,
the frequency should vary as a low-order polynomial.

\subsection{Searching for stochastic waves}
\label{sec:stochsearches}

As discussed in section~\ref{sec:stochsources}, stochastic 
gravitational waves arise from a superposition
of incoherent sources. Because the signal itself is (by
definition) random, it is difficult to separate \apriori\ 
from detector noise, especially given the low strain levels 
expected and especially for sources, such as primordial
background radiation, assumed to be stationary and isotropic.
With a single detector it is challenging to probe 
isotropic astrophysical strain noise much below its
purely instrumental noise.

Nonetheless, one can carry out searches with surprisingly good
sensitivity by exploiting cross-correlations among different 
detectors. By incoherent integration of cross correlated power
over a long observation time, one can probe a stochastic strain
noise power density well below that of a single detector.
This technique was well established and was used in most searches
for gravitational waves carried out with bar detectors and
prototype interferometers before the commissioning of the 1st-generation
interferometers\cite{bib:stochbar1,bib:stochifo1,bib:stochbar2,bib:stochbar3,
bib:stochbar4,bib:stochbar5}

\subsubsection{Searching for an isotropic stochastic background}
\label{sec:stochsearchesisotropic}

In searching for isotropic backgrounds, however,
there is an important constraint. The large geographical separations between detectors,
desirable for suppressing common terrestrial noise, also smear
out correlations of detector responses to random, isotropic signals
for gravitational wavelengths much shorter than the detector
separations.

\begin{figure}[tb]
\begin{center}
\epsfig{file=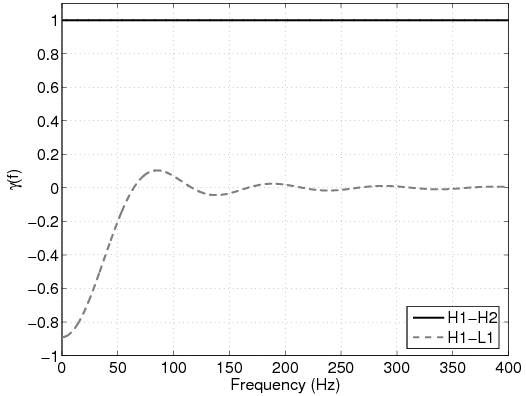,scale=0.7}
\caption{Overlap reduction function $\gamma(f)$ between the LIGO Livingston 
and the LIGO Hanford sites~\cite{bib:stochsearchs1}. The value of $|\gamma|$ is
slightly less than unity at 0 Hz because the interferometer arms are
not exactly co-planar and co-aligned between the two sites.
\label{fig:overlapreduction}}
\end{center}
\end{figure}

To be more quantitative, consider constructing a cross-correlation detection statistic
meant to favor detecting a particular shape to the background
radiation spectral density $S_{\rm GW}$ described in section~\ref{sec:stochsources}.
As above, express the detector outputs $x_i(t)$ as a sum of
instrumental noise $n$ and a gravitational wave background $h$:
\begin{equation}
x(t) \quad = \quad h_i(t) + n_i(t)
\end{equation}
and compute a general cross-correlation~\cite{bib:allenromano} between a pair of
detectors over an observation time $T$ centered on zero:
\begin{equation}
Y \quad \equiv \quad \int_{-T/2}^{T/2}dt_1\int_{-T/2}^{T/2}dt_2\>x(t_1)\,Q(t_1-t_2)\,s_2(t_2),
\end{equation}
where $Q(t_1-t_2)$ is a real filter function chosen to maximize
the SNR for $Y$ and where $Q$ is appreciably non-zero only for time
differences $|t_1-t_2|\ll T$. As has been the case for other gravitational wave
searches, working in the Fourier domain proves convenient. 
In the following, assume~\cite{bib:stochsearchs1} the detector noise 
is (i) stationary over the measurement time $T$; (ii) Gaussian; (iii)
uncorrelated among the detectors; (iv) uncorrelated with the stochastic
gravitational wave signal; and (v) much greater in power at any
frequency than the stochastic gravitational wave background.
With these assumptions, one can show~\cite{bib:allenromano} 
that the expectation value $\mu_Y$ of $Y$ depends only upon the stochastic
signal:
\begin{equation}
\mu_Y \quad \equiv \quad <\!Y\!> \quad = \quad {T\over2}\int_{-\infty}^\infty df\,\gamma(|f|)\,S_{\rm GW}(|f|)\,\tilde Q(f),
\end{equation}
where $\tilde Q(f)$ is the Fourier transform of $Q(t)$, and $\gamma(f)$ [real]
is known as the {\it overlap reduction function}~\cite{bib:christensen,bib:flanagan}, which characterizes
the reduction in sensitivity to an isotropic stochastic background arising
from the separation time delay and relative orientation of the detectors.
For co-located and co-aligned detectors (\eg, H1 and H2), 
$\gamma(f)=+1$ for all frequencies. For separated but co-aligned 
detectors, $\gamma(f)\rightarrow+1$ as $f\rightarrow0$. 
Figure~\ref{fig:overlapreduction} shows, for example, the overlap
reduction function between the LIGO interferometers at Hanford
and Livingston. In this figure $\gamma(f)$ approaches a negative
value as $f\rightarrow0$ because the Hanford $y$-arm is parallel
to the Livingston $x$-arm. It fails to reach $-1$ because
the curvature of the Earth prevents the Hanford $x$-arm from
being parallel to the Livingston $y$-arm by about 27$^\circ$.

The variance in $Y$ determines a detector pair's sensitivity to
a stochastic gravitational wave background:
\begin{equation}
\sigma_Y^2 \quad \equiv \quad <\!(Y-<\!Y\!>)^2\!> \quad \approx \quad 
{T\over4}\int_{-\infty}^{\infty}df\,P_1(|f|)\,|\tilde Q(f)|^2\,P_2(|f|),
\end{equation}
where $P_1(f)$ and $P_2(f)$ are the one-sided strain noise power spectra
of the two detectors.

The optimum shape of $\tilde Q(f)$ depends on the assumed shape of
the stochastic gravitational wave background:
\begin{equation}
\tilde Q(f) \quad \propto \quad {\gamma(|f|)\,S_{\rm GW}(|f|)\over P_1(|f|)\,P_2(|f|)}.
\end{equation}
Under the assumption (used in many LIGO and Virgo analyses for concreteness)
that over the detector's band of sensitivity the normalized 
energy density in stochastic gravitational waves 
[see equation~(\ref{eqn:omegadefinition})] is a constant: $\Omega(f) = \Omega_0$, then
\begin{equation}
\tilde Q(f) \quad \propto \quad {\gamma(|f|)\over |f^3|\,P_1(|f|)\,P_2(|f|)},
\end{equation}
and the signal noise ratio has an expectation value~\cite{bib:stochsearchs1}:
\begin{equation}
<\!\rho_Y\!> \quad = \quad {\mu_Y\over\sigma_Y} \quad \approx \quad 
{3H_0^2\over10\,\pi^2}\,\Omega_0\,\sqrt{T}\,
\left[\int_{-\infty}^\infty df {\gamma^2(|f|)\over f^6\,P_1(|f|)\,P_2(|f|)|}\right]^{1/2}.
\end{equation}

Implementing a search pipeline to calculate these quantities in interferometer time
series data involves a number of technical issues of coherence length choice,
windowing and mitigating correlated instrumental spectral artifacts, 
such as those that affect continuous wave sources discussed above. 
It should be noted that the time lag method of background estimation used
extensively in coalescence and burst searches is also used in 
stochastic searches as an independent cross check. 
A good introduction to these issues can be found in the first
LIGO publication~\cite{bib:stochsearchs1} reporting an upper limit on
an isotropic stochastic gravitational wave background from the 
S1 science run. That first limit was an astrophysically
uninteresting level of $\Omega_0<46$ (assuming recent $H_0$ determinations), 
but the search served as a valuable pioneering exercise in both data analysis
and detector diagnostics for the much more sensitive searches to follow.
With succeeding science runs, the limits on the quantity $\Omega_0$
in the LIGO band of best isotropic stochastic sensitivity 
(exact band is run-dependent because of frequency-dependent sensitivity
improvements) decreased 
to $8.4\times10^{-4}$ in the S3 data~\cite{bib:stochsearchs3},
to $6.5\times10^{-5}$ in the S4 data~\cite{bib:stochsearchs4}, and
to $6.9\times10^{-6}$ in the S5 data~\cite{bib:stochsearchs5}.
Figure~\ref{fig:stochasticlandscape} (discussed in section~\ref{sec:stochsources}) 
shows the S4 and S5 limits superposed together with expectation ranges from 
different possible backgrounds and limits derived from other measurements.
The S5 limit from LIGO interferometers improved for the first time upon
limits derived from big bang nucleosynthesis~\cite{bib:bbnlimits} and from measurements of
the cosmic microwave background~\cite{bib:cmbrlimits}.

\begin{figure}[tb]
\begin{center}
\includegraphics[width=13cm]{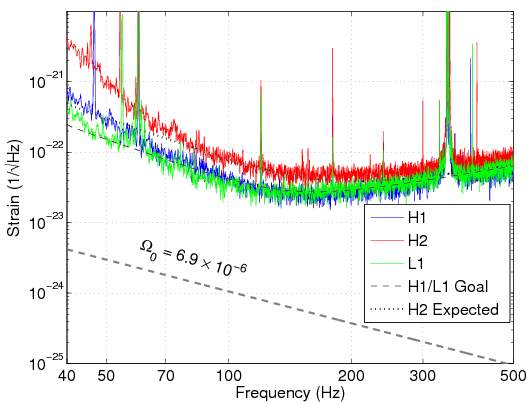}
\caption{Strain amplitude spectral noise densities of the individual LIGO interferometers (S5 data) compared
to the upper limits ($\sim$100 $\times$ lower) placed on a stochastic gravitational wave background by
cross-correlating the interferometer data~\cite{bib:stochsearchs5}. 
\vspace{-\abovedisplayskip}
\label{fig:omegasensitivity}}
\end{center}
\end{figure}

The dramatic improvement in sensitivity to a stochastic gravitational wave
background that comes from cross correlating detector outputs is illustrated
in figure~\ref{fig:omegasensitivity} which shows typical strain sensitivities
of the three LIGO interferometers during the S5 run, together with the much lower
strain noise corresponding to the limit $\Omega_0<6.9\times10^{-6}$.

In addition to the above searches carried out using LIGO data alone,
there have been two searches for isotropic stochastic radiation based on cross correlation with other
detectors: Allegro and Virgo. The Allegro bar detector~\cite{bib:allegro} at Louisiana
State University collected data simultaneously with the LIGO S4 run
and was located only 40 km away from the Livingston L1 interferometer.
This nearness permits an overlap reduction function as high as 95\%
in the high-frequency band ($\sim$900 Hz) where Allegro is sensitive.
Although the resulting limits~\cite{bib:stochligoallegro} 
on $\Omega_0$ ($<\sim1$) were much less sensitive than
those derived from LIGO H1-L1 correlations in the S4 data, the
LIGO-Allegro limits were derived from much higher-frequency data.

The LIGO S5 and Virgo VSR1 data sets were analyzed together~\cite{bib:stochligovirgo} in the
frequency range 600-1000 Hz, using the additional two detector-pair
baselines provided by Virgo (Hanford-Cascina and Livingston-Cascina).
Despite the small overlap reduction functions among these widely
separated detectors, the upper limit of 0.16 on $\Omega_0$ was significantly
better than that found from the earlier Livingston-Allegro measurement.

\subsubsection{Searching for an anisotropic stochastic background}
\label{sec:stochsearchesdirected}

In addition to isotropic stochastic backgrounds (primordial or from
a superposition of astrophysical sources), there may be point sources
or ``patches'' on the sky from which a stochastic background could
be detectable. Potentially interesting point sources include known neutron
stars and the galactic center. Potentially interesting patches include
our galactic plane or the Virgo galaxy cluster. 

Two distinct methods have been used to search in LIGO and Virgo data
for a stochastic gravitational background displaying anisotropy, a background
that could go undetected in the isotropic searches described above. 
The first, known as the Radiometer method~\cite{bib:radiometermethod}, 
is optimized for point sources. The second method, which uses spherical harmonic 
decomposition~\cite{bib:spherharmonicmethod},
is better suited to extended sources. The baseline separation between
a pair of detectors defines an effective aperture, which sets a frequency-dependent
limit on source resolution. For a large number of point sources or a small
number of closely spaced point sources, the radiometer method leads to
interference which can be negative. Hence it is not well suited to detecting
extended sources.

In the radiometer method, one computes a cross correlation similar to that
used in the isotropic search, but does so for a grid of points on the
sky, where for each point on the sky, an explicit correction (dependent
on sidereal time) is made for the time delay between the detectors.
Note that the overlap reduction function that degrades isotropic
searches at high frequencies does not affect the radiometer
search, as there is no averaging over different sky directions, although
antenna pattern corrections still must be applied (usually assuming an unpolarized
source, for simplicity).

Model-dependent upper limits can be placed on the strain power spectrum
from a given source over the search band. In the first publication using
this method on LIGO S4 data~\cite{bib:radiometersearchs4}, sky-dependent
upper limits were placed that ranged from 
$8.5\times10^{-49}$ to $6.1\times10^{-48}$ Hz$^{-1}$. 
In addition, a direction coinciding with the LMXB Scorpius X-1 was chosen
and frequency-dependent limits placed in terms of its RMS strain emission
$h_{\rm RMS} < \approx 3.4\times10^{-24} (f_{\rm GW}/{\rm 200\>Hz})$ for 
$f_{GW}>200$ Hz.

A recent search~\cite{bib:radiosphersearchs5} in the more sensitive LIGO S5 data used both
the Radiometer method and the spherical harmonic decomposition approach.
In the latter approach, one allows for an explicit (smoothly varying)
dependence of the stochastic background strain power spectral density $P(f,\hat\Omega)$ upon sky direction:
\begin{equation}
\Omega_{\rm GW}(f) \quad = \quad {2\pi^2\over3H_0^2}\,f^3\int_{S^2}d\hat\Omega\,P(f,\hat\Omega),
\end{equation}
where $\hat\Omega$ denotes sky direction. In the S5 search it was
assumed that $P(f,\hat\Omega)$ could be factored into an
angular power spectrum $P(\hat\Omega)$ and a frequency-dependent
factor $(f/f_0)^\beta$, where $f_0$ is a reference frequency
and $\beta$ is a spectral index. Two explicit values of $\beta$
were chosen: $\beta=0$ for an astrophysical source and
$\beta=-3$ for a cosmological source~\cite{bib:radiosphersearchs5}.

The spherical harmonic decomposition is taken to be:
\begin{equation}
P(\hat\Omega) \quad \equiv \quad \sum_{\ell,m}P_{\ell m}Y_{\ell m},
\end{equation}
where $Y_{\ell m}$ are normalized spherical harmonic functions
and $P_{\ell m}$ are the coefficients to be determined from the
data. Since the interferometer baselines place an implicit limit
on directional resolution, the sum over $\ell$ is truncated
at an $\ell_{\rm max}$ that depends on the assumed source power spectrum
and on the frequency-dependent effective aperture. In the S5 search,
$\ell_{\rm max}$ was chosen to be 7 for $\beta=-3$ and to be 12 for
$\beta=0$, the differences reflecting the greater importance
of higher frequencies to the astrophysically motivated $\beta=0$ search.
In deriving $P_{\ell m}$ from the data, technical complications
arise from the deconvolution of sky-dependent cross-correlations,
requiring regularization to eliminate low-eigenvalue contributions,
at some expense in signal-dependent bias~\cite{bib:radiosphersearchs5}.

\begin{figure*}[t!]
  \begin{tabular}{ccc}
      \psfig{file=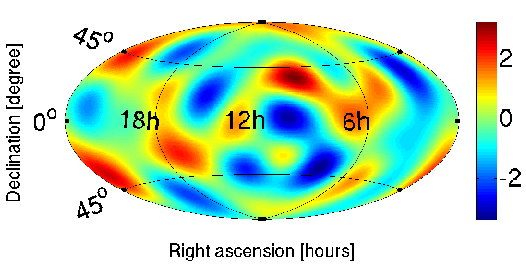,width=2.25in} & 
      \psfig{file=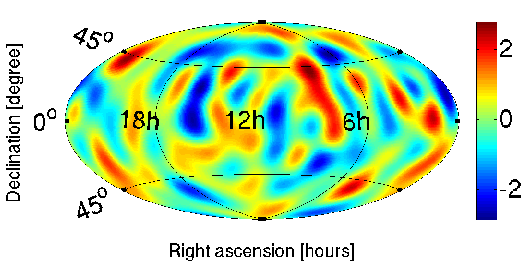,width=2.25in} & 
      \psfig{file=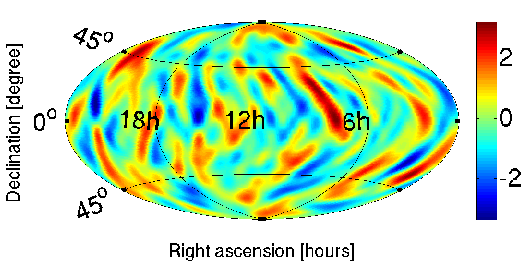,width=2.25in} \\ 
      \psfig{file=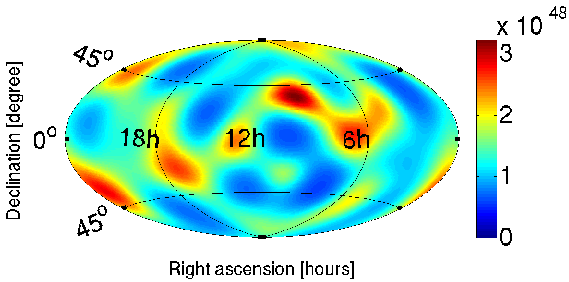,width=2.25in} &  
      \psfig{file=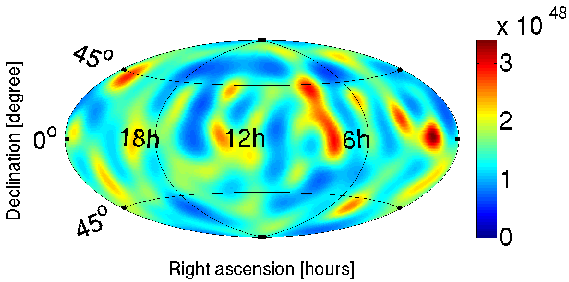,width=2.25in} & 
      \psfig{file=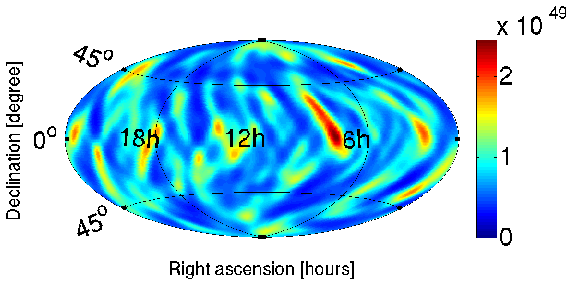,width=2.25in} \\ 
  \end{tabular}
  \caption{
    Top row: Signal-to-noise ratio sky maps for three different directional searches
    for stochastic gravitational radiation: spherical harmonic decomposition for $\beta=-3$ (left)
    and $\beta=0$ (center); and radiometer point-source search for $\beta=0$ (right)~\cite{bib:radiosphersearchs5}.
    Bottom row: The corresponding 90\% CL upper limit maps on strain power
    in units of
    $\unit[]{strain^2 Hz^{-1}sr^{-1}}$ for the spherical harmonic decomposition, and units of
    $\unit[]{strain^2 Hz^{-1}}$ for the radiometer search.
    \label{fig:radiosphermaps}
  }
\end{figure*}

\begin{figure*}[t!]
\begin{center}
  \begin{tabular}{cc}
      \psfig{file=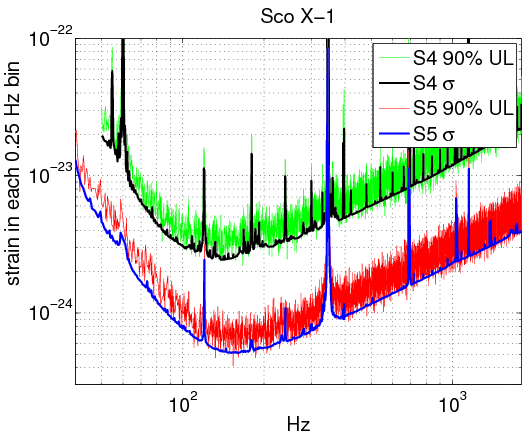,width=3.2in} &
      \psfig{file=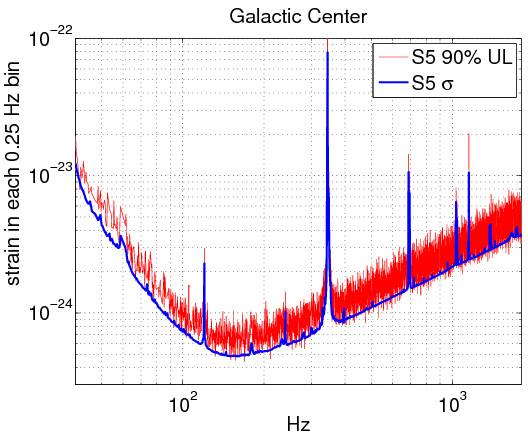,width=3.2in} \\
     \psfig{file=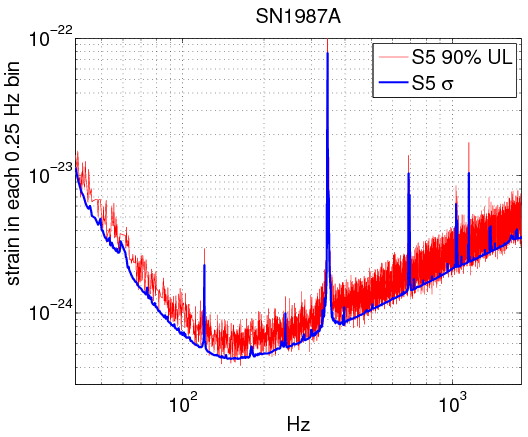,width=3.2in}
  \end{tabular}
  \caption{
    Upper limits (90\% CL) on RMS gravitational wave strain in each $\unit[0.25]{Hz}$ wide bin from the LIGO S5 
    radiometer search~\cite{bib:radiosphersearchs5}
    as a function of frequency for the directions of Scorpius X-1 (upper left), the Galactic Center (upper right) 
    and SN1987A (lower left).
    The previous S4 upper limits for Sco X-1 \cite{bib:radiometersearchs4} are also plotted in the
    upper left panel.
    \label{fig:radiolimits}
    }
\end{center}
\end{figure*}

Figure~\ref{fig:radiosphermaps} shows the SNR maps and resulting
upper limit maps on integrated strain power for the $\beta=(-3,0)$ spherical
harmonic searches and for the radiometer search.
Figure~\ref{fig:radiolimits} shows upper limit strain spectra ($h_{\rm RMS}$)
from the S5 radiometer search for the directions of Scorpius X-1, the 
galactic center and Supernova 1987A. The limits on Scorpius X-1 strain
radiation remain well above that expected from torque balance 
[equation~(\ref{eqn:torquebalance})], but represent the most sensitive
achieved to date.

\section{Summary and prospects}
\label{sec:summary}

To date no gravitational wave detections have been made. While the Hulse-Taylor
pulsar system gives us confidence that appreciable gravitational radiation is indeed emitted
by compact binary star systems, plausible gravitational wave sources are
expected to be weak, and it is unsurprising that detectors built to date have
so far failed to detect them. While not all of the data in hand has been thoroughly
analyzed, particularly for sources for which searches are heavily computationally bound,
such as unknown galactic neutron stars, it seems likely at this point that
2nd-generation (advanced) interferometers will be required for detection in
the 10-10,000 Hz band. 

The prospects for detection by these advanced interferometers are bright.
For the nominal Advanced LIGO design sensitivity, realistic NS-NS, NS-BH
and BH-BH coalescence detection rates are estimated to be 40, 10 and 20
per year, respectively~\cite{bib:cbcratespaper}.

Based on past experience with major interferometers, however,
it is nearly certain that these detectors will not simply turn on at design sensitivity.
Commissioning is likely to be a painstaking, multi-year endeavor interspersed
with short data runs taken with sensitivity intermediate between 1st-generation
and the design sensitivities for advanced detectors. If the optimistic coalescence rates presented
in section~\ref{sec:cbcsources} are accurate, the first detection could occur in
one of those early runs. If the pessimistic estimates are accurate, then discovery
may require several additional years of commissioning, as interferometers 
approach ultimate design sensitivities. 

Unfortunately, as of mid-2012 the funding prospects for space-based gravitational-wave
interferometers are uncertain, at best, despite their tremendous scientific
potential.

On a brighter note, there is a serious potential that stochastic or continuous 
gravitational waves can be detected at several-nHz frequencies using radio telescopes
already operating throughout the world, with discovery possible sooner in
this band than
in the higher-frequency band accessible to ground-based interferometers.

Finally, whether first detection occurs at low or high frequencies, it seems highly
likely that in this decade gravitational wave science will move
from its current status of placing occasionally interesting upper limits
on particular sources to making first discoveries and then beyond to becoming not
only a testing ground for fundamental physics, but also a 
full-fledged field of observational astronomy. 

\section{Acknowledgements}

The author is deeply grateful to colleagues in the LIGO Scientific Collaboration
and Virgo collaboration for years of stimulating discussions and presentations
from which he has benefited in preparing this article. The author also thanks 
Eric Howell for helpful suggestions concerning the manuscript prior to submission 
and thanks LIGO and Virgo for the use of many figures here. In addition, the
helpful corrections and suggestions of the anonymous journal referees are much
appreciated. This work was supported in part
by National Science Foundation Award PHY-0855422.

\medskip
\noindent {\it Cited LIGO reports can be obtained from
     the LIGO Document Control Center: \\{\tt https://dcc.ligo.org/}, \\
     and cited Virgo reports can be obtained from the
     Virgo Technical Documentation System: \\ {\tt https://tds.ego-gw.it/}}.

\end{document}